\documentclass[aps,amssymb,showpacs,11pt,superscriptaddress,eqsecnum,nofootinbib]{revtex4}
\usepackage{epsfig}
\usepackage{graphicx,psfrag}
\newcommand\Cont[1]{\hsp{.4}\vtop{\ialign{##\crcr$\hfil\displaystyle
                   {\hsp{-.4}#1}\hfil\hsp{-.4}$\crcr
                   \noalign{\kern-1.9pt\nointerlineskip\vskip2pt}
                   \leftrighthookfill\crcr}}\hsp{.4}}
\newcommand\hsp[1] {\mbox{\hspace{#1 em}}}
\def\lefthook      {{\vrule height5pt width0.4pt depth0pt}}
\def\righthook     {{\vrule height5pt width0.4pt depth0pt}}
\def\leftrighthookfill{$\mathsurround=0pt \mathord\lefthook
                   \hrulefill\mathord\righthook$}
                   
\newcommand{\maybeeq}[1]{Eq.~(\ref{#1})}
\newcommand{\maybeeqs}[1]{Eqs.~({#1})}

\newcommand{\figs}{Figs.}
\newcommand{\Fig}[1]{Fig.~{\ref{#1}}}
\newcommand{\fig}[1]{Fig.~{\ref{#1}}}
\newcommand{\N}{\mathcal{N}}
\newcommand{\Nt}{\tilde{\mathcal{N}}}
\newcommand{\order}[1]{\mathcal{O}({1\over m^{#1}})}
\newcommand{\pbm}{\protect\boldmath}

\newcommand{\be}{\begin{equation}}
\newcommand{\ee}{\end{equation}}
\newcommand{\bea}{\begin{eqnarray}}
\newcommand{\eea}{\end{eqnarray}}
\newcommand\egal{&\!\!\!=\!\!\!&}
\newcommand\plus{&\!\!\!+\!\!\!&}
\newcommand\minus{&\!\!\!-\!\!\!&}
\newcommand\txt[1]{{\textstyle #1}}
\def\d{{\rm d}}
\def\b{\beta}
\def\a{\alpha}
\def\l{\ell}
\def\O{{\cal O}}
\def\half{{\textstyle {1 \over 2}}}
\def\la{\langle}
\def\ra{\rangle}
\def\Z{\mathbb Z}
\def\eti{\Big|_{\rm ETI}}
\def\tit{\tilde \theta}
\newcommand\p{\partial}
\newcommand\bp{\bar\partial}
\newcommand\NP[1]{:\! #1 \!:}
\renewcommand\t{\theta}

\begin{document}

\title{Height variables in the Abelian sandpile model:\\
scaling fields and correlations}

\author{Monwhea Jeng}
\affiliation{Department of Physics,
Southern Illinois University Edwardsville,
Edwardsville, IL 62025}
\affiliation{Department of Physics, Syracuse University, Syracuse, NY 13244}
\author{Geoffroy Piroux}
\affiliation{Institut de Physique Th\'eorique,
Universit\'e catholique de Louvain, B-1348
Louvain-La-Neuve, Belgium}
\author{Philippe Ruelle}
\affiliation{Institut de Physique Th\'eorique,
Universit\'e catholique de Louvain, B-1348
Louvain-La-Neuve, Belgium}

\begin{abstract}
We compute the lattice 1-site probabilities, on the upper half-plane, of the four height 
variables in the two-dimensional Abelian sandpile model. We find their exact
scaling form when the insertion point is far from the boundary, and
when the boundary is either open or closed. Comparing with the predictions of
a logarithmic conformal theory with central charge $c=-2$, we find a full 
compatibility with the following field assignments: the heights 2, 3 and 4 behave like
(an unusual realization of) the logarithmic partner of a primary field with scaling
dimension 2, the primary field itself being associated with the height 1 variable.
Finite size corrections are also computed and successfully compared with numerical
simulations. Relying on these field assignments, we formulate a conjecture for the scaling 
form of the lattice 2-point correlations of the height variables on the plane,
which remain as yet unknown. The way conformal invariance is realized in this system
points to a local field theory with $c=-2$ which is different from the triplet theory.
\end{abstract}

\maketitle


\section{Introduction}

Sandpile models are open dynamical systems which generically show a very rich
spectrum of critical behaviours, both spatially and temporally. Along with many
other models, they are complex critical models. Because of their non-linear dynamics
and their non-local features, a fixed external driving can trigger events whose scales
follow critical distributions. The references \cite{bak,jensen,dhar-rev}
review the ideas and models involved.

One of the simplest and yet most challenging models is the two-dimensional model originally
proposed by Bak, Tang and Wiesenfeld \cite{btw}, now referred to as the BTW model, or
the unoriented Abelian sandpile model (ASM) after it was shown by Dhar to possess
an Abelian group \cite{dhar-prl}. It is this model we consider in this article.

The most natural formulation of the ASM is in terms of discrete height variables,
taking the four integral values 1 to 4 and located at the sites of a grid. The
configuration of the sandpile, formed by the values of the heights, evolves under a
stochastic dynamics, which eventually reaches a stationary regime, where the
configurations are weighted according to an invariant measure. Numerical simulations
showed that this model has critical properties \cite{btw}.

The first analytic treatment of the ASM was carried out by Dhar \cite{dhar-prl},
where general and important features of the ASM dynamics were obtained. That
work paved the way for numerous exact results.
Among these, one can mention the results on correlations of height 1 variables and
related clusters (the so-called weakly allowed sub-configurations)
\cite{majdhar,bip,mr,jeng3,marr}, on 1-site probabilities of height variables on the
whole plane \cite{priez}, on boundary correlations of height variables
\cite{iv,jeng2,piru2}, on avalanche distributions \cite{ikp1,dharman,ikp2,pik,kp}, on
specific effects of boundary conditions \cite{r,jeng1,piru1}, on finite size
corrections \cite{majdhar2,r}, and on the insertion of dissipation
\cite{mr,jeng1,piru1,jeng2}. Moreover, a number of these references analyze their
results in the light of a possible conformal field description. All available results
are compatible with such a description, and point to a specific logarithmic conformal
field theory, having central charge $c=-2$.

Quite recently, the 1-site probability for the height 2 variable on the upper
half-plane, in fact a disguised 2-point correlator, has been computed exactly in the
scaling limit, the results having been reported in \cite{piru3}. It
was proved, as had been suspected before, that the height 2 variable behaves
logarithmically, unlike the height 1: the height 2 correlations contain logarithms,
while those for the height 1 variables are rational functions of the distances.
Moreover, numerical simulations suggested that the height 3 and 4 variables
have the same behaviour and scale in the same way as the height 2. Finally, the
logarithmic conformal field to which these variables converge in the scaling limit was
identified, and then used to make predictions for the 2-site bulk correlations of
height variables, which, to this date, have not been computed on the lattice
(except for height 1). In the logarithmic conformal
field theory terminology, the scaling limit of the height 2 (and 3 and 4) is a
logarithmic field, which is the partner of the field to which the height 1 scales.

Our purpose in this article is two-fold. First, we want to give the details of the
analysis reported in \cite{piru3}. It involves a lattice part,
where the probability of interest is computed in the ASM, and a conformal theory
part, where the relevant 4-point correlation function with the intended scaling field
is calculated and then compared with the lattice results (and simulations). The
details on both, too long and therefore omitted in \cite{piru3}, will be given
here. Second, we have extended the analysis to include the height 3 and 4
variables. Our new exact results confirm the behaviour conjectured in \cite{piru3}.
Since the height 3 and 4 variables each scale like linear combinations of the
height 1 and 2 variables, any multisite height probability can be reduced, in the 
scaling limit, to multisite probabilities involving heights 1 and 2 only.  
These results should close the long chapter devoted to the height probabilities in
the ASM and their description by a field theory\footnote{With 
the usual reservations: we have identified the fields describing 
the height variables provided that indeed such an identification exists at all, i.e. is
compatible with every single lattice joint
probability.}.

That the critical properties of the ASM can indeed be described by a conformal
field theory remains for us a major motivation, for two reasons. First, the
description by an underlying conformal field theory can be considered as an actual 
and explicit solution of the lattice model. Second, the ASM would provide the first
instance where the correspondence between a logarithmic conformal field theory and a
lattice realization is most clearly exposed. We believe that such a correspondence, so
common for usual equilibrium lattice models, should inevitably clarify tricky
issues on both sides.

The article is organized as follows. In Section II, we define the sandpile model we
consider, review its basic features and recall the elements which prove useful for
what follows.

In Sections III and IV, we review the graph theoretic tools used for the derivation
of the 1-site height probabilities on the plane, in a way that allows the formalism to
be applied to the upper half-plane (UHP). The calculation presented in
Section III for the height 2 is identical to that of Priezzhev \cite{priez}, but our
derivation for the height 3 in Section IV is different. The method we use leads to
the same results as Priezzhev's but has an interesting corollary, since it leads to an
exact relation between the probabilities $P_2$ and $P_3$, a relation unnoticed so
far. Moreover, based on very strong numerical evidence, we conjecture exact and simple 
formulas for $P_2$ and $P_3$, similar to that for $P_1$, namely, simple polynomials in $1
\over \pi$. Although we could not provide a proof for them, we have checked that these
formulas are accurate to one part in $10^{12}$.

Section V applies the formalism developed in the previous two sections to the upper
half-plane, with two homogeneous boundary conditions on the real axis: open and
closed. The resulting formulas are then collected in Section VI: the four
probabilities $P_2^{\rm cl}(m)$, $P_3^{\rm cl}(m)$, $P_2^{\rm op}(m)$ and 
$P_3^{\rm op}(m)$, where $m$ is the distance of the reference site to the boundary,
are explicit but complicated expressions, involving multiple integrals and
summations.

In Section VII, we proceed to the asymptotic analysis of the integral expressions just
derived. These expressions have expansions in inverse powers of $m$, with coefficients
which are polynomials in $\log{m}$. Anticipating that the scaling fields
corresponding to the height variables have dimension 2, we restrict the analysis to
the terms of order 2 in $m^{-1}$. The dominant terms in the probabilities are of order 
$m^{-2}$, as expected, and are computed exactly. The final results are summarized in Section
VIII. The reader interested in the results, but not in the technical details of the
calculations, can safely go straight to Section VIII.

Section IX starts the conformal field theoretic side of our work. In order to assess
the effect of a change of boundary condition on the 1-site probabilities, we
first compute the correlation function $\la \mu(z_1) \mu(z_2) \psi(z,\bar z)\ra$ on
the UHP, where the $\mu$ fields change the boundary condition at two points $z_1$ and
$z_2$ on the real axis, and where the non-chiral field $\psi(z,\bar z)$ is supposed
to describe the scaling limit of the height variables (2, 3, or 4).
The above 3-point field correlator enables us to relate to each other the two UHP  
1-point functions $\la \psi\ra$ in front of a boundary which is either fully open or
fully closed. When $\psi$ is taken to be the logarithmic partner of a dimension 2
primary field $\phi$, this relation exactly reproduces the way the probabilities
$P_i^{\rm op}(m)$ and $P_i^{\rm cl}(m)$ are related on the lattice.

The identification with $\psi$ is further confirmed in Section X, where the finite
size corrections predicted by the conformal theory are compared to the results of
numerical simulations of the lattice model, showing an excellent agreement.

Section XI reconsiders the field correlation functions established before
from a purely conformal field theory point of view. The local logarithmic conformal
theory with $c=-2$, believed to be relevant here, has a well-known realization in
terms of free symplectic fermions $\t, \tilde \t$ \cite{gur,gk,k}. The primary
field $\phi$ describing the height 1 variable in the ASM has been well studied and
is known to be expressible as a local field in the $\t, \tilde \t$ theory. We show
that the field $\psi$, corresponding to the heights 2, 3 and 4, has no local
realization in the $\t, \tilde \t$ theory. It follows that the well-studied triplet
theory with $c=-2$ \cite{gk} is not the appropriate logarithmic conformal field theory 
to describe the sandpile model.

Finally, relying on the field identifications obtained before, we discuss in Section
XII the height joint probabilities on the plane, and formulate conjectures for the 2-site
lattice probabilities on the plane, unknown so far. These conjectures compare well
with the results obtained from simulations.


\section{The Abelian sandpile model}

We consider a finite, two-dimensional grid $\cal L$ 
and attach to each site $i$ a random variable $h_i$, representing 
the height of sand at that site and taking integer values 
1, 2, 3, ... A configuration $\cal C$ of the sandpile is 
the set of values $\{h_i\}$ for all sites. The probability 
distribution over the height variables or equivalently, over the configurations, evolves
dynamically from some initial distribution $P_0({\cal C})$. 

The discrete dynamics, itself stochastic, is completely defined in terms of a toppling
matrix $\Delta$. This matrix has integral entries, strictly positive on the
diagonal, negative or zero off the diagonal, and moreover satisfies the condition
that its row sums $\sum_j \Delta_{ij}$ are non-negative. A configuration is called stable if its height values satisfy $h_i \leq \Delta_{ii}$ for all $i$.  

At time $t$, the dynamics takes the stable configuration ${\cal C}_t$ into a new
stable configuration ${\cal C}_{t+1}$ as follows. To the configuration ${\cal
C}_t$, we first add a grain of sand at a random site $i$ by setting $h_i \to h_i + 1$
(we assume that the site $i$ is chosen randomly with a uniform distribution on $\cal
L$). This new configuration, if stable, defines ${\cal C}_{t+1}$. If it is not
stable, the site $i$ which has been seeded has now a height $h_i = \Delta_{ii} + 1 >
\Delta_{ii}$, and, as consequence, topples: it loses $\Delta_{ii}$ grains of sand,
the other sites $j$ each receive $-\Delta_{ij}$ grains, and $\sum_j \Delta_{ij}$
grains are dissipated. In other words, when the site $i$ topples, the heights
change according to
\be
h_j \to h_j - \Delta_{ij}, \qquad \forall j \in {\cal L}.
\label{update}
\ee

After the site $i$ has toppled, other sites can be unstable (have heights $h_j > \Delta_{jj}$),
in which case they too topple according to the same toppling rule (\ref{update}).
Once all unstable sites have been toppled, one is left with a stable configuration,
${\cal C}_{t+1}$. One can show \cite{dhar-prl} that provided there are dissipative
sites in $\cal L$ (namely, sites $k$ for which $\sum_j \Delta_{kj} > 0$),
this dynamics is well-defined:  the order in which the unstable
sites are toppled does not matter, and the new stable configuration ${\cal C}_{t+1}$
is reached after a finite number of topplings. 

In the usual BTW model, which we consider here, the grid $\cal L$ is a
rectangular portion of the discrete plane $\Z^2$, and $\Delta$ is the discrete
Laplacian on $\cal L$ with appropriate boundary conditions on $\p{\cal L}$. Each
boundary site $k$ can be either open or closed, depending on whether it is
dissipative or not. Explicitly, we take the toppling matrix to be
\be
\Delta_{ij} = \cases{z_i & if $i=j$ is either a bulk site or a closed boundary
site,\cr
4 & if $i=j$ is an open boundary site,\cr
-1 & if $i,j$ are nearest neighbours, \cr
0 & otherwise,}
\ee
where $z_i$ is the coordination number of $i$ in $\cal L$, equal to 4 for a bulk
site, to 3 for a boundary site and to 2 for a corner site. The toppling rule
(\ref{update}) then says that when a bulk or a closed boundary site topples, the sand
grains it loses are all transferred to the nearest neighbours, and when an open
boundary site topples, it loses 4 grains of sand, $z_i$ of them going to the
nearest neighbours, and the others falling off the sandpile. In the latter case, it is
convenient to think of the sand grains falling off the pile as being transferred to a
sink site, connected to all dissipative sites. Bulk and closed boundary sites are
conservative ($\sum_j \Delta_{kj} = 0$), while open boundary sites are dissipative 
($\sum_j \Delta_{kj} > 0$).

Because it involves adding a sand grain at a random site, the dynamics is
stochastic and can be written in terms of a probability transition matrix $p$. The
matrix $p$ gives rise to a master equation, $P_{t+1}({\cal C}) = \sum_{{\cal C}'}
p_{{\cal C},{\cal C}'} \, P_{t}({\cal C}')$, which specifies the time evolution of the
probability measure on the configurations. 

The long-time behaviour of the sandpile is described by the time-invariant
probability measures. For the BTW model as defined above, Dhar has shown that there
is a unique time-invariant measure $P^*_{\cal L}$ \cite{dhar-prl}. This measure
assigns all stable configurations their probability of occurrence, when the
dynamics has been applied for a sufficiently long time so that the system has set
in the stationary regime. The thermodynamic limit of $P^*_{\cal L}$
is what we want to compare with a conformal field theoretic measure.

The number of stable configurations is $\prod_i \Delta_{ii}$, but only a small
fraction of them keep reappearing under the dynamical evolution. The transient
configurations are not in the repeated image of the dynamics, and occur only a
finite number of times. As a consequence, they all have a zero measure with respect to
$P^*_{\cal L}$. A simple example of a transient configuration is $h_i=1$ for all
$i$. The non-transient configurations are called recurrent and asymptotically occur
with a non-zero probability. 

For the BTW model, the recurrent configurations all occur with equal probability under
the dynamics \cite{dhar-prl}. The invariant measure $P^*_{\cal L}$ is thus a simple
uniform distribution, but its support, the set of all recurrent configurations, is not
simple, which explains why exact calculations in the ASM are notoriously difficult. On
general grounds, it can be shown that the number of recurrent configurations is equal
to $\N = \det \Delta$ \cite{dhar-prl}. For the particular model discussed here, this
number is roughly equal to $3.21^{|{\cal L}|}$ (up to surface terms), an exponentially
small fraction of the $4^{|{\cal L}|}$ stable configurations.

Transient and recurrent configurations can be distinguished by a criterion based on forbidden  
sub-configurations (FSC's): a subset of sites ${\cal K} \subset {\cal L}$ is said to
form a forbidden sub-configuration if $h_{i}\le -\sum_{j  \in {\cal
K}\backslash\{i\}}\Delta_{ji} = \#\{\hbox{neighbours of $i$ in $\cal K$}\}$, for all
sites $i$ of ${\cal K}$. For instance, two neighbour sites with height 1 form an FSC. 
Then a configuration is recurrent if and only if it contains no FSC \cite{majdhar2}.  

A practical way to test whether a configuration is recurrent or transient is to use the
burning algorithm \cite{dhar-prl}. At time 0, the sink site is the only site 
to be burnt: we define ${\cal K}_{0}={\cal L}$ to be the set of unburnt sites at time
0. The sites $i$ of ${\cal K}_{0}$ such that $h_{i} > -\sum_{j\in{\cal K}_0\backslash
\{i\}} \Delta_{ji} =  \#\{\hbox{neighbours of $i$ in ${\cal K}_0$}\}$ are burnable at
time 1 (they can only be dissipative sites, namely, open boundary sites). So we burn
them, obtaining a set  ${\cal K}_1 \subseteq {\cal K}_0$ of unburnt sites at time 1.  
Then the sites of ${\cal K}_{1}$ which are burnable at time 2, {\it i.e.} those 
satisfying $h_{i} > -\sum_{j\in{\cal K}_1\backslash \{i\}} \Delta_{ji} =  
\#\{\hbox{neighbours of $i$ in ${\cal K}_1$}\}$, are burnt. This leaves 
a set ${\cal K}_2 \subseteq {\cal K}_1$ of unburnt sites at time 2. This burning
process is carried on until no more sites are burnable, which means that ${\cal 
K}_{T+1}={\cal K}_{T}$ for a certain $T$. Then the configuration is recurrent if and
only if all sites of ${\cal L}$ have been burnt (${\cal K}_{T}=\emptyset$). 
Otherwise ${\cal K}_{T}$ is an FSC, and the configuration is transient.

The burning algorithm allows the definition of a unique rooted spanning tree on a graph
${\cal L}^{\star}$, from the path followed by the fire in the lattice
\cite{majdhar2}. The graph ${\cal L}^{\star}$ has the sites of ${\cal L}$ and the sink as
vertices, and has links defined by $\Delta$: an off--diagonal entry $\Delta_{ij} = -n$
means there are $n$ bonds connecting the sites $i$ and $j$, and each site $i$ is
connected to the sink by a number of bonds equal to $\sum_{j \in {\cal L}} \Delta_{ij}
\geq 0$, namely the number of grains dissipated when  
$i$ topples. At time 0, the sink is the only burnt site and forms the root of the tree. In
the next steps, the fire propagates from the sink to those sites which are burnable at time
1, then from the sites which have been burnt at time 1 to those which are burnable at time
2, and so on. If a site burns at time $t$, it catches fire from one among its
neighbours that were burnt at time $t-1$ (or from the sink site at time 1). In case
there are more than one of these, a fixed ordering prescription is used to decide along
which bond the fire actually propagates \cite{majdhar2}. The collection of all
bonds forming the fire path defines a spanning tree, rooted in the sink, and growing
towards the interior of the lattice ${\cal L}$. Conversely, to every spanning tree
---from Kirchhoff's theorem, their number also equals $\N = \det\Delta$--- one can
associate a unique recurrent configuration if the ordering prescription 
mentioned above is known. 

Since there is a one-to-one mapping between the recurrent configurations and the
spanning trees, the latter provide alternative variables for the description of
the ASM. So we trade the random height variables for random spanning trees, on which
the invariant measure $P_{\cal L}^*$ is uniform. Although the local height variables
look more natural, they are strongly correlated over the whole lattice, since
their values have to make up a recurrent configuration. Their long-range correlations
are encoded in the global structure of the spanning trees. Thus height variables
are local but constrained while spanning trees are unconstrained but global. For
most practical calculations, like those presented here, spanning trees are more
convenient.
 
We want to compute one-site height probabilities, namely the four quantities
$P_{\cal L}^*(h_{i_0}=a)$ for $a=1,2,3,4$. On a finite grid $\cal L$, it amounts to
count the recurrent configurations with a height $a$ at position $i_0$. As we will
use the spanning tree description, the first task is to characterize those spanning
trees which correspond to recurrent configurations with a height $a$ at site $i_0$.
This was done for the first time in \cite{priez}, and was recently reviewed,
with two different derivations, in \cite{jeng2,piru2}. That part is not difficult,
and we only quote the results.

{}From the burning algorithm, the spanning tree grows from its root to the open boundary 
sites and then towards the interior of $\cal L$. A site $i$ is called a predecessor
of another site $i_0$ if the unique path on the spanning tree from $i$ to the root
passes through $i_0$ (equivalently, the fire has propagated from $i_0$ to $i$).
Then the probabilities that the site $i_0$ has height $a$ are given by
\be
P_{\cal L}^*(h_{i_0}=a) = P_{\cal L}^*(h_{i_0}=a-1) + {X_{a-1} \over (5-a)\N},
\qquad a=1,2,3,4,
\label{1siteprob}
\ee
where $P_{\cal L}^*(h_{i_0}=0) = 0$ by definition, and where $X_k$ is the number
of spanning trees on $\cal L$ where the site $i_0$ has exactly $k$ predecessors among
its nearest neighbours (so for $k=0$, $i_0$ is a leaf of the tree). 

While the calculation of $X_0$ is relatively easy (it was done first in
\cite{majdhar} without the spanning tree description), that of the other $X_1,X_2$ and
$X_3$ is much harder. The basic reason for this is that the spanning trees involved
in $X_0$ are subjected to a local constraint, while those counting for $X_k$,
$k>0$, must satisfy a global constraint. In the case of $X_1$ for instance, a spanning
tree is counted only if the firepath comes to $i_0$, then goes on straight to one
neighbour of $i_0$ but does not eventually come back to another neighbour of $i_0$
after a long tour through the lattice. 

Priezzhev has devised a technique to compute the numbers $X_k$ \cite{priez}. It
turns out that they can all be expressed ---in a complicated way--- in terms of
entries of the inverse toppling matrix $\Delta^{-1}$, that is, in terms of the
discrete Green function on $\cal L$ for appropriate boundary conditions. The
thermodynamic limit of the above probabilities, when the grid $\cal L$ becomes
infinite, then simply corresponds to taking the discrete Green function on the
corresponding infinite lattice. For simplicity, we denote the corresponding measure by
$P = \lim_{|{\cal L}| \to \infty} P_{\cal L}^*$.

The calculation of the 1-site probabilities, equivalently of the numbers $X_k$, was
carried out by Priezzhev in the case of the infinite plane ${\cal L} \to \Z^2$,
thereby obtaining the 1-site bulk probabilities $P_1, P_2, P_3$ and $P_4$. In the
following, we carry out the calculation for the upper half-plane ${\cal L} \to
\Z_+ \times \Z$, when the boundary is either fully open or fully closed. The
corresponding probabilities $P_a(m)$ depend on the distance $m$ of the reference
site $i_0$ to the boundary. That dependence is precisely what enables us to make
the comparison with field theory correlations, and to assess the scaling form of the
random height variables $\delta(h_{i_0}-a)$.


\section{Height two on the plane}

In this section we review Priezzhev's calculation of the higher height
probabilities on the plane, since these calculations are the starting point for our
calculations in the UHP \cite{priez}. We sketch the key points here; the reader is
referred to Priezzhev's original paper for details.

\subsection{General Methods}

Majumdar and Dhar \cite{majdhar} had already shown in 1991 that the height one
probability at a site $i_0$ is given by
\begin{equation}
P_1={{X_0}\over{4\N}},
\label{eq:P1.predecessor}
\end{equation}
where $X_0$ is the number of spanning trees in which the site $i_0$ is a leaf, and
$\N = \det\Delta$ (see Section II).

The restrictions on $X_0$ are local: a spanning tree is in $X_0$ if and only if no neighbour of
$i_0$ has an arrow pointing to $i_0$. Local restrictions on the spanning tree can be
computed as finite-dimensional matrix determinants by the Majumdar-Dhar method. 
We briefly describe the key points of this method, but the reader is directed to the
original paper for a detailed description \cite{majdhar}. 

The set of spanning trees is characterized by the toppling matrix $\Delta$, which
specifies the bonds on the lattice. When we want a subset of
spanning trees with local restrictions, those restrictions can be imposed by changes
in a finite number of entries of the toppling matrix. For example, to remove the bond
between neighbours $i$ and $j$, we simply change $\Delta_{ii}$ and $\Delta_{jj}$ to 3,
and set $\Delta_{ij}$ and $\Delta_{ji}$ to 0. Calling the new toppling matrix $\Delta'$, 
Kirchhoff's theorem implies that $\det\Delta'$ is the number of spanning trees on 
the lattice with the bond $ij$ removed, or equivalently, the number of spanning
trees on the original lattice which do not use the bond connecting $i$ and $j$.

\begin{psfrags}
\psfrag{i0}{$i_0$}
\psfrag{i1}{$i_1$}
\psfrag{j1}{$j_1$}
\psfrag{j2}{$j_2$}
\psfrag{j3}{$j_3$}
\psfrag{j4}{$j_4$}
\psfrag{a}{$a$}
\psfrag{b}{$b$}

\psfrag{=}{$=$}
\psfrag{-}{$-$}
\psfrag{+}{$+$}
\psfrag{=2}{$=\;2$}
\psfrag{-2}{$-\:2$}
\psfrag{+2}{$+\;2$}
\psfrag{-4}{$-\:4$}

\begin{figure}[tb]
\includegraphics{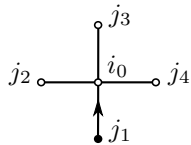}
\caption{Labelling of sites.}
\label{fig:Site.Labels}
\end{figure}

To compute $X_0$, we have to count the spanning trees in which $i_0$ is connected to
only one of its neighbours. The simplest way to do this is to remove the three bonds 
connecting $i_0$ to three neighbours, say $j_2,j_3,j_4$ (see Fig. \ref{fig:Site.Labels})
and to define the new matrix $\Delta' = \Delta + B$, with
\be
B = \pmatrix{-3 & 1 & 1 & 1 \cr 1 & -1 & 0 & 0 \cr 1 & 0 & -1 & 0 \cr 1 & 0 & 0 &
-1},
\ee
where the rows and columns are labeled by $i_0,j_2,j_3,j_4$ (all the other entries
of $B$ are zero). Then by symmetry, $X_0 = 4 \det\Delta'$.

$\Delta'$ is an infinite-dimensional matrix, so its determinant does not make much sense. However the probability $P_1$ only requires the ratio of two infinite determinants. Noting that $B$ has finite rank, and defining $G \equiv \Delta^{-1}$, one easily sees that 
\begin{equation}
P_1 = {{\det(\Delta')} \over {\det(\Delta)}} = \det (1 + BG)
\end{equation}
in fact reduces to a 4-by-4 determinant.

It is easy to compute this finite-dimensional determinant. In the limit of an
infinite planar lattice, $G$ is the Green matrix of the discrete Laplacian, whose values and properties are well-known \cite{Green.function}. Using them, Majumdar and Dhar were able to show that \cite{majdhar}
\begin{equation}
P_1 = {{2(\pi-2)}\over{\pi^3}} \simeq 0.07363.
\end{equation}

The other $X_k$ for $k>0$ are much harder to calculate. We discuss $X_1$ in
this Section, and $X_2$ in Section \ref{sec:HeightThreeBulk}.

\subsection{Decomposition into local, loop, and $\theta$-graphs}

\begin{figure}[tb]
\psfrag{X1/12}{$\displaystyle{X_1\over12}\:=$}
\psfrag{N1}{$\mathcal{N}_{\text{local}}$}
\psfrag{N2}{$\mathcal{N}_{\text{loop}}$}
\psfrag{N3}{$\;\mathcal{N}_{\theta}$}
\includegraphics{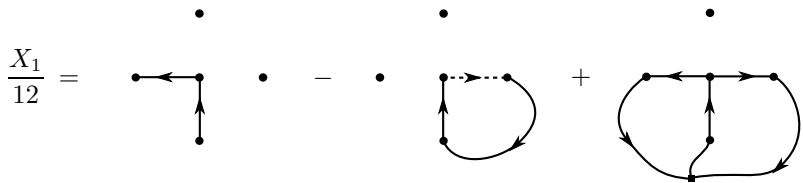}
\caption{Breaking $X_1$ into local, loop, and $\theta$-graph terms.}
\label{fig:X1.Decomposition}
\end{figure}

Since all four directions are symmetric, to find $X_1$, we can, without
loss of generality, assume that the single predecessor is the site to the south of 
$i_0$. \Fig{fig:Site.Labels} shows the labels of the 4 neighbouring sites, with the
predecessor site indicated by a large solid circle, and the non-predecessor sites by
large open circles. By enumerating possible arrow configurations in $X_1$, Priezzhev
showed that $X_1$ can be broken up into the terms in \fig{fig:X1.Decomposition},
or the following equation:
\begin{equation}
X_1=12\left( \N_{\rm local}-\N_{\rm loop}+\N_\theta \right),
\label{eq:X1}
\end{equation}
where the $\N$'s are positive (they count certain graphs). The reader is referred to
\cite{priez} for the derivation of (\ref{eq:X1}), or to Section \ref{sec:HeightThreeBulk} 
for derivations of similar decompositions for $X_2$. In this Section, we take the 
decomposition in Equation (\ref{eq:X1}) as given, and only discuss what the different
terms mean, and how to evaluate them.

In the first term in \fig{fig:X1.Decomposition}, the arrow from $j_1$ must point to
$i_0$, and the arrow from $i_0$ must point to $j_2$, while the sites $j_3$ and $j_4$
are disconnected from $i_0$. Since the conditions on the spanning tree are local,
the size of this set of spanning trees can be calculated with the Majumdar-Dhar
method (or Kirchhoff's theorem). The $B$ matrix that corresponds to these conditions is
\begin{eqnarray}
\nonumber
& & \hspace{0.2in}
\begin{array}{ccccccc}
i_0 & \ j_3 & \ j_4 & \ j_1\ & 
j_1-\hat{x} & j_1+\hat{x} & j_1-\hat{y}
\end{array}\\
B & = &
\left (
\begin{array}{ccccccc}
-3 &  1 & 1  & 1  & \quad 0 & \qquad 0 & \qquad 0 \quad \\
 1 & -1 & 0  & 0  & \quad 0 & \qquad 0 & \qquad 0 \quad \\
 1 &  0 & -1 & 0  & \quad 0 & \qquad 0 & \qquad 0 \quad \\
 0 &  0 & 0  & -3 & \quad 1 & \qquad 1 & \qquad 1 \quad
\end{array} \right)
\begin{array}{c}
i_0 \\ j_3 \\ j_4 \\ j_1
\end{array}
\end{eqnarray}

Then
\begin{equation}
{\N_{\rm local} \over \N}=\det (1 + BG) = {1\over{2\pi}} - {5\over{2\pi^2}} + {4\over{\pi^3}}
\label{eq:h2.local}
\end{equation}
gives the probability that a random spanning tree meets these conditions.

In the second term in \fig{fig:X1.Decomposition}, the arrow from $j_1$ must point to
$i_0$, and the arrow from $i_0$ must point to $j_4$, and there must be some chain of
arrows that takes us from $j_4$ to $j_1$, forming a closed loop. The sites $j_2$ and
$j_3$ are disconnected from $i_0$, and we also require that there be no other closed
loops. Because we have a closed loop of arrows, this term is not a subset of the set
of spanning trees. The condition on the graphs is not local, so this term cannot be
calculated by the Majumdar-Dhar method.

Priezzhev showed \cite{priez} how to adapt the Kirchhoff theorem to deal with closed
loop diagrams such as these. He showed that if $n$ bond weights in the toppling
matrix are set to $-\epsilon$, then
\begin{equation}
\lim_{\epsilon\to\infty} {{\det(\Delta')} \over {\epsilon^n}}
\end{equation}
counts the total number of arrow configurations where each of the $n$ bonds is
required to be in a closed loop, and each closed loop contains at least one of the
$n$ bonds; in this count, each configuration is weighted by $(-1)^c$, where $c$ is the
number of closed loops.

The second graph in \fig{fig:X1.Decomposition} can thus be calculated by giving the
bond from $i_0$ to $j_4$ weight $-\epsilon$, as represented by the dashed line, and
making other appropriate local modifications. The matrix $B$ is then given by
\eject
\begin{eqnarray}
\nonumber
& & \hspace{0.2in}
\begin{array}{cccc}
i_0 & j_4 & j_3 & \ j_2 
\end{array}\\
B & = &
\left (
\begin{array}{cccc}
0 & -\epsilon & 0 & 0 \\
1 & 0 & -1  & 0 \\
1 & 0 & 0 & -1 
\end{array} \right)
\begin{array}{c}
i_0 \\ j_3 \\ j_2
\end{array}
\end{eqnarray}
Since we are defining $\N_{\rm loop}$ to be the number of graphs, and each
configuration gets a factor of $(-1)^c$, $\N_{\rm loop}/\N$ is given by $-\det (1 + BG)$,
rather than $\det (1 + BG)$, and is found to be equal to
\begin{equation}
{\N_{\rm loop} \over \N} = -{1\over{4\pi^2}} + {{2(\pi-2)}\over{\pi^3}} G_{0,0}.
\end{equation}
It contains a term proportional to $G_{0,0}$, the inverse of the discrete Laplacian
at coincident sites, which is infinite, see below \maybeeq{eq:Lattice.Green} (for a finite lattice, $G_{0,0}$ diverges as $\log{L}$, where $L$ is the system size). The
divergence of this second term in \fig{fig:X1.Decomposition} reflects the fact that
the number of diagrams with loops is much greater than the number of spanning trees.
Since $X_1$ is a finite fraction of the total number of spanning trees, the factors of
$G_{0,0}$ must cancel in the second and third terms of \maybeeq{eq:X1}.

\subsection{\pbm{$\theta$}-graphs}

While the first and second terms in \fig{fig:X1.Decomposition} can be calculated by
finite matrix determinants, the third term is more complicated. Priezzhev called these
$\theta$-graphs. In these graphs we have two arrows from $i_0$ (instead of the usual
case of one arrow), to $j_2$ and $j_4$, and both arrows are parts of two separate
closed loops, that go through $j_1$ to get back to $i_0$; $j_3$ is disconnected from
$i_0$. The number $\N_\theta$ of $\theta$-graphs cannot be calculated by a single
matrix determinant, and the bulk of the work in calculating $P_2$ lies here. 
To do this, Priezzhev used an elegant technique, known as the ``bridge'' trick.

\begin{figure}[htb]
\includegraphics{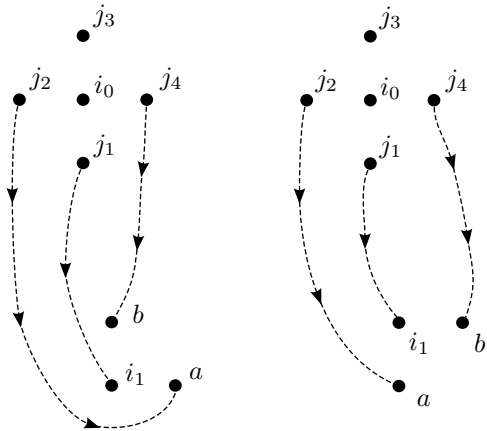}
\caption{The bridge trick: $L$ and $\Gamma$ configurations.}
\label{fig:Bridge.Trick.LAndGamma}
\end{figure}

The two paths from $i_0$ must meet at some point before returning to $i_0$---we call
that point $i_1$. Priezzhev then considered two possible graphs, labelled
the $L$ and $\Gamma$ graphs, each of which has three $-\epsilon$ bonds (``bridges'')
between non-adjacent sites, as shown in \fig{fig:Bridge.Trick.LAndGamma}. Looking 
at the $\Gamma$ graph on the right side, we see that there are several possible ways
to fulfill the condition that every $-\epsilon$ bond be in a closed loop. 
One possibility is to make three closed loops by separate paths from $a$ to
$j_2$, $i_1$ to $j_1$, and $b$ to $j_4$. The other possibility is to have one closed
path that contains all three of the $-\epsilon$ bonds. This second possibility can
occur in two possible ways. First, by paths from $a$ to $j_1$, $i_1$ to $j_4$, and $b$ to
$j_2$. And second, by paths from $a$ to $j_4$, $b$ to $j_1$, and
$i_1$ to $j_2$. This gives the three possible $\theta$-graphs shown in \fig{fig:Gamma.cases},
after reversing the direction of some of the paths, which does not affect the overall
counting. (Note that the topology forbids cases where the three $-\epsilon$ bonds
only produce two closed loops. If such cases existed, they would get a different
factor of $(-1)^c$, where $c$ is the number of closed loops, and ruin the bridge
trick.)

\begin{figure}[htb]
\includegraphics{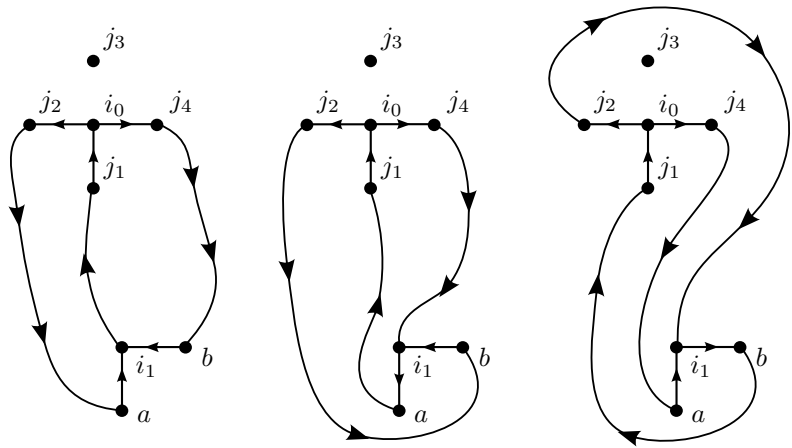}
\caption{Three possible $\theta$-graphs from the $\Gamma$ configuration of the bridge
trick.}
\label{fig:Gamma.cases}
\end{figure}

When all the graphs produced by the $L$ and $\Gamma$ bridge configurations are
considered, it is found that almost all the $\theta$-graphs are listed. Characterizing
the $\theta$-graphs by the three sites adjacent to $i_1$ that are connected by paths
to $j_1$, $j_2$, and $j_4$, it is found that we are lacking all cases where the three
sites are $\{ i_1-\hat{x},i_1+\hat{y},i_1-\hat{y} \}$, and have overcounted (by a
factor of 2) all cases where the three sites are $\{ i_1+\hat{x},i_1+\hat{y},i_1-\hat{y}\}$.
However, since the ASM is symmetric under horizontal flips, the overcounting and
undercounting exactly cancel.

The bridge diagrams shown in \fig{fig:Bridge.Trick.LAndGamma} are calculated with the
following $B$ matrix:
\begin{eqnarray}
\nonumber
& & \hspace{0.2in}
\begin{array}{ccccc}
j_3 & i_0\ \! & i_1 \ & a\ \ & b
\end{array} \\
B & = &
\left (
\begin{array}{ccccc}
-1 & 1 & 0 & 0 & 0 \\
0  & 0 & -\epsilon & 0 & 0 \\
0  & 0 & 0 & -\epsilon & 0 \\
0  & 0 & 0 & 0 & -\epsilon
\end{array} \right)
\begin{array}{c}
j_3 \\ j_1 \\ j_2 \\ j_4
\end{array}
\label{eq:Bridge.h2.B.Matrix}
\end{eqnarray}
Since $i_1$ can be anywhere, we need to sum over all possible locations of $i_1$ on
the plane,
\begin{equation}
\sum_{i_1} \Big[ \N_L (i_1) + \N_\Gamma (i_1) \Big].
\label{eq:h2.infinite.sum}
\end{equation}
$\N_L (i_1)$ and $\N_\Gamma (i_1)$ are both given by $\det(1 + BG)$, with $B$ as in
\maybeeq{eq:Bridge.h2.B.Matrix}. $\N_L (i_1)$ and $\N_\Gamma (i_1)$ differ only in the
locations of $a$ and $b$ with respect to $i_1$ (see \fig{fig:Bridge.Trick.LAndGamma}).
While $B$ has the same form in each case, the row and column indices associated with
$i_1$, $a$ and $b$ are different for different terms, which then involve different
entries of the matrix $G$.

However, \maybeeq{eq:h2.infinite.sum} does not exactly give $\N_\theta$. The bridge
trick (i.e. the equivalence of the bridge diagram and the relevant $\theta$-graphs)
requires that the sets $\{i_0,j_2,j_4\}$ and $\{i_1,a,b\}$ have no points
in common. When $i_1$ and $i_0$ are the same, or close, \maybeeq{eq:Bridge.h2.B.Matrix}
may or may not correspond to the subset of $\theta$-graphs shown in 
\fig{fig:Bridge.Trick.LAndGamma}. Priezzhev checked all such cases, and found 
what we will call ``special'' cases, where \maybeeq{eq:Bridge.h2.B.Matrix} either
produces graphs that are not $\theta$-graphs (in which case these terms need to be
subtracted from the infinite sum), or misses certain $\theta$-graphs (in which cases
new terms need to be added to the infinite sum). When these special terms are
accounted for, the total number of $\theta$-graphs is given by
\begin{eqnarray}
\N_\theta & = & \sum_{i_1} \left[ \N_{L}(i_1) + \N_{\Gamma}(i_1) \right] +
\N_L(j_4) + N_\Gamma(j_4)+2\N_T(j_4) - \N_L(j_2) - \N_\Gamma(j_2) \nonumber \\
& & \qquad - \N_L(i_0) - \N_\Gamma(i_0) - \N_\Gamma(j_3),
\label{eq:Theta.Graph.Sum}
\end{eqnarray}
where $\N_T(j_4)$ is the graph obtained by adding the three $-\epsilon$ bonds as shown
in \fig{fig:T.Graph}, and removing the bond between $j_3$ and $i_0$ as before.

\begin{figure}[tb]
\includegraphics{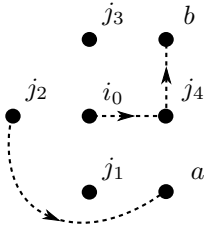}
\caption{The $T$ configuration used to compute $\N_T(j_4)$. The sites $a$ and $b$
are nearest neighbours of $j_4$.}
\label{fig:T.Graph}
\end{figure}

All of the individual terms in \maybeeq{eq:Theta.Graph.Sum} can
be calculated by taking finite-dimensional determinants (using  
\maybeeq{eq:Bridge.h2.B.Matrix} for all terms except $\N_T(j_4)$). The final
difficulty is in the infinite sum over $i_1$. The two matrix determinants in
$\N_L(i_1)$ and $\N_\Gamma(i_1)$ differ only in a single column, and can thus be
combined in a single matrix determinant:
\begin{eqnarray}
\sum_{i_1} \left[ {\N_{L}(i_1) \over \N} + {\N_{\Gamma}(i_1) \over \N} \right]  & = &
\sum_{k,\l \in \Z} \det \left(
\begin{array}{cccc}
G_{1,0}-G_{0,0} & G_{k,\l} & G_{k+1,\l} & G_{k,\l-1}-G_{k,\l+1} \\
G_{0,0}-G_{1,0}-1 & G_{k,\l-1} & G_{k+1,\l-1} & G_{k,\l-2}-G_{k,\l} \\
G_{1,1}-G_{1,0} & G_{k-1,\l} & G_{k,\l} & G_{k-1,\l-1}-G_{k-1,\l+1} \\
G_{1,1}-G_{1,0} & G_{k+1,\l} & G_{k+2,\l} & G_{k+1,\l-1}-G_{k+1,\l+1}
\end{array}
\right), \nonumber \\
\label{eq:h2.InfiniteSum}
\end{eqnarray}
where $G_{k,\l}$ stands for the entry $\Delta^{-1}_{i,j}$ with $(k,\l) = i-j$. A 
convenient integral representation of it is
\begin{equation}
G_{k,\l} = \int\!\!\!\!\int_{-\pi}^\pi {{\d\a\d\b} \over {8\pi^2}} \;
{{e^{i\a k+i\b \l}}\over{2-\cos{\a}-\cos{\b}}}.
\label{eq:Lattice.Green}
\end{equation}

\subsection{\pbm Final results for $h=2$ on the plane}

The Green function $G_{k,\l}$ can be computed explicitly for small values of $k$ and
$\l$, so that the first column of the matrix in \maybeeq{eq:h2.InfiniteSum}
contains known numbers. For the other three columns, the representation 
\maybeeq{eq:Lattice.Green} can be used to write each entry as a two-dimensional
integral, over the variables $(\a_3,\b_3)$ in the second column, $(\a_2,\b_2)$ in
the third column, and $(\a_1,\b_1)$ in the fourth column. The sum over $k$ and $\l$
produces two $\delta$-functions $\delta(\a_1+\a_2+\a_3) \, \delta(\b_1+\b_2+\b_3)$,
which allow to integrate over $\a_3,\b_3$. So we end up with a four-dimensional
integral (two further integrations can be carried out analytically, see Section VII).
The end result is the expression obtained by Priezzhev \cite{priez} for the height two
probability, namely
\begin{equation}
P_2 = {1\over2} - {3\over{2\pi}} - {2\over\pi^2} + {12\over\pi^3} + 
{1\over 4} I\left(1,{4\over\pi}-1,3\right),
\label{eq:exact.h2}
\end{equation}
where
\be
I(c_1,c_2,c_3) \equiv  \int\!\!\!\!\int_{-\pi}^\pi {{\d\a_1\d\b_1} \over {4\pi^2}}
\; \int\!\!\!\!\int_{-\pi}^\pi {{\d\a_2\d\b_2} \over {4\pi^2}} \; 
{{i \sin{\beta_1} \cdot \det M(c_1,c_2,c_3)}\over
{L(\alpha_1,\beta_1)L(\alpha_2,\beta_2) L(\alpha_1+\alpha_2,\beta_1+\beta_2)}},
\label{eq:4Dintegral.general}
\ee
with $L(\a,\b) = 2 - \cos{\a} - \cos{\b}$, and 
\be
M(c_1,c_2,c_3) = \left(
\begin{array}{cccc}
c_1 & 1 & e^{i\alpha_2} & 1 \\
c_3 & e^{i(\beta_1+\beta_2)} & e^{i(\alpha_2-\beta_2)} & e^{-i\beta_1} \\
c_2 & e^{i(\alpha_1+\alpha_2)} & 1 & e^{-i\alpha_1} \\
c_2 & e^{-i(\alpha_1+\alpha_2)} & e^{2i\alpha_2} & e^{i\alpha_1}
\end{array}
\right).
\ee

We now make an observation that will be useful for comparison with $P_3$ later. The
function $I(c_1,c_2,c_3)$ does not depend on $c_3$, because all the terms in the
integrand proportional to $c_3$ are antisymmetric under $\beta_1\to-\beta_1$,
$\beta_2\to-\beta_2$. Since the integral is linear in $c_1$ and $c_2$, we can write it
as the sum of two separate four-dimensional integrals:
\begin{equation}
I(c_1,c_2) = c_1 J_1 + c_2 J_2.
\end{equation}
$J_1$ and $J_2$ still need to be evaluated numerically, so this decomposition does not help 
us get a closed form expression for $P_2$, but we will see in the next section that it
does allow us to get a closed form relationship between $P_2$ and $P_3$. Numerical
integrations yield $J_1 = -0.26866$ and, quite unexpectedly, $J_2 = 0.5$
with a error smaller than $10^{-12}$, leading to the value $P_2 = 0.1739$, as quoted
in \cite{priez}. Although we could not evaluate it exactly, it is very tempting to
conjecture that $J_2$ is exactly equal to $\half$. As we will see in the next section,
this would imply an exact, closed and simple formula for the 1-site height
probabilities $P_i$ on the plane.


\section{Height three on the plane}
\label{sec:HeightThreeBulk}

While Priezzhev calculated the height two and height three probabilities on the
plane, he only gave the details of the height two calculation. In this Section, we
give the details of our derivation of the height three probability, both because
these details are necessary for the calculation of the height three probability in the
UHP, and to derive an exact relationship between $P_2$ and $P_3$.

\subsection{Height Three Decomposition}

For the height three probability, we need $X_2$, the number of spanning trees where 
exactly two neighbours of $i_0$ are predecessors. This can happen in three ways, as shown in
\fig{fig:X2.123}. Priezzhev showed that $X_2^{(2)}$ can be written as a linear combination 
of simple local graphs, as shown in \fig{fig:X22.Simple}, yielding (there appears to be a 
misprint in Eq.(32) of \cite{priez})
\begin{equation}
{{X_2^{(2)}}\over\N} = {8\over \pi} - {24\over\pi^2}.
\end{equation}

\begin{figure}[t]
\psfrag{X21}{$X_2^{(1)}:$}
\psfrag{X22}{$X_2^{(2)}:$}
\psfrag{X23}{$X_2^{(3)}:$}
\includegraphics{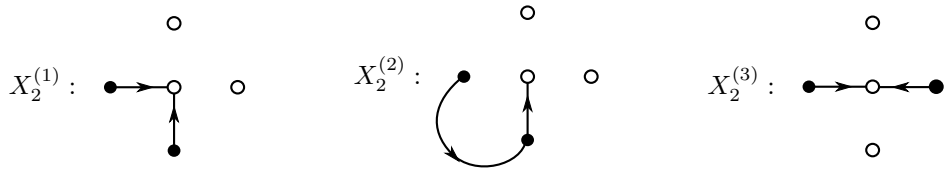}
\caption{Definitions of $X_2^{(1)}$, $X_2^{(2)}$, and $X_2^{(3)}$.}
\label{fig:X2.123}
\end{figure}

The calculation of $X_2^{(1)}$ and $X_2^{(3)}$ is more complicated. Just as Priezzhev broke 
$X_1$ into local, loop, and  more complicated terms (see \fig{fig:X1.Decomposition}),
by counting arrow configurations, we can break up $X_2^{(1)}$ and $X_2^{(3)}$ into
similar terms. We discuss the decomposition of $X_2^{(1)}$ in some detail. The
derivation is shown graphically in \figs~{\ref{fig:X21.Decomposition.Part1}},
\ref{fig:X21.Decomposition.Part2}, and~\ref{fig:X21.Decomposition.Part3}.

\begin{figure}[t]
\psfrag{X2/16}{$\displaystyle{X_2^{(2)}\over16}=$}
\includegraphics{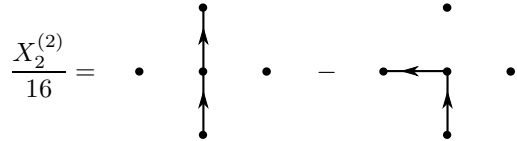}
\caption{A simple decomposition of $X_2^{(2)}$.}
\label{fig:X22.Simple}
\end{figure}

\begin{figure}[b]
\psfrag{X2/4}{$\displaystyle{X_2^{(1)}\over4}=$}
\includegraphics{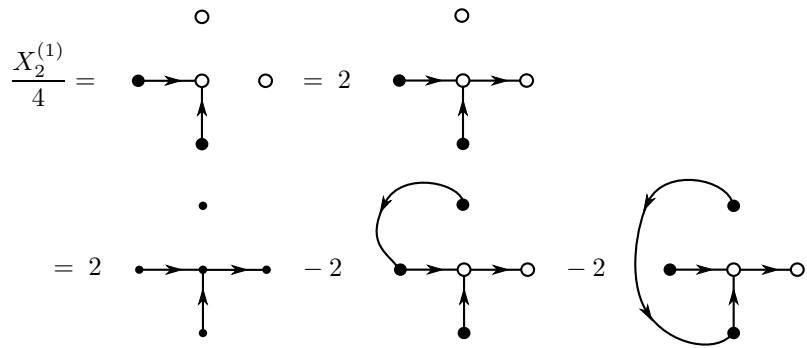}
\caption{Decomposition of $X_2^{(1)}$, Part 1.}
\label{fig:X21.Decomposition.Part1}
\end{figure}

We begin by explaining the derivation in \fig{fig:X21.Decomposition.Part1}. For
$X_2^{(1)}$, we get the same number of spanning trees whether the arrow from $i_0$
points to $j_3$ or $j_4$, so without loss of generality, we pick $j_4$. Next, recall
that the open circle on $j_3$ indicates that it is not a predecessor of $i_0$. We have
no simple matrix method for calculating subsets of spanning trees with such a
restriction. So we start off with the subset with no such restriction, and then
subtract off the different ways that $j_3$ can be a predecessor of $i_0$: $j_3$ can be
a predecessor of $i_0$ either by a path through $j_2$, or by a path through $j_1$.
This gives the last line in \fig{fig:X21.Decomposition.Part1}.

The first graph in the last line of \fig{fig:X21.Decomposition.Part1} can be 
calculated by the Majumdar-Dhar method. The last two graphs in this line require 
further manipulation to be put into a form that can be calculated with matrix
determinants. They are calculated in \figs~{\ref{fig:X21.Decomposition.Part2}
and \ref{fig:X21.Decomposition.Part3}}.

\begin{figure}[t]
\includegraphics{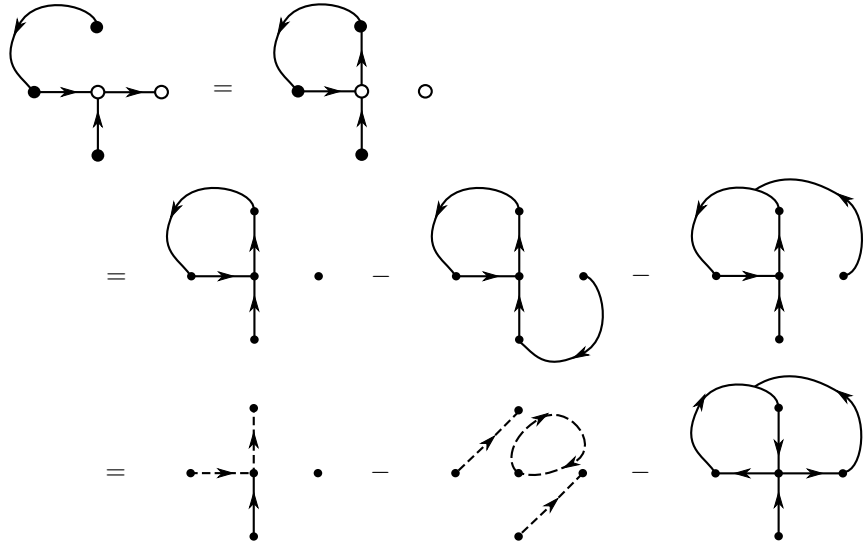}
\caption{Decomposition of $X_2^{(1)}$, Part 2.}
\label{fig:X21.Decomposition.Part2}
\end{figure}

\begin{figure}[b]
\includegraphics{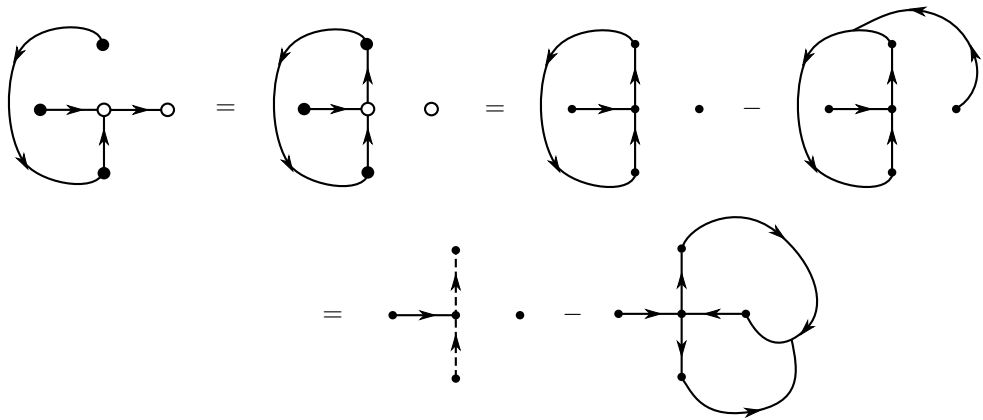}
\caption{Decomposition of $X_2^{(1)}$, Part 3.}
\label{fig:X21.Decomposition.Part3}
\end{figure}

In \fig{fig:X21.Decomposition.Part2}, we first use the fact that redirecting the
$i_0\to j_4$ arrow to an $i_0\to j_3$ arrow will not affect the total number of arrow
configurations. $j_4$ is not permitted to be a predecessor of $i_0$, and as before, we
deal with this condition by starting without this condition, and subtracting off the
different ways that $j_4$ can be a predecessor of $i_0$. This gives the second line of
\fig{fig:X21.Decomposition.Part2}. The first two graphs in this line can be calculated 
as matrix determinants, by adding $-\epsilon$ bonds. The last graph in 
\fig{fig:X21.Decomposition.Part2} is what we call a $\tilde\theta$-graph. 
In the last line of \fig{fig:X21.Decomposition.Part2} we no longer draw large filled
or large empty circles, because there are no predecessor conditions imposed, other
than those that follow naturally from the arrows drawn.

The analysis in \fig{fig:X21.Decomposition.Part3} is similar. In the last line of
that equation we again use the fact that dashed lines must be contained in closed
loops, and the fact that directions of paths of arrows can be reversed without
changing the total number of graphs.

The graphical decomposition of $X_2^{(3)}$ uses a similar logic, and is shown in 
\figs~{\ref{fig:X23.Decomposition.Part1} and \ref{fig:X23.Decomposition.Part2}}.
Defining the graphs as in \fig{fig:Ti.Definitions}, we have
\begin{eqnarray}
X_2^{(1)} & = & 8(T_1-T_3+T_4-T_5) + 16 \N_{\tilde\theta}, \\
X_2^{(3)} & = & 4(T_2-2T_3+2T_4) + 8 \N_{\tilde\theta}.
\end{eqnarray}

\begin{figure}[t]
\psfrag{X2/2}{$\displaystyle{X_2^{(3)}\over2}=$}
\includegraphics{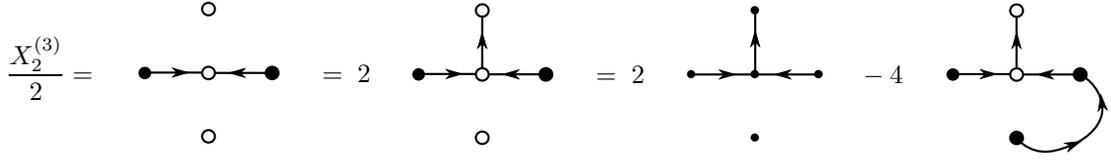}
\caption{Decomposition of $X_2^{(3)}$, Part 1.}
\label{fig:X23.Decomposition.Part1}
\end{figure}

\begin{figure}[t]
\includegraphics{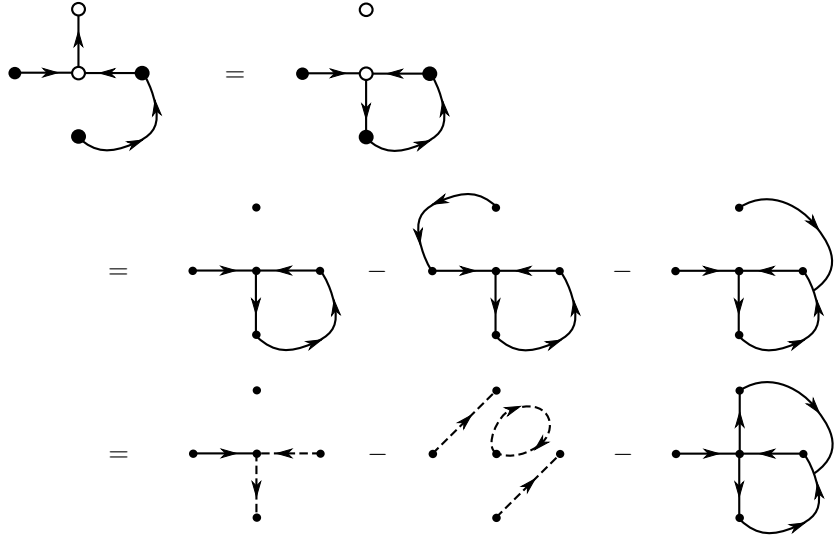}
\caption{Decomposition of $X_2^{(3)}$, Part 2.}
\label{fig:X23.Decomposition.Part2}
\end{figure}

The $\tilde\theta$-graph is defined similarly to the $\theta$-graph, and is shown in
\fig{fig:Ti.Definitions} rotated to its standard form. Just as with the $\theta$-graph,
there are two arrows from $i_0$, pointing to $j_2$ and $j_4$, which form two closed
loops that reach $i_1$ through $j_1$. The only difference from the $\theta$-graph is
that now, instead of removing the bond from $j_3$ to $i_0$, we require an arrow from
$j_3$ to $i_0$. (The paths from $j_4$ and $j_2$ now cannot go through $j_3$, since
they must still reach $i_0$ via $j_1$.) To represent the $\tilde{\theta}$-graphs, we
use the fact that if a $-\epsilon$ bond is added from a site to itself (i.e. along the
diagonal of the $B$-matrix), it effectively turns that site into a sink,
where arrows can terminate, but not begin. So by adding a $-\epsilon$ bond from $j_3$
to itself, we fulfill the conditions necessary for a $\tilde\theta$ graph.

\begin{figure}[t]
\psfrag{T1=}{$T_1 =$}
\psfrag{T2=}{$T_2 =$}
\psfrag{T3=}{$T_3 =$}
\psfrag{T4=}{$T_4 =$}
\psfrag{T5=}{$T_5 =$}
\psfrag{Ntt=}{$\mathcal{N}_{\tilde{\theta}}=$}
\includegraphics{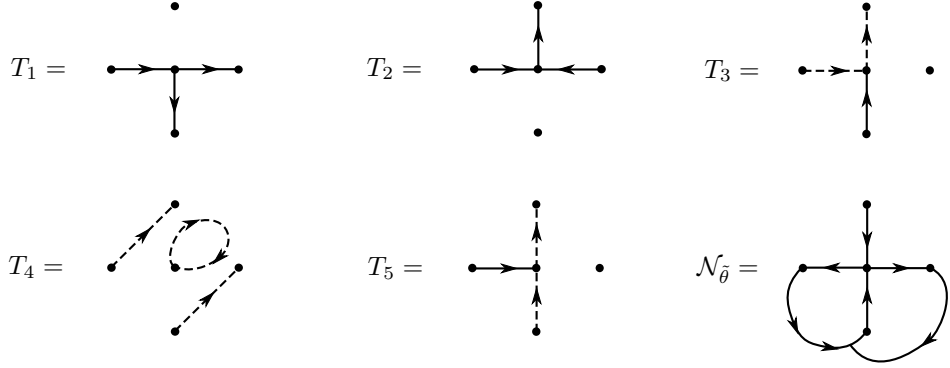}
\caption{Definitions of the $T_i$ and $\tilde\theta$ graphs.}
\label{fig:Ti.Definitions}
\end{figure}

\subsection{$\tilde\theta$-graphs}

Since the $\tilde\theta$-graph differs from the $\theta$-graph only in that a $-\epsilon$ bond
has been added from $j_3$ to itself, the only change from the $h=2$ infinite sum is
that instead of \maybeeq{eq:Bridge.h2.B.Matrix}, we have
\begin{eqnarray}
\nonumber
& & \quad\   
\begin{array}{ccccc}
j_3\ & i_1\ \ \! & a\ \ \! & b
\end{array} \\
B & = &
\left (
\begin{array}{ccccc}
-\epsilon & 0 & 0 & 0 \\
0 & -\epsilon & 0 & 0 \\
0 & 0 & -\epsilon & 0 \\
0 & 0 & 0 & -\epsilon
\end{array} \right)
\begin{array}{c}
j_3 \\ i_0 \\ j_2 \\ j_4
\end{array}
\label{eq:Bridge.h3.B.Matrix}
\end{eqnarray}

The analysis of the $\tilde\theta$-graphs then proceeds identically to that for the
$\theta$-graphs. There are again a number of special cases that arise when $i_1$
borders or is equal to $i_0$. Checking these cases by hand gives correction terms
analogous to those in \maybeeq{eq:Theta.Graph.Sum}:
\be
\N_{\tilde\theta} = \sum_{i_1} \left[ \Nt_{L}(i_1) + \Nt_{\Gamma}(i_1) \right] 
+ \Nt_L(j_4) + \Nt_\Gamma(j_4) + \Nt_T(j_4) - \Nt_L(j_2) - \Nt_\Gamma(j_2)
- \Nt_\Gamma(i_0).
\label{eq:ThetaTilde.Graph.Sum}
\ee
$\Nt_L(i_1)$ and $\Nt_\Gamma(i_1)$ are given by $\det (1 + BG)$, where $B$ is given by
\maybeeq{eq:Bridge.h3.B.Matrix}, and $a$ and $b$ are located as shown in  
\fig{fig:Bridge.Trick.LAndGamma}. The $\tilde{T}$ graph is defined similarly to the
$T$ graph in \fig{fig:T.Graph}, except that instead of forbidding the arrow from $j_3$
to $i_0$, we add a $-\epsilon$ bond from $j_3$ to itself.

\subsection{Final results on the plane}

As in the height two case, all the individual terms in \maybeeq{eq:ThetaTilde.Graph.Sum}
can be calculated as finite-dimensional matrix determinants. The infinite sum is
similar to the one in \maybeeq{eq:h2.InfiniteSum}. Because only one bond of the $B$ 
matrix has been changed, only the first column of the matrix in \maybeeq{eq:h2.InfiniteSum}
is changed, as follows:
\begin{equation}
\pmatrix{
G_{1,0}-G_{0,0} \cr G_{0,0}-G_{1,0}-1 \cr G_{1,1}-G_{1,0} \cr G_{1,1}-G_{1,0}}
\longrightarrow
-\pmatrix{G_{1,0} \cr G_{0,0} \cr G_{1,1} \cr G_{1,1}}
\end{equation}
Note that this column is no longer finite; instead, all its terms diverge as
$G_{0,0}$.

Just as with the height two probability, our end result for $P_3$ is a sum of 
local and loop graphs, which can be computed exactly, and an infinite sum, which
results in a four-dimensional integral. Since only the first column of
\maybeeq{eq:h2.InfiniteSum} is changed, the four-dimensional integral still has the
form in \maybeeq{eq:4Dintegral.general}, and we get the same integrals $J_1$ and
$J_2$, but with different coefficients. We obtain explicitly
\begin{eqnarray}
{{X_2^{(1)}}\over\N} & = & 1 - {6\over\pi} + {24\over\pi^2} - {32\over\pi^3}
- \left(J_1 + {4\over\pi} J_2 \right) + G_{0,0} \left(4(J_1 + J_2) - {8\over{\pi^2}}(\pi-2)
\right), \\
{{X_2^{(3)}}\over\N} & = & -{3\over 2} + {7\over{\pi^2}} - {16\over\pi^3}
- {1\over 2}\left(J_1 + {4\over\pi} J_2 \right) + G_{0,0} \left(2 (J_1 + J_2) - 
{4\over\pi^2}(\pi-2) \right).
\end{eqnarray}

Even for an infinite lattice, the ratios $X_2^{(1)}/\N$ and $X_2^{(3)}/\N$ must be
finite since they represent fractions of all spanning trees. $G_{0,0}$ being infinite,
this immediately requires that the coefficients of $G_{0,0}$ must cancel, giving
\begin{equation}
J_1 + J_2 = {2\over\pi^2}(\pi-2).
\label{eq:exact.J1.J2}
\end{equation}
While we could not evaluate $J_1$ and $J_2$ analytically, not even 
their sum, the numerical values mentioned before confirm \maybeeq{eq:exact.J1.J2}.

We find $P_3$ from \maybeeq{1siteprob}, with $X_2=X_2^{(1)}+X_2^{(2)}+X_2^{(3)}$, and
then rewrite both $P_2$ and $P_3$ with \maybeeq{eq:exact.J1.J2}, reproducing the
values found in \cite{priez}, namely
\begin{eqnarray}
P_2 & = & {1\over 2} - {1\over\pi} - {3\over\pi^2} +
{12\over\pi^3} - {\pi - 2 \over 2\pi} J_2 \simeq 0.1739,\label{p2j2}\\
\noalign{\medskip}
P_3 & = & {1\over 4} + {2 \over \pi} - {12\over\pi^3} - {8 - \pi \over 4\pi} J_2
\simeq 0.3063.
\label{p3j2}
\end{eqnarray}
These two equations give us an exact relationship between $P_2$ and $P_3$:
\be
(\pi-8) P_2 + 2(\pi-2) P_3 = \pi - 2 - {3\over\pi} + {12\over\pi^2} -
{48\over\pi^3}.
\ee
To our knowledge, this exact relation is new. 

Using the value of $P_1 = {2(\pi-2) \over \pi^3}$, we can rewrite the previous
relation as
\be
{48 - 12 \pi + 5\pi^2 - \pi^3 \over 2(\pi-2)}\, P_1 + (\pi - 8)\,
P_2 + 2(\pi - 2)\, P_3 = {(\pi-2)(\pi-1) \over \pi}.
\label{relprob}
\ee
The significance of this new relation will be made clear in Section VIII.

Coming back to the conjecture we made at the end of the previous section, namely
that $J_2 = \half$ exactly (and checked to 12 decimal places), we see that,
together with (\ref{p2j2}) and (\ref{p3j2}), it yields a very simple and exact formula
for $P_2$ and $P_3$, similar to the formula for $P_1$:
\be
P_2 = {1 \over 4} - {1 \over 2\pi} - {3 \over \pi^2} + {12 \over \pi^3},\qquad
P_3 = {3 \over 8} + {1 \over \pi} - {12 \over \pi^3}.
\label{p2p3}
\ee
Remarkably these values imply an even simpler formula for the mean height in the
bulk,
\be
\la h \ra = P_1 + 2P_2 + 3P_3 + 4P_4 = {25 \over 8},
\ee
a value conjectured by Grassberger \cite{dhar-rev}. The striking simplicity of this
result clearly calls for a better explanation than just long calculations. 

The formulae (\ref{p2p3}) rely on $J_2$ being exactly equal to $\half$. An integral
representation for $J_2$ has been given at the end of Section III. By carrying out two
of the four integrations, we can rewrite it in a simpler form, as a two-fold
integral,
\bea
J_2 &=& {4 \over \pi^2} - {14 \over \pi} - 8 - {4\sqrt{2} \over \pi^2} \int_0^\pi {\d\b_1 \over 
\sqrt{3-\cos{\b_1}}} \int_{-\pi}^\pi {\d\b_2 \over 1-t_1t_2t_3} \sin{\b_1-\b_2 \over 2}\; 
\Big[\cos{\b_1-\b_2 \over 2} - 2\cos{\b_1+\b_2 \over 2}\Big] \nonumber\\
\noalign{\medskip}
&& \hspace{3cm} \times \Big[(3 - \cos{\b_1} + \cos{\b_2}) \cos{\b_1 \over 2} - 2 
\sin{\b_2} \sin{\b_1 \over 2} \Big], 
\eea
where $t_i = y_i - \sqrt{y_i^2-1}$, $y_i = 2 - \cos{\b_i}$ and $\b_3 = -(\b_1 +
\b_2)$. This integral expression has been used for the numerical evaluation of $J_2$, 
yielding $J_2 = 0.5 + o(10^{-12})$. 


\section{Heights two and three on the upper half-plane}

The calculations of the height two and three probabilities on the UHP use the same 
formalism as the infinite plane calculations. The relation between the 1-site height
probabilities in the ASM, and the probabilities of numbers of predecessors in spanning
trees, stated in \maybeeq{1siteprob}, still holds in the presence of a boundary.
The corresponding equation for the height 1, \maybeeq{eq:P1.predecessor}, has been
used by Brankov {\it et al.} to obtain the 1-site height one probabilities on the UHP
in the presence of an open or a closed boundary \cite{bip}.

On the UHP, we will use integer coordinates $(m,n)$, where $n \in \Z$ runs along the
horizontal axis, $m \in \Z_+$ along the vertical axis, and with the boundary
located at $m=1$. We want to compute the 1-site height probabilities $P_a(m)$ for
having a height $a$ at the site $(0,m)$, and when the boundary is either fully open or
fully closed. By horizontal translation invariance, the coordinate $n$ plays no role.

When calculating $X_1$ on the plane, we obtained a factor of 12 in \fig{fig:X1.Decomposition}
because of the symmetry between the four possible predecessors of $i_0$, and between
the three possible directions of the arrow from $i_0$. In the UHP, we still have the
latter factor of three, but the four choices of predecessors are no longer symmetric 
in the presence of a boundary.  We thus need to break $X_1$ up into the $A$, $B$, and
$C$ configurations in \fig{fig:X1.Rotations}, where the dashed line indicates the
boundary. (The cases with $j_2$ and $j_4$ as predecessors are still symmetric.)

Just as in the bulk, each of these configurations can be broken up into local, loop,
and $\theta$-diagrams (see \fig{fig:X1.Decomposition}). The $\theta$-graph can then
be written as an infinite sum, and some special correction graphs where $i_1$ is close
to $i_0$.

\begin{figure}[t]
\psfrag{X1=}{$X_1 =$}
\psfrag{N2a}{$\mathcal{N}_{2,A}$}
\psfrag{N2b}{$\mathcal{N}_{2,B}$}
\psfrag{N2c}{$\mathcal{N}_{2,C}$}
\includegraphics{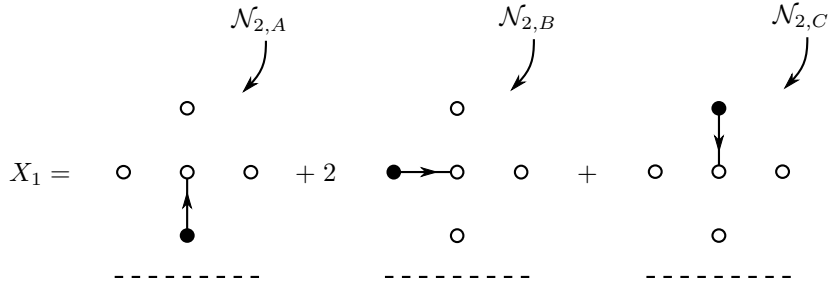}
\caption{Inequivalent rotations of $X_1$ with a boundary to the south.}
\label{fig:X1.Rotations}
\end{figure}

The local, loop, and special correction graphs can all again be calculated as matrix
determinants. In these terms, the only effect of the boundary is to modify the
lattice Green function. The lattice Green functions in the presence of open or
closed boundaries can be found in terms of that on the plane using the image method:
\begin{eqnarray}
\label{eq:Green.open}
G^{\rm open}_{m_1,m_2,n_1-n_2} & \equiv & G^{\rm open}_{(n_1,m_1),(n_2,m_2)} =
G_{n_1-n_2,m_1-m_2}-G_{n_1-n_2,m_1+m_2}, \\ \label{eq:Green.closed}
G^{\rm cl}_{m_1,m_2,n_1-n_2} & \equiv & G^{\rm cl}_{(n_1,m_1),(n_2,m_2)} =
G_{n_1-n_2,m_1-m_2}+G_{n_1-n_2,m_1+m_2-1}.
\end{eqnarray}

For the local, loop, and special correction graphs, we only need the lattice Green
function between two sites near $(0,m)$, and thus simply expand out the above Green
functions for $m_1 \approx m_2 \approx m \gg 1$ and $n_1\approx n_2$. The relevant
expansion in inverse powers of $m$ reads ($k,n \ll m$)
\begin{equation}
G_{k,2m+n} = G_{0,0} - {1\over{2\pi}}\log(2m) - {n\over{4\pi m}} - {1\over\pi}
\left({\gamma\over 2}+{3\over 4} \log 2\right) + {1 + 6(n^2-k^2) \over 96\pi m^2}
+\order{3} \ ,
\end{equation}

\noindent where $\gamma=0.57721\dots$ is the Euler-Mascheroni constant.

The local graphs have the same values as on the plane, up to terms of $\mathcal{O}(m^{-1})$.
For example, $\N_{\rm local}$ in \fig{fig:X1.Decomposition} becomes three terms
corresponding to the three orientations in \fig{fig:X1.Rotations}, also shown in
\fig{fig:X1.Rotations.more}. For the closed boundary case we have
\begin{eqnarray}
{\N_{2,{\rm cl},A}^{\rm local} \over \N} & = & {\N_{2,{\rm local}} \over \N} +  
{{\pi-2}\over{4\pi^3 m}} - {{\pi^2-10\pi+18}\over{16\pi^3 m^2}} + \order{3}, \\
\noalign{\medskip}
{\N_{2,{\rm cl},B}^{\rm local} \over \N} & = & {\N_{2,{\rm local}} \over \N} + 
{{\pi-2}\over{4\pi^3 m}} - {{\pi^2-10\pi+22}\over{16\pi^3 m^2}} + \order{3}, \\
\noalign{\medskip}
{\N_{2,{\rm cl},C}^{\rm local} \over \N} & = & {\N_{2,{\rm local}} \over \N} - 
{{\pi-2}\over{4\pi^3 m}} - {{\pi^2-6\pi+10}\over{16\pi^3 m^2}} + \order{3}.
\end{eqnarray}
where $\N_{2,{\rm local}}$ refers to the plane value in \maybeeq{eq:h2.local}.

The loop graphs are also matrix determinants with exactly the same $B$ matrices as on
the plane. Recall that on the plane, loop graphs had terms proportional to $G_{0,0}$,
the constant diverging part of the Green function. When there are boundaries, we
define
\begin{eqnarray}
g(m) & \equiv & -{1\over{2\pi}} \Big[\log{m} + \gamma + {5\over 2} \ln 2 \Big]
+ 2 G_{0,0} = G_{0,0} + G_{2m,0} + \order{2}, \\
\noalign{\medskip}
\tilde{g}(m) & \equiv & -{1\over{2\pi}} \Big[\log{m}+\gamma + {5\over 2} \ln
2\Big] = G_{0,0} - G_{2m,0} + \order{2},
\end{eqnarray}
to be the dimension zero terms of the Green functions in the closed and open cases
respectively. One then finds that the loop graphs in the closed case are given by
\begin{eqnarray}
\nonumber
{\N_{2,{\rm closed},A}^{\rm loop} \over \N} & = & {\N_{2,{\rm closed},B}^{\rm loop}
\over \N} = - {1\over{4\pi^2}} + {{2(\pi-2)}\over{\pi^3}} g(m) + {1\over \pi^2 m}
\Big({1\over 4} - {1\over{\pi^2}} \Big) \nonumber\\
\noalign{\medskip}
&& \hspace{2cm} + \; {1\over \pi^2 m^2} \Big({3 \over 16} + {1 \over 48\pi} -
{13\over{21\pi^2}} - {\pi-2 \over 2\pi} g(m) \Big) + \order{3}, \\
\noalign{\medskip}
{\N_{2,{\rm closed},C}^{\rm loop} \over \N} & = &
- {1\over{4\pi^2}} + {{2(\pi-2)}\over{\pi^3}} g(m) - {1\over \pi^2 m}
\Big({1\over{4}} - {1\over\pi} + {1\over\pi^2} \Big) \nonumber\\
\noalign{\medskip}
&& \hspace{2cm} - \; {1\over \pi^2 m^2} \Big({1\over{16}} - {25\over{48\pi}} +
{13\over{21\pi^2}} + {\pi-2 \over 2\pi} g(m) \Big) + \order{3}.
\end{eqnarray}

\begin{figure}[t]
\psfrag{Nla=}{$\mathcal{N}_{2,{\text{cl}},A}^{\text{local}}=$}
\psfrag{NLa=}{$\mathcal{N}_{2,{\text{cl}},A}^{\text{loop}}=$}
\psfrag{Nlb=}{$\mathcal{N}_{2,{\text{cl}},B}^{\text{local}}=$}
\psfrag{NLb=}{$\mathcal{N}_{2,{\text{cl}},B}^{\text{loop}}=$}
\psfrag{Nlc=}{$\mathcal{N}_{2,{\text{cl}},C}^{\text{local}}=$}
\psfrag{NLc=}{$\mathcal{N}_{2,{\text{cl}},C}^{\text{loop}}=$}
\includegraphics{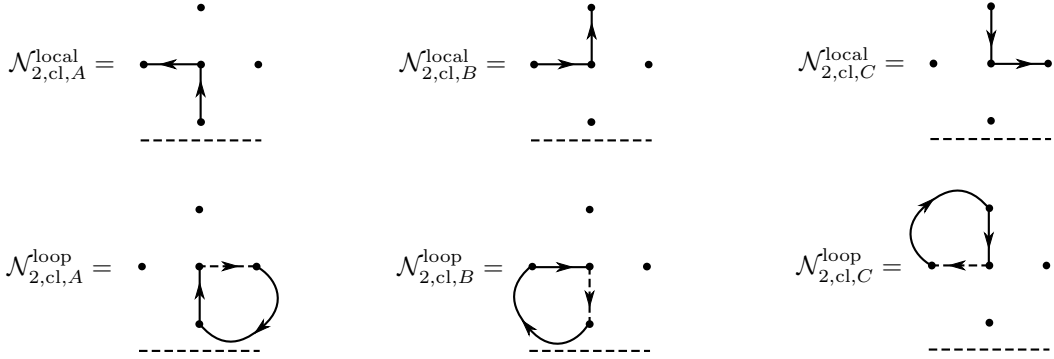}
\caption{Rotations of graphs used to calculate $X_1$ on the UHP.}
\label{fig:X1.Rotations.more}
\end{figure}
\end{psfrags}

Once the local, loop, and special correction graphs have all been computed as series
in $1/m$, all that remains to evaluate is the infinite sum contribution to
$\N(\theta)$, which is the most difficult part to analyze. The infinite sum 
contribution can be set up similarly to the bulk case, with two differences. First,
the boundary Green functions must be used; second, the summation over $i_1$ must cover
the UHP, rather than the whole plane. For example, for the $A$-orientation of $\N_2$
in the closed case, we obtain the sum ($(k,\l)$ are the coordinates of $i_1$)
\begin{equation}
I_A^{h=2,{\rm cl}} = \sum_{k=-\infty}^{\infty} \sum_{\l=1-m}^{\infty}
[{\N_L(i_1) \over \N} + {\N_\Gamma(i_1) \over \N}],
\end{equation}
with
\be
\label{eq:UHP.matrix.L}
{\N_L \over \N} = \left|
\begin{array}{cccc}
G^{\rm cl}_{m+1,m+1,0}-G^{\rm cl}_{m+1,m,0}-1 & 
G^{\rm cl}_{m+1,m,0}-G^{\rm cl}_{m,m,0} & G^{\rm cl}_{m+1,m,1}-G^{\rm cl}_{m,m,1} 
& G^{\rm cl}_{m+1,m,1}-G^{\rm cl}_{m,m,1} \\
G^{\rm cl}_{m+\l,m+1,k} & G^{\rm cl}_{m+\l,m,k} & G^{\rm cl}_{m+1,m,k+1} &
G^{\rm cl}_{m+\l,m,k-1} \\
G^{\rm cl}_{m+\l,m+1,k+1} & G^{\rm cl}_{m+\l,m,k+1} & G^{\rm cl}_{m+\l,m,k+2} &
G^{\rm cl}_{m+\l,m,k} \\
G^{\rm cl}_{m+\l+1,m+1,k} & G^{\rm cl}_{m+\l+1,m,k} & G^{\rm cl}_{m+\l+1,m,k+1} &
G^{\rm cl}_{m+\l+1,m,k-1}
\end{array}
\right|,
\ee
and
\be
\label{eq:UHP.matrix.Gamma}
{\N_\Gamma \over \N} = \left|
\begin{array}{cccc}
G^{\rm cl}_{m+1,m+1,0}-G^{\rm cl}_{m+1,m,0}-1 & 
G^{\rm cl}_{m+1,m,0}-G^{\rm cl}_{m,m,0} & G^{\rm cl}_{m+1,m,1}-G^{\rm cl}_{m,m,1} &
G^{\rm cl}_{m+1,m,1}-G^{\rm cl}_{m,m,1} \\
G^{\rm cl}_{m+\l,m+1,k} & G^{\rm cl}_{m+\l,m,k} & G^{\rm cl}_{m+1,m,k+1} &
G^{\rm cl}_{m+\l,m,k-1} \\
G^{\rm cl}_{m+\l,m+1,k+1} & G^{\rm cl}_{m+\l,m,k+1} & G^{\rm cl}_{m+\l,m,k+2} &
G^{\rm cl}_{m+\l,m,k} \\
-G^{\rm cl}_{m+\l-1,m+1,k} & - G^{\rm cl}_{m+\l-1,m,k} & - G^{\rm cl}_{m+\l-1,m,k+1} &
-G^{\rm cl}_{m+\l-1,m,k-1}
\end{array}
\right|.
\ee

The end result for the $\theta$-graph contribution to the $A$-orientation of $\N_2$
in the closed case is (including the special correction graphs)
\begin{eqnarray}
\nonumber
{\N_{2,{\rm closed},A}^{\theta} \over \N} & = & - {1\over 8} + {7\over{8 \pi}}-{5\over
{4\pi^2}} - {{2(\pi-2)}\over{\pi^3}} g(m) + {1\over m} \Big({1\over\pi^4} -
{3\over{2\pi^3}} + {3\over{16\pi}} \Big) \\
\noalign{\medskip}
& & + \; {1\over m^2} \Big({13\over{24\pi^4}} - {43\over{48 \pi^3}} -
{3\over{32\pi^2}} + {13\over{64\pi}} - {(\pi-1)(\pi-2) \over 2\pi^3} g(m) \Big)
+ I_A^{h=2,{\rm cl}}.
\end{eqnarray}

The overall result for the $A$-orientation is then
\begin{eqnarray}
\noindent
{\N_{2,{\rm closed},A} \over \N}  & = & {\N_{2,{\rm closed},A}^{\rm local} \over \N} +
{\N_{2,{\rm closed},A}^{\rm loop} \over \N} - {\N_{2,{\rm closed},A}^{\theta} \over \N}  
= {1\over 8}-{3\over{8\pi}} - {1\over{\pi^2}} + {4\over{\pi^3}} +
{1\over m}\Big({1\over\pi^3} - {3\over{16\pi}}\Big) \nonumber\\
\noalign{\medskip}
&& \hspace{5mm} + \; {1\over{m^2}} 
\Big(-{1\over{4\pi^3}} + {17\over{32\pi^2}} - {17\over{64\pi}} +
{\pi-2 \over 2\pi^2} g(m) \Big) - I_A^{h=2,{\rm cl}}.
\end{eqnarray}

Terms $I_B^{h=2,{\rm cl}}$ and $I_C^{h=2,{\rm cl}}$ are defined similarly to
$I_A^{h=2,{\rm cl}}$, but with the sites in the Green functions rotated appropriately.
We then get the probability that a site a distance $m$ from the boundary has height
two:
\begin{eqnarray}
P_2^{\rm cl} (m) = P_1^{\rm cl}(m) + {\N_{2,{\rm closed},A} \over \N} + 2 {\N_{2,{\rm
closed},B} \over \N} + {\N_{2,{\rm closed},C} \over \N}.
\end{eqnarray}

Putting everything together, $P_2^{\rm cl}(m)$ can be written as the sum of a power
series in $1/m$ and three infinite sums of matrix determinants. It appears in
the next section, along with corresponding expressions for other heights and boundary
conditions. 

The analysis of $P_3(m)$ in the UHP is similar to that of $P_2(m)$. The local and loop
contributions, and the correction terms for the $\tilde\theta$-graph, can again be
calculated as matrix determinants, and expanded in $1/m$. Just as on the plane, the
key change is that for the $\tilde\theta$-graphs, we have an $-\epsilon$ bond from
$j_3$ to itself, which changes the first row of the matrix determinants in
\maybeeqs{\ref{eq:UHP.matrix.L}} and (\ref{eq:UHP.matrix.Gamma}).


\section{Graph theoretical results}

We collect in this section the results obtained by putting together the various
pieces corresponding to the local, loop and $\theta,\tilde\theta$-graphs
contributions to the 1-site probabilities $P_2(m)$ and $P_3(m)$, when the
reference site lies at a distance $m$ from the boundary. For each probability, we
consider a closed and an open boundary. As in the previous section, $I_B$ refers to
the $b$-orientation, while $I_{AC} = I_A + I_C$ is the sum of the other two
orientations.

When the boundary is closed, the probabilities are given by the following
expressions,
\bea
P_2^{\rm cl}(m) &=&
{\pi^3 - 3\pi^2 - 4\pi + 24 \over 2\pi^3} + {3\pi^2 - 6\pi - 8 \over 8\pi^3m}
- {4\pi^2 - 9\pi + 20 - 16 \pi(\pi-2) g(m) \over 8\pi^3m^2} \nonumber\\
\noalign{\medskip}
&-& I^{h=2,{\rm cl}}_{AC} - 2 I^{h=2,{\rm cl}}_B + \ldots \\
\noalign{\bigskip}
P_3^{\rm cl}(m) &=& P_2^{\rm cl}(m) - {\pi^3 - 18\pi^2 + 96 + 24\pi(\pi-2)g(m) \over
4\pi^3} + {9(\pi-4) \over 16\pi^2m} \nonumber\\
&+& {47\pi^2 - 128\pi + 52 + 24 \pi(20 - 7\pi) g(m) \over 32\pi^3m^2} 
+ 3 I_{AC}^{h=3,{\rm cl}} + 6 I_B^{h=3,{\rm cl}} + \ldots
\eea
with $g(m) = 2G_{0,0} - {1 \over 2\pi}[\log{m} + \gamma + {5 \over
2} \log{2}] = G_{2m,0} + G_{0,0} + {\cal O}(m^{-2})$.

When the boundary is open, the expressions are slightly different and read
\bea
P_2^{\rm open}(m) &=&
{\pi^3 - 3\pi^2 - 4\pi + 24 \over 2\pi^3} - {3\pi^2 - 6\pi - 8 \over 8\pi^3m}
+ {11\pi^2 - 40\pi + 64 + 32 \pi(\pi-2) \tilde g(m) \over 16\pi^3m^2} \nonumber\\
\noalign{\medskip}
&-& I^{h=2,{\rm op}}_{AC} - 2 I^{h=2,{\rm op}}_B + \ldots \\
\noalign{\bigskip}
P_3^{\rm open}(m) &=& P_2^{\rm open}(m) - {\pi^3 - 18\pi^2 + 96 + 24\pi(\pi-2)
\tilde g(m) \over 4\pi^3} - {3(3\pi^2 - 4\pi - 16) \over 16\pi^3m} \nonumber\\
&-& {19\pi^2 - 94\pi + 158 - 12\pi(20 - 7\pi) \tilde g(m) \over 16\pi^3m^2} 
+ 3 I_{AC}^{h=3,{\rm op}} + 6 I_B^{h=3,{\rm op}} + \ldots
\eea
with $\tilde g(m) = - {1 \over 2\pi}[\log{m} + \gamma + {5 \over 2} \log{2}] = 
G_{2m,0} - G_{0,0} + {\cal O}(m^{-2})$.

In the previous four expressions, the explicit terms represent the local and loop
contributions, as well as the special correction graphs encountered in the
$\theta$ and $\tilde\theta$ graphs. All expansions are made to order 2 in $m^{-1}$,
the ellipses standing for higher order terms. Finally, the remaining $I_B$ and
$I_{AC}$ pieces correspond to the infinite sum contributions of the $\theta$ and
$\tilde\theta$ graphs, which we now specify. As we have seen above, these terms 
include a summation over the discrete UHP, and also multiple integrals from the
integral representation of the Green matrix entries. They are functions of the
single variable $m$.

The functions $I_{AC}$ in the various cases are equal to
\bea
I_{AC}(m) &=& \sum_{k \in \Z} \sum_{\l=1-m}^\infty
\int\!\!\!\!\int_{-\pi}^\pi {\d\alpha_1 \d\beta_1 \over 8\pi^2}
{e^{-i\l\alpha_1} \pm e^{i(2m+\l-1)\alpha_1} \over
2-\cos{\alpha_1}-\cos{\beta_1}} \;
\int\!\!\!\!\int_{-\pi}^\pi {\d\alpha_2 \d\beta_2 \over 8\pi^2}
{e^{-i\l\alpha_2} \pm e^{i(2m+\l-1)\alpha_2} \over 2-\cos{\alpha_2}-\cos{\beta_2}}
\nonumber \\
\noalign{\medskip}
&& \hspace{-1.4cm}
\times \int\!\!\!\!\int_{-\pi}^\pi {\d\alpha_3 \d\beta_3 \over
8\pi^2} (-2i\sin{\alpha_3}) {e^{-i\l\alpha_3} \mp e^{i(2m+\l-1)\alpha_3} \over
2-\cos{\alpha_3}-\cos{\beta_3}} \; e^{ik(\beta_1+\beta_2+\beta_3)} \;
e^{i\beta_2} [\det M_A - \det M_C],
\label{IAC}
\eea
where the matrices $M_A$ and $M_C$ take the form
\be
M_A = \pmatrix{c+e & e^{i\alpha_1} & e^{i\alpha_2} & e^{i\alpha_3} \cr
\noalign{\medskip} {a+c \over 2} & 1 & 1 & 1 \cr \noalign{\medskip} 
{b+c \over 2} & e^{i\beta_1} & e^{i\beta_2} & e^{i\beta_3} \cr \noalign{\medskip}
{b+c \over 2} & e^{-i\beta_1} & e^{-i\beta_2} & e^{-i\beta_3}}, \qquad
M_C = \pmatrix{-d+e & e^{-i\alpha_1} & e^{-i\alpha_2} & e^{-i\alpha_3} \cr
\noalign{\medskip} {a-c \over 2} & 1 & 1 & 1 \cr \noalign{\medskip} 
{b-c \over 2} & e^{i\beta_1} & e^{i\beta_2} & e^{i\beta_3} \cr \noalign{\medskip}
{b-c \over 2} & e^{-i\beta_1} & e^{-i\beta_2} & e^{-i\beta_3}}.
\ee
In the integrand, the upper signs are chosen for a closed boundary, and the lower ones
for an open boundary, independently of the height value ($h=2$ or $h=3$) one considers. 
The dependence on $h$ is entirely contained in the values of the parameters 
$a,b,c,d$ and $e$, which also depend on the type of boundary condition: 

\bea
&& \hbox{height 2, closed} \;:\; \left\{\begin{array}{l} \displaystyle
a = -{1 \over 2} + {1 \over 8\pi m^2}, \quad \displaystyle c = -{2m+1 \over 4\pi
m^2}, \\
\noalign{\medskip}
\displaystyle  b = {1 \over 2} - {2 \over \pi} + {1 \over 8\pi m^2}, \quad d = 0,
\quad \displaystyle e = -{3 \over 4} + {1 \over 4\pi m} + {5 \over 16\pi m^2},
\end{array} \right. \\
\noalign{\bigskip}
&& \hbox{height 2, open} \;:\; \left\{\begin{array}{l} \displaystyle
a = -{1 \over 2} - {1 \over 8\pi m^2}, \quad \displaystyle c = {1 \over 2\pi
m}, \\
\noalign{\medskip}
\displaystyle  b = {1 \over 2} - {2 \over \pi} - {1 \over 8\pi m^2}, \quad d = 0,
\quad \displaystyle e = -{3 \over 4} - {1 \over 4\pi m} - {3 \over 16\pi m^2},
\end{array} \right. \\
\noalign{\bigskip}
&& \hbox{height 3, closed} \;:\; \left\{\begin{array}{l} \displaystyle
a = 2g(m) - {1 \over 2} + {1 \over 2\pi m} + {13 \over 48\pi m^2}, \quad
\displaystyle c = d = -{2m+1 \over 4\pi m^2}, \\ 
\noalign{\medskip}
\displaystyle  b = 2g(m) - {2 \over \pi} + {1 \over 2\pi m} + {7 \over 48\pi m^2},
\quad  e = g(m) + {1 \over 4\pi m} + {31 \over 96\pi m^2},
\end{array} \right. \\
\noalign{\bigskip}
&&\hbox{height 3, open} \;:\; \left\{\begin{array}{l} \displaystyle
a = -2\tilde g(m) - {1 \over 2} - {7 \over 48\pi m^2}, \quad
\displaystyle c = d = {1 \over 2\pi m}, \\ 
\noalign{\medskip}
\displaystyle  b = -2\tilde g(m) - {2 \over \pi} - {1 \over 48\pi m^2},
\quad  e = -\tilde g(m) - {25 \over 96\pi m^2}.
\end{array} \right.
\eea

In a similar way, the other functions $I_B$ can be written
\bea
I_B(m) &=& \sum_{k \in \Z} \sum_{\l=1-m}^\infty
\int\!\!\!\!\int_{-\pi}^\pi {\d\alpha_1 \d\beta_1 \over 8\pi^2}
{e^{-i\l \alpha_1} \pm e^{i(2m+\l-1)\alpha_1} \over 2-\cos{\alpha_1}-\cos{\beta_1}}
\; \int\!\!\!\!\int_{-\pi}^\pi {\d\alpha_2 \d\beta_2 \over 8\pi^2}
{e^{-i\l\alpha_2} \pm e^{i(2m+\l-1)\alpha_2} \over 2-\cos{\alpha_2}-\cos{\beta_2}}
\nonumber \\
\noalign{\medskip}
&& \hspace{-1cm} \times \int\!\!\!\!\int_{-\pi}^\pi {\d\alpha_3 \d\beta_3 \over
8\pi^2} {[e^{i(1-\l)\alpha_3} \pm e^{i(2m+\l-2)\alpha_3}]e^{-i\b_2} +
[e^{-i(\l+1)\alpha_3} \pm e^{i(2m+\l)\alpha_3}]e^{i\b_2}
\over 2-\cos{\alpha_3}-\cos{\beta_3}} \nonumber\\
\noalign{\medskip}
&& \hspace{7.5cm} \times \;
e^{ik(\beta_1+\beta_2+\beta_3)} \; \det M_B\,,
\eea
where the matrix $M_B$ has the form
\be
M_B = \pmatrix{a' & e^{-i\b_1} & e^{-i\b_2} & e^{-i\b_3} \cr \noalign{\smallskip}
b' & 1 & 1 & 1 \cr \noalign{\smallskip} 
c'-d' & e^{i\a_1} & e^{i\a_2} & e^{i\a_3} \cr \noalign{\smallskip}
c'+d' & e^{-i\a_1} & e^{-i\a_2} & e^{-i\a_3}}.
\ee
The choice of the signs is the same as for $I_{AC}$, and again the
actual values of the parameters $a',b',c'$ and $d'$ depend on which height variable
and which boundary condition one considers,
\bea
&& \hbox{height 2, closed} \;:\; \left\{\begin{array}{l} \displaystyle
a' = -{3 \over 4} + {1 \over 16\pi m^2}, \quad \displaystyle c' = {1 \over 4}
- {1 \over \pi} - {1 \over 16\pi m^2}, \\
\noalign{\medskip}
\displaystyle  b' = -{1 \over 4} - {1 \over 16\pi m^2}, \quad d' = 0,
\end{array} \right.\\
\noalign{\bigskip}
&& \hbox{height 2, open} \;:\; \left\{\begin{array}{l} \displaystyle
a' = -{3 \over 4} - {1 \over 16\pi m^2}, \quad \displaystyle c' = {1 \over 4}
- {1 \over \pi} + {1 \over 16\pi m^2}, \\
\noalign{\medskip}
\displaystyle  b' = -{1 \over 4} + {1 \over 16\pi m^2}, \quad d' = 0,
\end{array} \right.\\
\noalign{\bigskip}
&& \hbox{height 3, closed} \;:\; \left\{\begin{array}{l} \displaystyle
a' = g(m) + {1 \over 4\pi m} + {7 \over 96\pi m^2}, \quad 
\displaystyle c' = g(m) - {1 \over \pi} + {1 \over 4\pi m} + {7 \over 96\pi m^2}, \\
\noalign{\medskip}
\displaystyle  b' = g(m) - {1 \over 4} + {1 \over 4\pi m} + {1 \over 96\pi m^2},
\quad \displaystyle d' = {2m+1 \over 8\pi m^2},
\end{array} \right.\\
\noalign{\bigskip}
&& \hbox{height 3, open} \;:\; \left\{\begin{array}{l} \displaystyle
a' = -\tilde g(m) - {1 \over 96\pi m^2}, \quad 
\displaystyle c' = -\tilde g(m) - {1 \over \pi} - {1 \over 96\pi m^2},\\
\noalign{\medskip}
\displaystyle  b' = -\tilde g(m) - {1 \over 4} + {5 \over 96\pi m^2},
\quad \displaystyle d' = -{1 \over 4\pi m}.
\end{array} \right.
\eea

In order to make the expressions of $P_2(m)$ and $P_3(m)$ completely explicit, the
remaining task is to compute the asymptotic value of the functions $I_{AC}$
and $I_B$, for the different heights and boundary conditions. As we will see, these
functions have an expansion in inverse powers of $m$, times logarithmic
corrections, which we need to compute to order $m^{-2}$.


\section{Asymptotic analysis}

We will not give all the details of the calculation of $I_{AC}$ and $I_B$ for the
open and closed boundaries, as the analysis is technical and rather long. 
However, as an illustration of how this analysis can actually be carried out, we give
here some details of the analysis of $I_{AC}$ for the closed boundary. The
reader who is not interested in these technical details may skip this section and go
straight to the next one, where the final results for the probabilities 
$P_2(m)$ and $P_3(m)$ are given.

The analysis proceeds by a long sequence of otherwise elementary steps. As a very 
first step, the similarity of $M_A$ and $M_C$ allows us to write the difference of
their determinants (appearing in $I_{AC}^{cl}(m)$) in a simpler form:
\be
\det M_A - \det M_C = \left|\matrix{
c+d & \cos{\alpha_1} & \cos{\alpha_2} & \cos{\alpha_3} \cr \noalign{\smallskip}
c & 1 & 1 & 1 \cr \noalign{\smallskip}
c & e^{i\beta_1} & e^{i\beta_2} & e^{i\beta_3} \cr \noalign{\smallskip}
c & e^{-i\beta_1} & e^{-i\beta_2} & e^{-i\beta_3}} \right|
\; + \; \left|\matrix{
0 & i\sin{\alpha_1} & i\sin{\alpha_2} & i\sin{\alpha_3} \cr\noalign{\smallskip}
a & 1 & 1 & 1 \cr\noalign{\smallskip}
b & e^{i\beta_1} & e^{i\beta_2} & e^{i\beta_3} \cr \noalign{\smallskip}
b & e^{-i\beta_1} & e^{-i\beta_2} &
e^{-i\beta_3}} \right|,
\ee
which no longer depends on $e$. We leave the parameters $a,b,c$ and $d$ free to
cover the two cases $h=2$ and $h=3$. We denote by $U_1$ and $U_2$ the two new matrices
in the previous equation, and accordingly we decompose $I_{AC}$ into two pieces,
$I_{AC}^{\rm cl}(m) = I_1^{\rm cl}(m) + I_2^{\rm cl}(m)$. They can be handled in
the same way, so we focus on the contribution of $U_2$, that of $U_1$ being
simpler.

In (\ref{IAC}), the summation over $k$ and a number of integrations can be
performed: the summation over $k$ produces a delta function $2\pi \delta(\b_1 +
\b_2 + \b_3)$, which then makes trivial the integration over $\b_3$;
the contour integrations over the three variables $\a_1,\a_2,\a_3$ may also be carried
out. This eventually leaves an expression which involves one semi-infinite
summation over $\l$, and two integrations over $\b_1,\b_2$. Before giving the result
of these easy steps, we introduce the following notation. We set
\be
y(\b) = 2 -\cos{\b}, \qquad t(\b) = y-\sqrt{y^2-1}, \qquad u(\b) = t^{-1} =
y+\sqrt{y^2-1},
\ee
and the same functions with indices $y_j, t_j, u_j$ if the argument is $\b_j$.
They are even periodic functions of $\b_j$, with $t_j$ (resp. $u_j$) being the root of
$z^2-2y_jz+1=0$ inside (resp. outside) the unit circle, for $y>1$. The range of
values of $y(\b)$ is the interval $[1,3]$, but more importantly for what follows,
$t(\b)$ is strictly positive, smaller than 1, and takes the value 1 at $\b=0$ only
(or a multiple of $2\pi$). The functions $t(\b)$ and $u(\b)$ are continuous but
have discontinuous derivatives at $\b=0$.

By expanding the determinant of $U_2$ on the first row, we obtain
\be
\det U_2 = -i \sin{\a_1} D(\b_2,\b_3) + i \sin{\a_2} D(\b_1,\b_3) - i
\sin{\a_3} D(\b_1,\b_2),
\ee
where the function $D(x,y)$ is a 3-by-3 subdeterminant, given by
\be
D(x,y) = 4i \, \Big[
a \cos{x - y \over 2} - b  \cos{x + y \over 2} \Big] \, \sin{x-y \over 2}.
\ee

By using the following formulas, in which $n \in \Z$,
\bea
&& \int_{-\pi}^{\pi} {\d \alpha \over 2\pi} {e^{in\alpha} \over
y-\cos{\alpha}} = {t^{|n|} \over \sqrt{y^2-1}}\,, \qquad
\int_{-\pi}^{\pi} {\d \alpha \over 2\pi} {e^{in\alpha} \sin{\alpha} \over
y-\cos{\alpha}} = i \, {\rm sgn}(n) \, t^{|n|}\,, \quad \hbox{(sgn(0) = 0)}
\hspace{5mm}\\
\noalign{\medskip}
&&\int_{-\pi}^{\pi} {\d \alpha \over 2\pi} {e^{in\alpha} \sin^2{\alpha} \over
y-\cos{\alpha}} = \Big[-\sqrt{y^2-1} + {u \over 2} \,  \delta_{n^2,1} + y
\, \delta_{n,0}\Big] \, t^{|n|}\,,
\eea
the summation over $k$ and the integration over $\a_1,\a_2,\a_3$ and $\b_3$ yield
\bea
I_2^{\rm cl}(m) &=& {1 \over 16\pi^2} \int\!\!\!\!\int_{-\pi}^\pi
\d\beta_1 \d\beta_2 \; {1 \over \sqrt{y_2^2-1}} \, (e^{i\b_2} - e^{i\b_1}) \,
D(\b_2,\b_3) \nonumber\\
\noalign{\medskip}
&& \hspace{1.5cm} \sum_{\l=1-m}^\infty \; [-{\rm sgn}(\l) t_1^{|\l|} +
t_1^{2m+\l-1}] \, [t_2^{|\l|} + t_2^{2m+\l-1}]
\, [{\rm sgn}(\l) t_3^{|\l|} + t_3^{2m+\l-1}] \nonumber\\
\noalign{\medskip}
&+& {1 \over 16\pi^2} \int\!\!\!\!\int_{-\pi}^\pi
\d\beta_1 \d\beta_2 \; {\sqrt{y_3^2-1} \over \sqrt{(y_1^2-1)(y_2^2-1)}} \,
e^{i\b_2} \, D(\b_1,\b_2) \nonumber\\
\noalign{\medskip}
&& \hspace{1.5cm} \sum_{\l=1-m}^\infty \; [t_1^{|\l|} +
t_1^{2m+\l-1}] \, [t_2^{|\l|} + t_2^{2m+\l-1}]
\, [t_3^{|\l|} - t_3^{2m+\l-1}] \nonumber\\
\noalign{\medskip}
&-& {1 \over 16\pi^2} \int\!\!\!\!\int_{-\pi}^\pi
\d\beta_1 \d\beta_2 \; {1 \over \sqrt{(y_1^2-1)(y_2^2-1)}} \,
e^{i\b_2} \, D(\b_1,\b_2) \Big\{y_3 \, (1 + t_1^{2m-1}) \, (1 +
t_2^{2m-1}) \nonumber\\
\noalign{\medskip}
&& \hspace{2cm} + \half (t_1 + t_1^{2m}) \, (t_2 + t_2^{2m}) +
\half (t_1 + t_1^{2m-2}) (t_2 + t_2^{2m-2}) \Big\},
\label{i2}
\eea
where we have kept $\b_3$ as a shorthand notation for $-\b_1-\b_2$.

In the third double integral, which has no summation over $\l$, the two variables
can be decoupled, and the two simple integrals can be carried out by using
\be
\int_{-\pi}^\pi \d\b \; {t^n \, e^{ik\b} \over \sqrt{y^2-1}}  =
4\pi G_{n,k}.
\ee
The result of the integrals is then expanded for large $m$ by using the following
asymptotic expansion of the Green matrix ($k,\ell \ll x$)
\be
G_{x+\l,k} = G_{x,0} - {\l \over 2\pi x} +{\l^2-k^2 \over 4\pi x^2} + \ldots
\ee
The calculations are straightforward, and one finds that the third double integral in
(\ref{i2}) equals
\bea
&& \hspace{-5mm} (a-b)\,\Big({1 \over 2} - {4 \over \pi} - {1 \over 2\pi
m^2}\Big) \; [G_{0,0}+G_{2m,0}] + a \Big({1 \over 2} - {3 \over 2\pi} + {2
\over \pi^2}\Big) + b \Big({1 \over 8} - {2 \over \pi^2}\Big) \nonumber\\
\noalign{\medskip}
&& \hspace{2cm} + \; {a-b \over 2\pi m} \Big({1 \over 4} - {2 \over \pi}\Big)
+ {1 \over 16\pi m^2} \Big\{a\,\Big({9 \over 2} - {4 \over \pi}\Big) + b\,
\Big(-3 + {8 \over \pi}\Big)\Big\} + \ldots
\label{third}
\eea

The summations in the remaining two double integrals are similar: for $\l >0$, they
are equal up to a sign and the permutation $\b_1 \leftrightarrow \b_3$;
for $\l < 0$, they are equal up to a sign. Separating the three parts $\l>0$,
$\l=0$ and $\l<0$, which we call respectively $Y_1$, $Y_2$ and $Y_3$, we can rewrite
the first two integrals in (\ref{i2}) as
\bea
\plus {1 \over 16\pi^2} \int\!\!\!\!\int_{-\pi}^\pi
\d\beta_1 \d\beta_2 \; {D(\b_1,\b_2) \over \sqrt{(y_1^2-1)(y_2^2-1)}} \,
\Big[\sqrt{y_3^2-1} \, e^{i\b_2} - \sqrt{y_1^2-1} \, (e^{i\b_2} -
e^{i\b_3})\Big]
\nonumber\\
&& \hspace{2.5cm} \sum_{\l=0}^\infty \; (t_1^\l + t_1^{2m+\l-1}) \,
(t_2^\l + t_2^{2m+\l-1}) \, (t_3^\l - t_3^{2m+\l-1}) \nonumber\\
\noalign{\medskip}
\minus {1 \over 16\pi^2} \int\!\!\!\!\int_{-\pi}^\pi
\d\beta_1 \d\beta_2 \; {D(\b_2,\b_3) \over \sqrt{y_2^2-1}} \, (e^{i\b_2} - e^{i\b_1}) \,
(1 + t_2^{2m-1}) (1 - t_1^{2m-1} + t_3^{2m-1}) \nonumber\\
\noalign{\medskip}
\plus {1 \over 16\pi^2} \int\!\!\!\!\int_{-\pi}^\pi
\d\beta_1 \d\beta_2 \; {e^{i\b_2} D(\b_1,\b_2) \sqrt{y_3^2-1} + (e^{i\b_2} -
e^{i\b_1}) D(\b_2,\b_3) \sqrt{y_1^2-1} \over \sqrt{(y_1^2-1)(y_2^2-1)}}
\nonumber\\
&& \hspace{2.5cm} \sum_{\l=1-m}^{-1} \; (t_1^{-\l} + t_1^{2m+\l-1}) \,
(t_2^{-\l} + t_2^{2m+\l-1}) \, (t_3^{-\l} - t_3^{2m+\l-1}).
\eea

$Y_2$, the second integral in the previous expression, contains no summation, and
is the simplest one to evaluate. If one expands the integrand, each term gives rise to
a double integral which can be decoupled, computed exactly, and finally expanded by
the procedure explained above. $Y_1$ and $Y_3$ are more complicated. We will give
the details for $Y_1$ only, as it contains all the necessary ingredients to compute
not only $Y_3$ but also $I_1^{\rm cl}$, and in fact all the integrals $I_{AC}$ and 
$I_B$ needed to compute the 1--site probabilites on the UHP. 

The summation in $Y_1$ is straightforward, and yields
\bea
Y_1 &=& {1 \over 16\pi^2} \int\!\!\!\!\int_{-\pi}^\pi
\d\beta_1 \d\beta_2 \; {D(\b_1,\b_2) \over \sqrt{(y_1^2-1)(y_2^2-1)}} \,
\Big[\sqrt{y_3^2-1} \, e^{i\b_2} - \sqrt{y_1^2-1} \, (e^{i\b_2} -
e^{i\b_3})\Big] \nonumber\\
&& \hspace{2.5cm} {1 \over 1 - t_1 t_2 t_3} \, (1 + t_1^{2m-1}) \,
(1 + t_2^{2m-1}) \, (1 - t_3^{2m-1}).
\eea

Because $D(\b_1,\b_2)$ is odd under $\b_1,\b_2 \to -\b_1,-\b_2$, the exponentials can
be replaced by their imaginary parts,
\bea
&& \hspace{-.7cm} Y_1 = -{1 \over 2\pi^2} \int\!\!\!\!\int_{-\pi}^\pi
\d\beta_1 \d\beta_2 \; {Q_1(\b_1,\b_2) \over \sqrt{(y_1^2-1)(y_2^2-1)}} \,
{1 \over 1 - t_1 t_2 t_3} \times \nonumber\\ \noalign{\medskip}
&& [1 + t_1^{2m-1} + t_2^{2m-1} - t_3^{2m-1} + (t_1 t_2)^{2m-1} -
(t_1 t_3)^{2m-1} - (t_2 t_3)^{2m-1} - (t_1 t_2 t_3)^{2m-1}],
\label{y1}
\eea
where the function $Q_1$ is defined as
\be
Q_1(\b_1,\b_2) = -{i \over 8} \, D(\b_1,\b_2) \, \Big[\sqrt{y_3^2-1}
\, \sin{\b_2} - \sqrt{y_1^2-1} \, (\sin{\b_2} - \sin{\b_3})\Big].
\ee
In this expression for $Y_1$, we distinguish three pieces according
to the number of $t_j$ to a large power that they contain, namely 1, then $t_1^{2m-1}
+ t_2^{2m-1} - t_3^{2m-1}$, and finally $(t_1 t_2)^{2m-1} - (t_1 t_3)^{2m-1} - (t_2
t_3)^{2m-1} - (t_1 t_2 t_3)^{2m-1}$.

\subsection{\pbm First piece in $Y_1$: no large power of $t_j$}

This first term is singular: $1/\sqrt{(y_1^2-1)(y_2^2-1)}$ has a single pole on the
lines $\b_1=0$ and $\b_2=0$, where $Q_1(\b_1,\b_2)$ does not vanish (moreover $(1-t_1
t_2 t_3)^{-1}$ is singular at the origin $\b_1=\b_2=0$, though that singularity
alone is integrable). In order to isolate the singular part, we write this first
piece as
\bea
&& - {1 \over 2\pi^2} \int\!\!\!\!\int_{-\pi}^\pi
{\d\beta_1 \d\beta_2 \over \sqrt{(y_1^2-1)(y_2^2-1)}} \,
\Big[{Q_1(\b_1,\b_2) \over 1 - t_1 t_2 t_3} - {Q_1(0,\b_2) \over 1-t_2^2} -
{Q_1(\b_1,0) \over 1-t_1^2}\Big] \nonumber\\
\noalign{\smallskip}
&& \hspace{1cm} - {1 \over 2\pi^2} \int\!\!\!\!\int_{-\pi}^\pi
{\d\beta_1 \d\beta_2 \over \sqrt{(y_1^2-1)(y_2^2-1)}} \,
\Big[{Q_1(0,\b_2) \over 1 - t_2^2} + {Q_1(\b_1,0) \over 1-t_1^2}\Big].
\eea

The second integral contains the singularity, proportional to $G_{0,0}$. Using 
$Q_1(\b,0) = Q_1(0,\b) = -{a-b \over 4} \, \sqrt{y^2-1} \, \sin^2{\b}$, it is equal
to
\be
{1 \over 2\pi^2} {a-b \over 2} \int_{\pi}^\pi {\d\b_1 \over
\sqrt{y_1^2-1}} \int_{-\pi}^\pi \d\b_2 \, {\sin^2{\b_2} \over 1-t_2^2} =
{a-b \over \pi} (2+{\pi \over 2}) G_{0,0}.
\ee

The first integral is convergent by construction, and separating into parts
proportional to $a$ and $b$, can be written as $-{a \over 4\pi^2} T_1 + {b \over
4\pi^2} T_2$, where $T_1,T_2$ are numerical integrals:
\bea
T_1 \egal \int_0^\pi {\d\b_1 \over \sqrt{y_1^2-1}} \int_{-\pi}^\pi
{\d\b_2
\over \sqrt{y_2^2-1}} \; \Big\{ {\sin{(\b_1 - \b_2)} \over 1 - t_1t_2t_3}
\left[\sqrt{y_3^2-1} \sin{\b_2} - \sqrt{y_1^2-1} (\sin{\b_2} -
\sin{\b_3})\right] \nonumber\\
&& \hspace{4cm} + \; {\sqrt{y_1^2-1} \, \sin^2{\b_1} \over 1 - t_1^2} + 
{\sqrt{y_2^2-1} \, \sin^2{\b_2} \over 1 - t_2^2}\Big\}, \label{num1}\\
T_2 \egal \int_0^\pi {\d\b_1 \over \sqrt{y_1^2-1}} \int_{-\pi}^\pi
{\d\b_2
\over \sqrt{y_2^2-1}} \; \Big\{ {\sin{\b_1} - \sin{\b_2} \over 1 - t_1t_2t_3}
\left[\sqrt{y_3^2-1} \sin{\b_2} - \sqrt{y_1^2-1} (\sin{\b_2} -
\sin{\b_3})\right] \nonumber\\
&& \hspace{4cm} + \; {\sqrt{y_1^2-1} \, \sin^2{\b_1} \over 1 - t_1^2} + 
{\sqrt{y_2^2-1} \, \sin^2{\b_2} \over 1 - t_2^2}\Big\}.\label{num2}
\eea
These integrals cannot be factorized, and therefore not easy to evaluate
analytically, but a numerical integration yields the values
\be
T_1 = 16.0897, \qquad  T_2 = 17.4056.
\ee
The contribution of the first term is thus
\be
{\rm first\ term\ of\ }Y_1 = (a-b) {\pi+4 \over 2\pi} G_{0,0} -
{a \over 4\pi^2} T_1 + {b \over 4\pi^2} T_2.
\label{y1fir}
\ee

\subsection{\pbm Second piece in $Y_1$: one large power of a $t_j$}

The second piece contains the three terms in $Y_1$ that involve only
one $t_j$ to a large power, see (\ref{y1}). Making appropriate permutations of the
variables and using the reflection symmetry $Q_1(-\b_1,-\b_2) = Q_1(\b_1,\b_2)$, it
may be written as
\be
-{1 \over 2\pi^2} \int\!\!\!\int_{-\pi}^\pi {\d\beta_1 \d\beta_2 \over
\sqrt{y_2^2-1}} \,
{t_1^{2m-1} \over 1 - t_1 t_2 t_3} \Big\{2{Q_1^{\rm sym}(\b_1,\b_2) \over
\sqrt{y_1^2-1}} - {Q_1^{\rm sym}(\b_2,\b_3) \over
\sqrt{y_1^3-1}} \Big\} \equiv -{1 \over 2\pi^2} \int_0^\pi \d\b_1 \, t_1^{2m-1} \,
f(\b_1),
\label{g1}
\ee
where $Q_1^{\rm sym}(x,y) = \half[Q_1(x,y) + Q_1(y,x)]$ is the symmetrized function.

The function $t_1 = t(\b_1)$ equals 1 at $\b_1=0$ and decreases monotonically as
$\b_1$ increases towards $\pi$. Therefore for $m$ large, $t_1^m$ is exponentially
decreasing, so that the dependence in $m$ of the integral is controlled by the way the
function $f$ behaves near $\b_1=0$. 

To see this more precisely, let us first establish the following bound, which will
prove extremely useful later,
\be
\Big|t(\b)^m - e^{-m|\beta|}(1 + {m|\beta|^3 \over 12} - {m|\beta|^5
\over 96})\Big| \leq {162 \, e^{-6} \over m^4} + {\cal O}(m^{-6}).
\label{bound}
\ee
To prove it, one may notice that the function on the l.h.s. (i.e. the difference)
vanishes at $\beta=0$, and has a unique maximum which moves towards the
origin and decreases in height as $m$ increases. The location of the maximum,
$\beta^*$, can be solved iteratively as a series in $1 \over m$, yielding $\beta^* =
{6 \over m} + {237 \over 70m^3} + {2607 \over 196m^5} + \ldots$. One then simply
bounds the l.h.s. by its value at $\beta^*$. The function $e^{-m\beta}(1 + {m\beta^3
\over 12} - {m\beta^5 \over 96})$ was chosen so that its Taylor expansion around zero
matched that of $t(\beta)^m$ to order 5.

Assume that the power expansion of $f(\b_1)$ contains a term of dimension $k \geq -1$,
namely a term $\b_1^k$, times possible powers of $\log{\b_1}$. Then at the
order $m^{-2}$, and  for $k>-1$,
\be
\int_0^\pi \d\b_1 \, t_1^m \b_1^k \simeq \int_0^\infty \d\b_1 \, 
e^{-m\b_1}(1 + {m\b_1^3 \over 12} - {m\b_1^5 \over 96}) \b_1^k \simeq \int_0^\infty
\d\b_1 \, e^{-m\b_1}\b_1^k = \O(m^{-1-k}).
\ee
We see that we may simply replace $t_1^m$ by $e^{-m\b_1}$, and that the terms in
the expansion of $f$ of degree 2 or higher in $\b_1$ do not contribute at order
$m^{-2}$. A simple dimensional analysis shows that powers of $\log{\b_1}$ simply bring
powers of $\log{m}$ in the result of the integral.

When $k=-1$, the previous result essentially goes through (up to a divergent 
constant term and a logarithmic term). Since $\sqrt{y_1^2-1} = \b_1 + \ldots$ near
the positive origin, one has
\be
\int_0^\pi \d\b_1 \, {t_1^{2m} \over \b_1} = \int_0^\pi \d\b_1 \, {t_1^{2m} \over 
\sqrt{y_1^2-1}} + \O(m^{-1}) = 2\pi G_{2m,0} + \O(m^{-1}).
\label{minus1}
\ee

Coming back to (\ref{g1}), the function $f(\b_1)$ is itself given by an integral
\be
f(\b_1) = 2\int_{-\pi}^\pi {\d\beta_2 \over \sqrt{y_2^2-1}} \,
{1 \over 1 - t_1 t_2 t_3} \Big\{2{Q_1^{\rm sym}(\b_1,\b_2) \over
\sqrt{y_1^2-1}} - {Q_1^{\rm sym}(\b_2,\b_3) \over \sqrt{y_1^3-1}} \Big\}
\equiv \int_{-\pi}^\pi \d\b_2 \, {G_1(\b_1,\b_2) \over 1 - t_1 t_2 t_3}.
\ee

To evaluate the dimension of $f(\b_1)$ near $\b_1=0$, one can make a Laurent
expansion in $\b_1$ inside the integral, that is, of the integrand $G_1(\b_1,\b_2)/(1
- t_1 t_2 t_3)$. One easily checks that it starts with a dimension $-1$ term near
$\b_1=0$, and so does the expansion of $f(\b_1)$. It implies that the integral
(\ref{g1}) will contribute to order $m^0$ (a divergent term), and to orders $m^{-1}$
and $m^{-2}$. 

Although this procedure of making the Laurent expansion go through the integral---we
call it the ETI procedure, Expand Then Integrate---gives the right dimension of $f$, 
it is not valid in general when the actual series expansion is required. In
particular, it can lead to wrong results when the expansion in powers of $\b_1$ of the
integrand has coefficients which are non-integrable functions of $\b_2$, thereby
producing potentially spurious divergences which the original integral does not have.
These spurious divergences are the signal that the integral defining the function $f$
in fact does not have a Laurent expansion. In our case, this phenomenon will manifest
itself by the presence of logarithms.  

We decompose the integration domain of $\b_2$ into three pieces, $[0,\pi], [-\b_1,0]$
and $[-\pi,-\b_1]$, and first consider the region $[0,\pi]$. The expansion of $G_1$
in $\b_1$ is well--defined, with regular coefficients,
\be
G_1(\b_1,\b_2) = {c_{-1}(\b_2) \over \b_1} + c_0(\b_2) + c_1(\b_2) \b_1 +
\ldots 
\ee
Since $1-t_1t_2t_3$ does not vanish (for $\b_1>0$), the previous expansion may be
inserted in the integral over $\b_2$, leading to a sum of terms of the form
$\int {\rm d}\b_2 \; c(\b_2)/(1-t_1t_2t_3)$, where $c(\b_2)$ is one of the
coefficients in the expansion of $G_1$. These in turn can be written
\bea
\int_0^\pi \d\b_2 \, {c(\b_2) \over 1 - t_1t_2t_3} \egal 
\int_0^\pi \d\b_2 \, \Big\{{c(\b_2) \over 1 - t_1t_2t_3} - {c(0)-\b_1 c'(0) 
+ \b_1^2 [c(0)/8+c''(0)/2] \over 2\sqrt{y_3^2-1}}
\Big\} \nonumber\\
\noalign{\medskip}
\plus  \int_0^\pi \d\b_2 \, {c(0)-\b_1 c'(0) + \b_1^2 [c(0)/8+c''(0)/2] \over
2\sqrt{y_3^2-1}}.
\eea
The virtue of this decomposition is that the first integral, with its subtraction
term, has a $\b_1$ expansion with coefficients which are integrable functions of
$\b_2$ (up to order 2 in $\b_1$), for which the ETI procedure may be used. Thus
\bea
\int_0^\pi \d\b_2 \, {c(\b_2) \over 1 - t_1t_2t_3} &=& 
\int_0^\pi \d\b_2 \, {c(\b_2) \over 1 - t_1t_2t_3}\eti \nonumber\\
\noalign{\medskip}
&& \hspace{-3.2cm} +\; {1 \over 2} \Big(c(0)-\b_1 c'(0) + \b_1^2 \Big[{c(0) \over 8} +
{c''(0) \over 2}\Big] + \ldots\Big) \int_0^\pi \d\b_2
\, \Big\{{1 \over \sqrt{y_3^2-1}} - {1 \over \sqrt{y_3^2-1}}\eti \Big\}.
\eea
{}From the explicit expansion of the coefficients near the positive origin,
$\beta_2 \sim 0^+$,
\be
c_{-1}(\b_2) = -(a-b) \b_2^2 + \ldots, \quad c_0(\b_2) = {a-b \over 4}
\b_2^3 + \ldots, \quad c_1(\b_2) = {3(a-b) \over 2} + \ldots,
\ee
we obtain
\be
\int_0^\pi \d\b_2 {G_1(\b_1,\b_2) \over 1 - t_1 t_2 t_3} = 
\int_0^\pi \d\b_2 {G_1(\b_1,\b_2) \over 1 - t_1 t_2 t_3}\eti
+ {a-b \over 4} \b_1 \int_0^\pi \d\b_2
\Big\{{1 \over \sqrt{y_3^2-1}} - {1 \over \sqrt{y_3^2-1}}\eti \Big\} + \ldots
\ee

The last two integrals can be computed explicitely to order 0 in $\b_1$, 
\bea
\int_0^{\pi} {\d\b_2 \over \sqrt{y_3^2-1}} \egal 
\int_{\b_1}^{\pi+\b_1} {\d\b_2 \over \sqrt{y_2^2-1}} \simeq 
\int_{\b_1}^{\pi} {\d\b_2 \over \sqrt{y_2^2-1}} =  
{\rm arcth}{\sqrt{2} \cos{\b_1 \over 2} \over \sqrt{3+\cos{(\pi-\b_1)}}}
\nonumber\\
\noalign{\medskip}
\egal -\log{\b_1} + \txt{{3 \over 2}} \log{2} + \ldots
\eea
whereas the other one yields $2\pi G_{0,0}$ to order 0. We find
\be
\int_0^\pi \d\b_2 \, {G_1(\b_1,\b_2) \over 1 - t_1 t_2 t_3} = 
\int_0^\pi \d\b_2 \, {G_1(\b_1,\b_2) \over 1 - t_1 t_2 t_3}\eti
-{a-b \over 4} \b_1 \, [2\pi G_{0,0} + \log{\b_1} - \txt{{3 \over 2}} \log{2}] + \ldots
\ee

The region $\b_2 \in [-\b_1,0]$ is left invariant under the change of variable $\b_2
\to -\b_1-\b_2=\b_3$. Thus the integrand may be symmetrized in $\b_2
\leftrightarrow \b_3$,
\be
\int_{-\b_1}^0 \d\b_2 \, {G_1(\b_1,\b_2) \over 1 - t_1 t_2 t_3} = {1 \over 2}
\int_{-\b_1}^0 \d\b_2 \, {G_1(\b_1,\b_2) + G_1(\b_1,\b_3) \over 1 - t_1 t_2
t_3}.
\ee
The two variables $\b_1$ and $\b_2$ being small, a double series expansion can be
used which shows that the integrand has dimension 4 (in the two variables).
Therefore the integral over $\b_2$ brings a contribution at order $\b_1^5$ and higher,
and can be neglected.

The third and last portion $[-\pi,-\b_1]$ is sent onto $[0,\pi-\b_1]$ by the change
of variable $\b_2 \to \b_3$, which is then further extended to $[0,\pi]$, since the
integrand is of order $\b_1$ around $\b_2 = \pi$:
\be
\int_{-\pi}^{-\b_1} \d\b_2 \, {G_1(\b_1,\b_2) \over 1 - t_1t_2t_3} =
\int_0^{\pi-\b_1} \d\b_2 \,{\tilde G_1(\b_1,\b_2) \over 1 - t_1t_2t_3} \simeq
\int_0^{\pi} \d\b_2 \,{\tilde G_1(\b_1,\b_2) \over 1 -
t_1t_2t_3},
\ee
where $\tilde G_1(\b_1,\b_2) = G_1(\b_1,\b_3)$.

Repeating the same calculation as with $G_1$, we obtain
\bea
\int_{-\pi}^{-\b_1} \d\b_2 {G_1(\b_1,\b_2) \over 1 - t_1t_2t_3} \egal
\int_0^\pi \d\b_2 {\tilde G_1(\b_1,\b_2) \over 1 - t_1 t_2 t_3}\Big|_{\rm
ETI} - {7(a-b) \over 4} \b_1 [2\pi G_{0,0} + \log{\b_1} - \txt{{3 \over 2}} \log{2}] + \ldots
\nonumber\\
\eea

All together, we find that the function $f$ is given by
\be
f(\b_1) = \int_0^\pi \d\b_2 \, {G_1 + \tilde G_1 \over 1-t_1t_2t_3}\Big|_{\rm
ETI} - 2(a-b) \b_1 \, [2\pi G_{0,0} + \log{\b_1} - \txt{{3 \over 2}} \log{2}] +
\ldots
\ee

The remaining integral must be computed by expanding the integrand in powers of
$\b_1$, and then integrating term by term. Given the expansion 
\be
{G_1 \over 1 - t_1t_2t_3} = {d_{-1}(\b_2) \over \b_1} +
d_0(\b_2) + \b_1 d_1(\b_2) + \ldots,
\ee
we find
\bea
{G_1 + \tilde G_1 \over 1-t_1t_2t_3} &=& {d_{-1}(\b_2) + d_{-1}(-\b_2) \over
\b_1} + [d_0(\b_2) + d_0(-\b_2) - d_{-1}'(-\b_2)] \nonumber\\
&+& \b_1 \,  [d_1(\b_2) + d_1(-\b_2) - d_0'(-\b_2) + \half d_{-1}''(-\b_2)]
+ \ldots 
\label{expan}
\eea

The symmetrized coefficients read
\bea
&& \hspace{-8mm} d_{-1}(\b_2) + d_{-1}(-\b_2) = - (a-b) \, \sin^2{\b_2} \,
{2-\cos{\b_2} - \sqrt{y_2^2-1} \over \sqrt{y_2^2-1}}\,, \\
\noalign{\medskip}
&& \hspace{-8mm} d_0(\b_2) + d_0(-\b_2) = {\cos^2{\b_2 \over 2} \over (3-\cos{\b_2})
\sqrt{y_2^2-1}} \, \Big[P_3(\cos{\b_2}) + P_2(\cos{\b_2}) \sqrt{y_2^2-1}\Big] \,, \\
\noalign{\medskip}
&& \hspace{-8mm} d_1(\b_2) + d_1(-\b_2) =
{1 \over (3-\cos{\b_2}) (y_2^2-1)} \, \Big[P_5(\cos{\b_2}) +
P_4(\cos{\b_2}) \sqrt{y_2^2-1}\Big]\,,
\eea
where $P_2,P_3,P_4$ and $P_5$ are some polynomials of degree 2 up to 5.

The integrals of these coefficients over $[0,\pi]$ can be done exactly.
The first two are integrable, but the third one produces a divergent term due to
the non-zero value of $P_4(1) = 4(a-b)$. One finds
\bea
&& \hspace{-8mm} \int_0^\pi \;
[d_{-1}(\b_2) + d_{-1}(-\b_2)] = -(a-b){\pi+4 \over 2},\\
\noalign{\medskip}
&& \hspace{-8mm} \int_0^\pi \;
[d_0(\b_2) + d_0(-\b_2)] = 4a + a {(5\sqrt{2}-8)\pi \over 2} -
b {(\sqrt{2}-1)\pi \over 2}, \\
\noalign{\medskip}
&& \hspace{-8mm} \int_0^\pi \;
[d_1(\b_2) + d_1(-\b_2)] = 8\pi (a-b) G_{0,0} + {7b-a \over 3} + a
{(39\sqrt{2}-29)\pi \over 24} - b{(15\sqrt{2}+13)\pi \over 24}.
\eea

The integral of the derivative terms in the expansion (\ref{expan}) poses no
problem and simply yields ${a-b \over 4} \b_1$. Collecting the various results, the
function $f(\b_1)$ is found to be equal to
\bea
f(\b_1) \egal - {(a-b) \over \b_1}{\pi+4 \over 2} + 4a + a {(5\sqrt{2}-8)\pi \over 2}
- b {(\sqrt{2}-1)\pi \over 2} + \b_1 \Big\{2(a-b) \,[2\pi G_{0,0} \nonumber\\
&-& \log{\b_1} + \txt{{3 \over 2}} \log{2}] 
+ {7b-a \over 3} + a {(39\sqrt{2}-29)\pi \over 24} - b{(15\sqrt{2}+13)\pi \over
24}\Big\} + \ldots
\eea

It remains to multiply this function by $t_1^{2m-1}$ and to integrate over
$\b_1$. The integrals are carried out by using the following formulas:
\bea
&& \hspace{-8mm} \int_0^\pi \d\b_1 \, {t_1^{2m-1} \over \b_1} = 2\pi G_{2m,0} +
{1 \over 2m} + {7 \over 48m^2} + \ldots \\
&& \hspace{-8mm} \int_0^\pi \d\b_1 \, t_1^{2m-1} = {1 \over 2m} + {1 \over 4m^2}
+ \ldots \qquad \qquad \int_0^\pi \d\b_1 \, t_1^{2m-1}\,\b_1 = {1 \over 4m^2} +
\ldots \\
&& \hspace{-8mm} \int_0^\pi \d\b_1 \, t_1^{2m-1}\,\b_1 \,\log{\b_1} = {1 \over
4m^2} [1 - \gamma - \log{2m}] + \ldots 
\eea
We eventually find
\bea
\hbox{second piece of }Y_1 &=& - {1 \over 2\pi^2} \int_0^\pi \d\b_1 \, t_1^{2m-2} \,
f(\b_1) \nonumber\\
\noalign{\medskip}
&& \hspace{-34mm} = (a-b){\pi+4 \over 2\pi} G_{2m,0} - {1 \over 4\pi^2 m}
\Big\{2(a+b) + {(5a-b)\pi \over \sqrt{2}} + {(2b-9a)\pi \over 2}\Big\} \nonumber \\
\noalign{\medskip}
&& \hspace{-34mm} - \;{a-b \over 2\pi m^2} \, [2G_{0,0} -
G_{2m,0}] - {1 \over 16\pi^2 m^2} \Big\{a + 11b + {(33a-9b)\pi \over 2\sqrt{2}} -
{(22a-b)\pi \over 2}\Big\} + \ldots \qquad
\label{y1sec}
\eea

\subsection{\pbm Third piece in $Y_1$: two and three large powers of $t_j$}

{}From (\ref{y1}), the last piece reads
\be
-{1 \over 2\pi^2} \int\!\!\!\!\int_{-\pi}^\pi \d\beta_1 \d\beta_2
{Q_1^{\rm sym}(\b_1,\b_2) \over \sqrt{(y_1^2-1)(y_2^2-1)}( 1 - t_1 t_2
t_3)} \, [(t_1t_2)^{2m-1} - 2 (t_1t_3)^{2m-1} - (t_1t_2t_3)^{2m-1}].
\ee
It is simpler to evaluate then the previous one because it contains at least two $t_j$
to a large power. This implies that the value of the integral is determined by the
behaviour of the rest of the integrand around the origin, so that a double series expansion
 around $\b_1 = \b_2 = 0$ can be used. The same dimensional analysis as before shows
that the expansion of ${Q_1^{\rm sym}(\b_1,\b_2)/\sqrt{(y_1^2-1)(y_2^2-1)}( 1 - t_1
t_2 t_3)}$ may be stopped to terms of global dimension 0 (in the two variables).

Special attention must be paid to the double expansions. Each of the three 
functions $\sqrt{y_j^2-1}$ has discontinuous derivatives on the
line $\b_j=0$. As the integrand involves them all, it has itself discontinuities when
one of the lines $\b_1=0$, $\b_2=0$ and $\b_1+\b_2=0$ is crossed. As a consequence,
the partial derivatives no longer commute, and a series expansion must be
computed separately in each of the six regions obtained when the domain
$[-\pi,\pi]^2$ is cut by the three lines. 

Since the integrand is even under $\b_1,\b_2 \to -\b_1,-\b_2$, the integration
can be restricted to half the square $\b_2 \geq 0$. This leaves three separate regions
which we denote by (A), (B) and (C):
\be
({\rm A}) = [0,\pi]^2, \qquad ({\rm B}) = \{\b_1 \leq 0,\, \b_2 \geq 0,\, \b_3 \leq 0\},
\qquad ({\rm C}) = \{\b_1 \leq 0,\, \b_2 \geq 0,\, \b_3 \geq 0\}.
\ee
In (A), the expansion is carried out for positive values of the two variables, in
any order we want. In (B) and (C), the expansion is made for negative values of
$\b_1$, but the order matters: the expansion in $\b_1$ is made first in (B), last
in (C). 

One may easily check that both the numerator $Q_1^{\rm sym}(\b_1,\b_2)$ and the
denominator $\sqrt{(y_1^2-1)(y_2^2-1)}(1 - t_1 t_2 t_3)$ have dimension 3 around
the origin, in the three regions (A), (B) and (C). Thus the ratio has
dimension zero, so that the above integral has a single contribution, at order
$m^{-2}$. The explicit expansions read 
\be
Q_1^{\rm sym}(\b_1,\b_2) = \cases{
-\txt{{1 \over 4}}(a-b) \, (\b_1+\b_2)(\b_1-\b_2)^2 + \ldots & in region (A),\cr
\txt{{1 \over 4}}(a-b) \, \b_2(\b_1-\b_2)(2\b_1+\b_2) + \ldots & in region (B),\cr
\txt{{1 \over 4}}(a-b) \, \b_1(\b_1-\b_2)(\b_1+2\b_2) + \ldots & in region (C),}
\ee
and
\be
\sqrt{(y_1^2-1)(y_2^2-1)}( 1 - t_1t_2t_3) = \cases{
2\b_1\b_2(\b_1+\b_2) + \ldots & in region (A), \cr
-2\b_1\b_2^2 + \ldots & in region (B), \cr
2\b_1^2\b_2 + \ldots & in region (C).}
\ee

The integrals can now be computed in the three regions (the formula (\ref{minus1})
is useful). The integration over (A) yields (the exponents $2m-1$ of the $t_j$
variables have been replaced by $2m$ since this makes no difference at order $m^{-2}$)
\bea
&& {a-b \over 16\pi^2} \int\!\!\!\!\int_{\rm (A)} \d\b_1\d\b_2 {(\b_1 - \b_2)^2
\over \b_1\b_2} [e^{-2m(\b_1+\b_2)} - 2 e^{-2m(2\b_1+\b_2)} - e^{-4m(\b_1+\b_2)}]
\nonumber\\
&& \hspace{2cm} = \; {a-b \over 64\pi^2m^2} \Big[-2\pi G_{2m,0} + {1 \over 2} + {5
\over 2} \log{2}\Big] + \ldots
\eea
The other two are similar,
\bea
&& {a-b \over 16\pi^2} \int\!\!\!\!\int_{\rm (B)} \d\b_1\d\b_2 {(\b_1 - \b_2)(2\b_1
+ \b_2) \over \b_1\b_2} [e^{2m(\b_1-\b_2)} - 2 e^{-2m\b_2} - e^{-4m\b_2}]
\nonumber\\
&& \hspace{2cm} = \; {a-b \over 64\pi^2m^2} \Big[-{5\pi \over 2} G_{2m,0} + {13 \over
4} - {11 \over 4} \log{2}\Big] + \ldots \\
\noalign{\medskip}
&& -{a-b \over 16\pi^2} \int\!\!\!\!\int_{\rm (C)} \d\b_1\d\b_2 {(\b_1 - \b_2)(\b_1
+ 2\b_2) \over \b_1\b_2} [e^{2m(\b_1-\b_2)} - 2 e^{2m(2\b_1+\b_2)} - e^{4m\b_1}]
\nonumber\\
&& \hspace{2cm} = \; {a-b \over 64\pi^2m^2} \Big[{\pi \over 2} G_{2m,0} + {21 \over 4}
 - {27 \over 4} \log{2}\Big] + \ldots
\eea

Adding these three contributions, each taken twice to cover the full domain $[-\pi,\pi]^2$,
one eventually arrives at
\be
\hbox{third piece of }Y_1 = {a-b \over 8\pi^2 m^2} \; \Big[{9 \over 4} -
{7 \over 4} \log{2} - \pi G_{2m,0}\Big] + \ldots
\label{y1thi}
\ee

\subsection{Final results}

Adding the three contributions (\ref{y1fir}), (\ref{y1sec}) and (\ref{y1thi})
yields our final result for the quantity $Y_1$ defined in (\ref{y1}),
\bea
Y_1 &=& (a-b) {\pi+4 \over 2\pi} [G_{0,0} + G_{2m,0}] - {a \over 4\pi^2} T_1 + 
{b \over 4\pi^2} T_2 \nonumber\\
\noalign{\medskip}
&-& {1 \over 4\pi^2 m} \Big\{2(a+b) + {(5a-b)\pi \over \sqrt{2}} + {(2b-9a)\pi \over
2}\Big\} - {a-b \over 8\pi m^2} \, [8G_{0,0} - 3G_{2m,0}] \nonumber\\
\noalign{\medskip}
&+& {1 \over 32\pi^2 m^2} \Big\{7a - 31b - {(33a-9b)\pi \over \sqrt{2}} + (22a-b)\pi
- 7(a-b)\log{2} \Big\} + \ldots
\label{y1fin}
\eea

As mentioned earlier, the other terms $Y_2$ and $Y_3$ can be computed in a similar
way, with analogous results. Together with (\ref{third}), $Y_1$, $Y_2$ and $Y_3$
complete the calculation for $I_2^{\rm cl}$. The evaluation of $I_1^{\rm cl}$
proceeds the same way and, with $I_2^{\rm cl}$, yield $I_{AC}^{\rm cl}$. The 
three other integrals, namely $I_B^{\rm cl}$, $I_{AC}^{\rm op}$ and $I_B^{\rm op}$,
are handled by the same techniques. The calculations become fairly long, but each
step is straightforward if one follows the methods used in the sample calculation
detailed above. The next section summarizes the final results.


\section{Single height probabilities on the discrete UHP}

In Section VI, explicit formulas were given for the single site probabilities
$P_2(m), P_3(m)$ on the UHP, at a site located a distance $m$ from the boundary,
chosen to be open or closed. We have shown in the previous section how to evaluate the
asymptotic expansion of the four functions, $I_{AC}^{\rm cl}$, $I_B^{\rm cl}$,
$I_{AC}^{\rm op}$ and $I_B^{\rm op}$, to second order in $m^{-1}$. We now put the
various pieces together and give the final results for the probabilities. The
divergences found in the individual terms, like in (\ref{y1fin}), all cancel out,
as they should, leaving finite probabilities.

We obtain for the closed boundary, 
\bea
P_2^{\rm cl}(m) &=& P_2 - {1 \over 8\pi^2 m^2} \left\{-11 + {34 \over \pi} +
{4(\pi - 2) \over \pi}\Big(\gamma + {5 \over 2}\log 2\Big) + {4(\pi - 2) \over \pi}
\log{m} \right\},\\
P_3^{\rm cl}(m) &=& P_3 - {1 \over 8\pi^2m^2} \left\{
\pi + {5 \over 2} - {44 \over \pi} + {2(8-\pi) \over \pi}\Big(\gamma + {5 \over 2}\log
2\Big) + {2(8-\pi) \over \pi} \log{m} \right\},
\eea
and for the open boundary,
\bea
P_2^{\rm op}(m) &=& P_2 + {1 \over 8\pi^2 m^2} \left\{-9 + {30 \over \pi} +
{4(\pi - 2) \over \pi}\Big(\gamma + {5 \over 2}\log 2\Big) + {4(\pi - 2) \over \pi}
\log{m} \right\}, \\ 
P_3^{\rm op}(m) &=& P_3 + {1 \over 8\pi^2m^2} \left\{
\pi + {3 \over 2} - {36 \over \pi} + {2(8-\pi) \over \pi}\Big(\gamma + {5 \over 2}\log
2\Big) + {2(8-\pi) \over \pi} \log{m}\right\}.
\eea
The first terms, independent of $m$, are the bulk probabilities, and
so can be written in terms of the results from Section IV.

{}From the above expressions, it is not difficult to see that the probabilities take
the form
\bea
P_i^{\rm cl}(m) &=& P_i - {1 \over m^2} (a_i + b_i \, \log{m}) + \ldots,
\label{pcl} \\
P_i^{\rm op}(m) &=& P_i + {1 \over m^2} (a_i + {b_i \over
2} + b_i \, \log{m}) + \ldots,
\label{pop}
\eea
with the following exact values for the coefficients $a_i$ and $b_i$,
\bea
a_2 &=& {\pi - 2 \over 2\pi^3} \Big(\gamma + {5 \over 2} \log{2}\Big) - {11\pi - 34
\over 8\pi^3}, \qquad b_2 = {\pi - 2 \over 2\pi^3}, \label{a2b2} \\
\noalign{\bigskip}
a_3 &=& {8 - \pi \over 4\pi^3} \Big(\gamma + {5 \over 2} \log{2}\Big) + {2\pi^2 + 5\pi
- 88 \over 16\pi^3}, \qquad b_3 = {8 - \pi \over 4\pi^3}.
\eea
For completeness, we recall that the 1-site probabilities $P^{\rm cl}_1(m)$ and $P^{\rm
op}_1(m)$ have the same form, with the following coefficients \cite{bip},
\be
a_1 = {\pi - 2 \over 2\pi^3}, \qquad b_1 = 0,
\ee
while those for $P^{\rm cl}_4(m)$ and $P^{\rm op}_4(m)$ are obtained by subtraction, 
\be
a_4 = -(a_1 + a_2 + a_3), \qquad b_4 = -(b_1 + b_2 + b_3).
\ee
As we shall see in the next sections, the identity $a_1=b_2$ is not accidental, but 
follows from the field assignment for the height 1 and 2 variables, namely the height
2 scaling field is the logarithmic partner of the height 1 field. Similar relations
will hold for $b_3$ and $b_4$.

The numerical values of these constants are
\bea
&& a_1 = 0.0184091, \quad a_2 = 0.0402789, \quad a_3 = -0.0154395, \quad a_4 = -0.0432485,\\
&& b_1 = 0, \hspace{1.9cm} b_2 = 0.0184091, \hspace{4mm} b_3 = 0.0391728,
\hspace{7.5mm} b_4 = -0.0575818.
\eea

The values of $a_2$ and $b_2$ have been announced in \cite{piru3}, though not in an
exact form for $a_2$. In the same reference, numerical values had also been given
for $a_3,b_3,a_4$ and $b_4$, and obtained by fitting numerical simulations of the
sandpile model with an Ansatz for the height 3 and 4 conformal fields. The values
quoted in \cite{piru3}, namely $a_3=-0.01243,\, b_3 = 0.03810, \, a_4 = -0.04636$ and
$b_4 = -0.05667$ ($a_4$ and $b_4$ were obtained from an independent fit, not by
subtraction), are in good agreement with the exact values given above.

The forms of (\ref{pcl}) and (\ref{pop}) make it manifest that the probabilities
$P_i(m)-P_i$ for $i=3,4$ are linear combinations of those for $i=1,2$, at order
$m^{-2}$, since they are all linear combinations of the two independent functions
$m^{-2}$ and $m^{-2}\log{m}$,
\bea
P_3(m) - P_3  &=& \alpha_3 \: [P_2(m) - P_2] + \beta_3 \: [P_1(m) - P_1],\label{p3}\\
P_4(m) - P_4  &=& \alpha_4 \: [P_2(m) - P_2] + \beta_4 \: [P_1(m) - P_1],
\label{p4}
\eea
for the two boundary conditions. The coefficients $\alpha_i$ and $\beta_i$ satisfy
$a_i = \alpha_i a_2 + \beta_i a_1$ and $b_i = \alpha_i b_2$ for $i=3,4$, and take the
actual values
\bea
\alpha_3 &=& {8-\pi \over 2(\pi-2)} \simeq 2.12791, \qquad \beta_3 = 
{\pi^3-5\pi^2+12\pi-48 \over 4(\pi-2)^2} \simeq -5.49453, \label{ab3} \\
\alpha_4 &=& -{\pi+4 \over 2(\pi-2)} \simeq -3.12791, \qquad \beta_4 = 
{32 + 4\pi + \pi^2 - \pi^3 \over 4(\pi-2)^2} \simeq 4.49453. \label{ab4}
\eea
Remarkably, it follows from these values that the linear relation we have derived
above in (\ref{relprob}) for the probabilities at infinity, actually holds everywhere
in the UHP (for large $m$ and at order $m^{-2}$, that is, in the scaling limit)
\be
{48 - 12 \pi + 5\pi^2 - \pi^3 \over 2(\pi-2)}\, P_1(m) + (\pi - 8)\,
P_2(m) + 2(\pi - 2)\, P_3(m) = {(\pi-2)(\pi-1) \over \pi}.
\ee
The constant part, corresponding to $m=\infty$, is the relation we had derived at the
end of Section IV. It is truly remarkable that the coefficients of the non-constant
parts, proportional to $1/m^2$ and $\log{m}/m^2$, obtained after a long and painful
asymptotic analysis, so conspire to fit the same linear relation. 

In fact, we argue in the next section that the relations (\ref{p3}) and (\ref{p4}) 
actually hold at the level of the fields representing the four height variables, so we
expect the last formula to hold in all generality, that is, in any geometry and for
any sort of boundary condition. Multisite probabilities would be linearly related in
the same way, leaving the probabilities for heights 1 and 2 as the only independent
ones (again in the scaling limit). The physical origin of this relation remains to be
understood.


\section{Conformal fields}

The conformal field theory interpretation of the previous results has been 
given in \cite{piru3}, which resulted in the field identification of the height 
variables in the scaling limit. We recall here the salient features of this
identification, give computational details that were omitted in \cite{piru3}, and
make further comments. 

The conformal field theory believed to describe the scaling limit of the sandpile
model has central charge $c=-2$. It is non-unitary, and logarithmic \cite{gur},
meaning that the theory possesses reducible but indecomposable representations of the
Virasoro algebra. This characteristic property implies that correlation functions
contain logarithms. 

We first review the basic ingredients we need to analyze the lattice results given
in the previous section, referring to \cite{gk} for a more complete description of
the $c=-2$ conformal theory (see also \cite{flohr} and \cite{gab} for more
general reviews on logarithmic conformal theories).

The 1-site probabilities $P_i(z) = \la \delta(h(z)-i) \ra$ are expectation values of
the random variables $\delta(h(z)-i)$, computed with respect to the stationary measure
of the sandpile model. In the continuum limit, the subtracted random variables
$\delta(h(z)-i) - P_i$, normalized to have a zero expectation value on the plane, are
expected to converge to conformal fields $h_i(z,\bar z)$, such that the multipoint
field correlators reproduce the scaling limit of the multisite probabilities. 

The lattice results show that the four fields $h_i$ have a scaling dimension equal to
2. The height 1 field $h_1(z,\bar z)$ was the first one to have been studied
\cite{majdhar,bip,mr,jeng3}, and turns out to be a primary scalar field
$\phi(z,\bar z)$ with conformal weights (1,1). This field is a true primary field,
with rational multipoint correlators.

It has been argued in \cite{piru3} that the other three fields $h_2(z,\bar
z),h_3(z,\bar z)$ and $h_4(z,\bar z)$ are all associated with another field
$\psi(z,\bar z)$, of conformal dimensions (1,1), the logarithmic partner of
$\phi(z,\bar z)$. The fields $\phi$ and $\psi$ form a rank 2 Jordan block under
dilations since their transformations read
\be
L_0 \, \phi = \phi, \qquad L_0 \, \psi = \psi - \half \phi.
\ee

The precise statement made in \cite{piru3} was that the lattice calculations for
the height 2, as well as the results of the simulations for the heights 3 and 4,
now evidenced by the exact results presented above, are all compatible with the
following assignments,
\bea
&& h_1(z,\bar z) = \phi(z,\bar z), \qquad h_2(z,\bar z) = \psi(z,\bar z), \\
&& h_3(z,\bar z) = \alpha_3 \psi(z,\bar z) + \beta_3 \phi(z,\bar z), \qquad
h_4(z,\bar z) =  \alpha_4 \psi(z,\bar z) + \beta_4 \phi(z,\bar z),
\label{h34}
\eea
with the coefficients $\alpha_i,\b_i$ given in the previous section.

The general form of the 1-point functions of the fields $h_i(z,z^*)$ on the UHP match
the dominant terms of the 1-site probabilities $P_i(m)-P_i$, when $z-z^*=2im$. The
normalizations of $h_1$ and $h_2$ can be uniquely fixed by comparing with, for
instance, the closed boundary results for $P_1$ and $P_2$. However we have to make
sure that the same normalizations would be obtained had we chosen to make the
comparisons with the open boundary results. This is a highly non-trivial
consistency check because much more of the structure of the $c=-2$ theory is
needed to prove that indeed the same normalizations would follow.

The probabilities $P_i^{\rm cl}(m)$ and $P_i^{\rm op}(m)$ refer to a different
boundary condition, and correspondingly the two 1-point functions $\la h_i(z,z^*)
\ra^{\rm cl}$ and $\la h_i(z,z^*) \ra^{\rm op}$ are related to each other by the
change of boundary condition, itself implemented by the insertion, in the
correlator, of a specific boundary field $\mu$ \cite{cardy}. In the present case,
we know from \cite{r} that the field which effects a change of boundary condition
from closed to open, and vice-versa, is a primary chiral field with conformal weight
$h=-1/8$. 

It may therefore be used to compute the exact relationship between the open
and closed results, and verify that it agrees with the lattice results (\ref{pcl})
and (\ref{pop}). The general form should be retained, but the coefficients of $1/m^2$
and $\log{m}/m^2$ should be related as $(-a_i,-b_i)^{\rm cl} \leftrightarrow (a_i+{b_i
\over 2},b_i)^{\rm op}$. This check probes correlators with two fields $\mu$, and
therefore the  operator product (OPE) of $\mu$ with itself. This OPE contains another
logarithmic field $\omega$, which itself proves extremely important to make the
conformal picture fully consistent with the lattice results. 

An even stronger and independent support for the above field identifications will be
provided by comparing conformal predictions with actual numerical simulations of the
sandpile model. Using again the $\mu$ field, one may compute the expectation values
$\la h_i(z,z^*) \ra$ on the UHP with the negative real axis closed and the positive
real axis open. They may then be conformally mapped onto an infinite strip with open
and closed boundary conditions on either side of the strip, and eventually compared
with the 1-site probability profiles across the strip, measured from numerical
simulations on a finite but sufficiently long strip. This is done in the next
section, where we will observe a very good agreement between the two sets of curves.

In order to relate the open and closed boundary expectation values of the fields
$h_i$, we consider the more general situation of an open boundary condition on the 
real axis with a finite stretch $I=[z_1,z_2]$ which is closed. The corresponding 1-site
probabilities $P_i^{{\rm op},I}(z)$, which are no longer translation invariant in the
horizontal direction, ought to be given in the scaling limit by
\be
P_i^{{\rm op},I}(z) - P_i = \la h_i(z,z^*) \ra_{{\rm op},I} = 
{\la \mu(z_1)\, \mu(z_2) \, h_i(z,z^*) \ra_{\rm op} \over \la \mu(z_1)\, \mu(z_2)
\ra_{\rm op}}.
\label{PopI}
\ee
The notation $\la \ldots \ra_{\rm op}$ implies that the boundary condition far out,
namely around the point at infinity, is open. Hence the product $\mu(z_1)\,
\mu(z_2)$ really means $\mu^{\rm op,cl}(z_1)\, \mu^{\rm cl,op}(z_2)$, since the
closed stretch is in between $z_1$ and $z_2$. The all open or all closed cases is
recovered upon taking the limit $z_1,z_2 \to 0$ or $-z_1,z_2 \to +\infty$
respectively. While the denominator is simple and can be taken to be $\la \mu(z_1)\,
\mu(z_2) \ra_{\rm op} = z_{12}^{1/4}$, the numerator is more complicated and is
computed in the next subsection.


\subsection{Conformal correlation functions}

The non-chiral 3-point function is equal to a 4-point chiral correlator, in which
two chiral fields $h_i$ are inserted at $z$ and at the mirror image point $z^*$, 
namely the left chiral field at $z$, and the right chiral field at $z^*$
\cite{cardy}. As $h_i$ is a combination of $\phi$ and $\psi$, we have to compute
the chiral 4-point functions $\la \mu \mu \phi \phi \ra$ and $\la \mu \mu \psi \psi
\ra$, the latter requiring in any case the former. So we concentrate on $\la
\mu(z_1) \mu(z_2) \psi(z_3) \psi(z_4) \ra$.

Because the field $\mu$ is degenerate at level 2, the chiral 4-point function can
be computed in the usual way, by solving first the conformal Ward identities, and
then an ordinary second order differential equation. A first complication 
arises because the field $\psi$ not only is the logarithmic
partner of a primary field $\phi$, but is also not quasi-primary. Under special
conformal transformations ($L_{1}$ and $\bar L_{1}$), it transforms into the
fields $\rho$ and $\bar \rho$, of conformal weights (0,1) and (1,0)
respectively. We will take the OPE of $\psi$ with the left and right stress-energy
tensors to be
\bea
T(z)\psi(w,\bar w) &=& {\rho(w,\bar w) \over (z-w)^3} + 
{\psi(w,\bar w) - {1 \over 2} \phi(w,\bar w) \over (z-w)^2} + 
{\partial \psi(w,\bar w)\over z-w} + \ldots \label{Tpsi}\\
\bar T(\bar z)\psi(w,\bar w) &=& {\bar\rho(w,\bar w) \over (\bar z-\bar w)^3} + 
{\psi(w,\bar w) - {1 \over 2} \phi(w,\bar w) \over (\bar z-\bar w)^2} + 
{\bar\partial \psi(w,\bar w)\over \bar z-\bar w} + \ldots.\label{Tbarpsi}
\eea
The factor $-1/2$ in front of $\phi$ is conventional (when $\psi$ is normalized 
to be the height 2 variable in the sandpile model, it ensures that $\phi$ has the
right normalization to be the height 1 variable). For more generality, we replace it
by $\lambda$ in what follows. 

A second complication comes from the fact that the fields $\rho$ and $\bar \rho$ are
themselves not fully primary: $\rho$ is a left primary (with left weight 0), but is
not quasi-primary under antiholomorphic conformal maps, 
\bea
T(z)\rho(w,\bar w) &=& {\partial \rho(w,\bar w)\over z-w} + \ldots \\
\bar{T}(\bar z)\rho(w,\bar w) &=& {\kappa \over (\bar z-\bar w)^3} + 
{\rho(w,\bar w) \over (\bar z-\bar w)^2} + {\bar\partial \rho(w,\bar w)\over 
\bar z-\bar w} + \ldots
\eea
It is the converse situation for $\bar \rho$, which is right primary but not left
quasi-primary, 
\bea
T(z)\bar \rho(w,\bar w) &=& {\kappa \over (z-w)^3} + 
{\bar \rho(w,\bar w) \over (z-w)^2} + {\partial \bar \rho(w,\bar w)\over z-w} +
\ldots\\
\bar T(z)\bar \rho(w,\bar w) &=& {\bar\partial \bar \rho(w,\bar w)\over \bar z-\bar w}
+
\ldots
\eea
The coefficients of the cubic terms are equal (to $\kappa$), on account of
$\bar L_1 \rho = \bar L_1 L_1 \psi = L_1 \bar L_1 \psi = L_1 \bar \rho$.

The inhomogeneous transformations of $\psi$ implies that the correlator $\la \mu
\mu \psi(z) \psi(z^*)\ra$ depends on eight other 4-point functions, in which at
least one $\psi$ has been replaced by either $\phi$ or $\rho$. (Note that the fields
$\psi(z)$ resp. $\psi(z^*)$ are subjected to left resp. right conformal transformations
only, so that the chiral $\rho$ and $\bar\rho$ behave as primary fields; since
$\bar\rho$ behaves under right transformations as $\rho$ does under left
transformations, we use the same notation $\rho$ for both.) As $\rho$ and $\phi$ are
chiral primary fields, the correlators $\la \mu\mu\rho\rho \ra$, $\la \mu\mu\phi\rho
\ra$, $\la \mu\mu\rho\phi \ra$ and $\la \mu\mu\phi\phi \ra$ can be computed separately
as solutions of a second or first order homogeneous differential equation. Once these
are known, the four correlators containing one $\psi$ field can be computed, namely
$\la \mu\mu\rho\psi \ra$,  $\la \mu\mu\psi\rho \ra$, $\la \mu\mu\psi\phi \ra$ and 
$\la \mu\mu\phi\psi \ra$, which depend on the first four. Finally $\la \mu\mu
\psi\psi \ra$ can be computed as a solution of a second order inhomogeneous
differential equation. More details on the procedure may be found in \cite{flohr}.

As usual, the conformal Ward identites are used the fix the kinematical factors in
the correlators, leaving an unknown function of the cross ratio $x$, which we will
take to be $x = {z_{12}z_{34} \over z_{13}z_{24}}$. The Ward identities read ($\la
\ldots \ra$ is any one of the nine correlators)
\bea
&& \Big\{\sum_{i} \p_{i} \Big\} \la \ldots \ra = 0,\\
&& \Big\{\sum_{i} (z_{i}\p_{i} + h_{i}) + 
\hat\delta_i \Big\} \la \ldots \ra = 0,\\
&& \Big\{\sum_{i} (z_{i}^{2}\p_{i} + 2h_{i}z_{i}) + 
2z_i \hat\delta_i + \hat\varepsilon_i \Big\} \la \ldots \ra = 0.
\eea
The operator $\hat\delta$ and $\hat\varepsilon$ take into account the inhomogeneous 
parts in the conformal transformations of $\psi$ under $L_0$ and $L_1$: $\hat\delta
\psi = \lambda\phi$ and $\hat \varepsilon \psi = \rho$, whereras $\hat\delta$ and
$\hat\varepsilon$ annihilate all the other fields.

All nine correlators contain a $\mu$ field and as a consequence of
its degeneracy at level 2, they satisfy a second order partial differential
equation,
\be
\Big\{2 \p_1^2 - \sum_{i=2}^{4} \Big[{\p_{i} \over z_{1}-z_{i}} + 
{h_{i} + \hat\delta_{i}\over (z_{1}-z_{i})^{2}} + {\hat \varepsilon_i \over
 (z_1-z_i)^{3}}\Big] \Big\} \la \ldots \ra = 0.
 \label{pdemu}
\ee
A correlator which contains a $\phi$ satisfies a different second
order differential equation, since $\phi$ is also degenerate at level 2. If the
field $\phi$ is at $z_{a}$, it reads
\be
\Big\{{1 \over 2} \p_a^2 - \sum_{i \neq a} \Big[{\p_{i} \over z_{a}-z_{i}} + 
{h_{i} + \hat\delta_{i}\over (z_{a}-z_{i})^{2}} + {\hat \varepsilon_i \over
 (z_a-z_i)^{3}}\Big] \Big\} \la \ldots \ra = 0.
 \label{pdephi}
\ee

As they may be of interest for other calculations, we give some details about the
derivation of the relevant 4-point correlators. The easiest correlators are those
without $\psi$. As their calculation is standard, we merely give the results,
\bea
\la \mu(1) \mu(2) \rho(3) \rho(4) \ra &=& z_{12}^{1/4} \, F_{\rho\rho}(x),
\qquad F_{\rho\rho} = A_{1} + A_{2} \log{1-\sqrt{1-x} \over 
1+\sqrt{1-x}},\label{rhorho}\\
\la \mu(1) \mu(2) \phi(3) \phi(4) \ra &=& {z_{12}^{1/4} \over 
z_{34}^{2}} \, F_{\phi\phi}(x), \qquad F_{\phi\phi} = B {x-2 \over 
\sqrt{1-x}}, \\
\la \mu(1) \mu(2) \phi(3) \rho(4) \ra &=& {z_{12}^{5/4} \over 
z_{13}z_{23}} \, F_{\phi\rho}(x), \qquad F_{\phi\rho} = C {\sqrt{1-x} 
\over  x},\\
\la \mu(1) \mu(2) \rho(3) \phi(4) \ra &=& {z_{12}^{5/4} \over 
z_{14}z_{24}} \, F_{\rho\phi}(x), \qquad F_{\rho\phi} = D {\sqrt{1-x} 
\over x}.
\eea
The five coefficients $A_{1},A_{2},B,C$ and $D$ are integration constants.

The conformal Ward identities satisfied by the correlator $\la \mu \mu \psi \rho 
\ra$ require that it be of the following general form,
\be
\la \mu(1) \mu(2) \psi(3) \rho(4) \ra = {z_{12}^{5/4} \over 
z_{13}z_{23}} \Big[ F_{\psi\rho} - \lambda F_{\phi\rho} \, \log\Big({z_{13}z_{23} 
\over z_{12}}\Big) + {z_{23} \over z_{12}} \, F_{\rho\rho}\Big].\label{psirho}
\ee
When one inserts this form into (\ref{pdemu}), one finds that the unknown function
$F_{\psi\rho}$ satisfies an inhomogeneous ordinary differential equation, 
\be
x(1-x)F''_{\psi\rho} + (3-{7x \over 2})F'_{\psi\rho}+ {1-x \over x}F_{\psi\rho} = 
2(1-x)F_{\rho\rho}' + {1-x \over 2x}F_{\rho\rho} + \lambda F_{\phi\rho},
\ee
where we have used the equations satisfied by $F_{\rho\rho}$ and $F_{\phi\rho}$ 
to simplify the right-hand side. The general solution involves two additional
arbitrary coefficients,
\bea
F_{\psi\rho}(x) &=& E_{1} {\sqrt{1-x} \over x} + E_{2} \, \Big\{{2 
\over x} + {\sqrt{1-x} \over x} \log{1-\sqrt{1-x} \over 1+\sqrt{1-x}}
\Big\} - {2\lambda C \over 3} {\sqrt{1-x} \over x} \log{x} 
\nonumber\\
&+& {A_{1} \over 2} - A_{2} \, \Big\{-{\sqrt{1-x} \over x} + {2
\over 3} {\sqrt{1-x} \over x} \log{x} - {1 \over 2} 
\log{1-\sqrt{1-x} \over 1+\sqrt{1-x}}\Big\}.
 \eea

The 4-point function $\la \mu \mu \rho \psi \ra$ is related to the previous one by
the permutation $3 \leftrightarrow 4$, which induces on $x$ the transformation $x
\to {x \over x-1}$. One thus obtains
\be
\la \mu(1) \mu(2) \rho(3) \psi(4) \ra = {z_{12}^{5/4} \over 
z_{14}z_{24}} \Big[ F_{\rho\psi} - \lambda F_{\rho\phi} \, 
\log\Big({z_{14}z_{24} \over z_{12}}\Big) + {z_{24} \over z_{12}} \,
F_{\rho\rho}\Big],\label{rhopsi}
\ee
and
\bea
F_{\rho\psi}(x) &=&  F_{1} {\sqrt{1-x} \over x} + F_{2} \, \Big\{{2(1-x) 
\over x} + {\sqrt{1-x} \over x} \log{1-\sqrt{1-x} \over 1+\sqrt{1-x}}
\Big\} + {2\lambda D \over 3} {\sqrt{1-x} \over x} \log{1-x \over x} 
\nonumber\\
&+& {A_{1} \over 2} - A_{2} \, \Big\{{\sqrt{1-x} \over x} + {2
\over 3} {\sqrt{1-x} \over x} \log{1-x \over x} - {1 \over 2} 
\log{1-\sqrt{1-x} \over 1+\sqrt{1-x}}\Big\}.
\eea

The next case is the 4-point function $\la \mu \mu \psi \phi \ra$. From the Ward
identities, its general form is
\be
\la \mu(1) \mu(2) \psi(3) \phi(4) \ra = {z_{12}^{1/4} \over 
z_{34}^{2}} \Big[F_{\psi\phi} - \lambda F_{\phi\phi} \, 
\log\Big({z_{13}z_{23} \over z_{12}}\Big) - {x \over x-1} {z_{34} \over z_{24}} \,
F_{\rho\phi}\Big].\label{psiphi}
\ee

When it is inserted into (\ref{pdemu}) and (\ref{pdephi}), one finds that
$F_{\psi\phi}$ satisfies two second order ordinary differential equations. 
By combining them, one obtains a first order equation, namely
\be
2(1-x)(2-x) F'_{\psi\phi} - x F_{\psi\phi} = -{4(4-6x+3x^{2}-x^{3}) \over 3x(2-x)}
\lambda F_{\phi\phi} + {2 \over 3} (4-5x+x^{2}) F_{\rho\phi}.
\ee
The general solution reads
\be
F_{\psi\phi}(x) = G{x-2 \over \sqrt{1-x}} - {(4\lambda B-D) \over
3} \sqrt{1-x} - {2\lambda B+D \over 3} {x-2 \over \sqrt{1-x}} 
\log{x}.
\ee

Similarly, one finds
\be
\la \mu(1) \mu(2) \phi(3) \psi(4) \ra = {z_{12}^{1/4} \over 
z_{34}^{2}} \Big[F_{\phi\psi} - \lambda F_{\phi\phi} \, 
\log\Big({z_{14}z_{24} \over z_{12}}\Big) + x {z_{34} 
\over z_{23}} \, F_{\phi\rho}\Big],\label{phipsi}
\ee
with
\be
F_{\phi\psi}(x) = H{x-2 \over \sqrt{1-x}} - {(4\lambda B+C) \over 
3\sqrt{1-x}} + {2\lambda B-C \over 3} {x-2 \over \sqrt{1-x}} 
\log{1-x \over x}.
\ee

We finally come to the last correlator $\la \mu \mu \psi \psi \ra$. From
the Ansatz 
\bea
\la \mu(1) \mu(2) \psi(3) \psi(4) \ra &=& {z_{12}^{1/4} \over z_{34}^{2}} \Big[
F_{\psi\psi} - \lambda F_{\phi\psi} \, f_{1}(z_{i}) - \lambda F_{\psi\phi} \,
f_{2}(z_{i}) - F_{\rho\psi} \, f_{3}(z_{i}) 
- F_{\psi\rho} \, f_{4}(z_{i}) \nonumber\\
&& \qquad + \lambda^{2} F_{\phi\phi} \, f_{5}(z_{i}) +
\lambda F_{\phi\rho} \, f_{6}(z_{i}) + \lambda F_{\rho\phi} \, f_{7}(z_{i}) 
+ F_{\rho\rho} \, f_{8}(z_{i}) \Big],
\eea
we find that the Ward identities are satisfied for the following choice of the
functions $f_{j}$:
\bea
&& f_{1}(z_{i}) = \log\Big({z_{13}z_{23} \over z_{12}}\Big), \quad 
f_{2}(z_{i}) = \log\Big({z_{14}z_{24} \over z_{12}}\Big), \quad
f_{3}(z_{i}) = {x \over x-1} {z_{34} \over z_{24}}, \quad 
f_{4}(z_{i}) = -x {z_{34} \over z_{23}}, \qquad\\
&& f_{5}(z_{i}) = f_{1}(z_{i}) f_{2}(z_{i}), \quad f_{6}(z_{i}) = f_{1}(z_{i})
f_{4}(z_{i}), \quad f_{7}(z_{i}) = f_{2}(z_{i}) f_{3}(z_{i}) , \quad
f_{8}(z_{i}) = {z_{34}^{2} \over z_{13}z_{14}}.
\eea

Inserting this form in (\ref{pdemu}) yields an ordinary differential equation for
$F_{\psi\psi}$,
\bea
x(1-x)F''_{\psi\psi} &\!\!\!+\!\!\!& (1-{3x \over 2})F'_{\psi\psi} - 
{x \over 2(1-x)}F_{\psi\psi} = -{x \over 2-x} 
\lambda F_{\phi\psi} - {x \over (1-x)(2-x)} \lambda F_{\psi\phi} \nonumber\\
&& \hspace{-1.5cm} - {x \over 2}\,  F_{\rho\psi} - {x \over 2(1-x)^{2}} F_{\psi\rho} +
 {2(4-4x+5x^{2}) \over 3x(2-x)^{2}} \lambda^{2} F_{\phi\phi} \nonumber\\
&& \hspace{-1.5cm} + {2(4-3x) \over 3(1-x)(2-x)} \, \lambda F_{\phi\rho} - 
{2(4-x) \over 3(2-x)} \, \lambda F_{\rho\phi} -
{x(6-6x+x^{2}) \over 2(1-x)^{2}} F_{\rho\rho} - {2x^{2}(2-x) \over 1-x}
F'_{\rho\rho},
\label{psipsi}
\eea
in terms of the eight functions derived earlier. 

The general solution to this equation would be fairly complicated to write down and
is in any case not needed. We are primarily interested in the above correlators 
when the two $\mu$ fields form a pair $\mu^{\rm op,cl}(z_1) \mu^{\rm cl,op}(z_2)$. 
When $z_1$ goes to $z_2$, the operator product expansion closes on fields which
live on an open boundary. As shown in \cite{piru1}, it reads 
\be
\mu^{\rm op,cl}(z_1) \mu^{\rm cl,op}(z_2) = z_{12}^{1/4} \, {\mathbb I} + \ldots
\label{mumuop}
\ee
Therefore the limit $z_{12} \to 0$ cannot produce any logarithmic singularity,
whatever the other fields in the correlators are. Imposing this condition on the
first eight correlators, brings five constraints on the coefficients, namely
\be
A_{2} = 0, \qquad C = -D = -\lambda B, \qquad E_2 = -F_2 = {\lambda^{2}B \over 3}.
\label{param1}
\ee

These relations reduce the number of arbitrary constants in the right-hand side of 
(\ref{psipsi}) from 11 to 6, namely $A_{1},\, B,\, E_{1},\, F_{1},\, G$ and $H$.
The most general function $F_{\psi\psi}$ contains the general solution of its
homogeneous equation, and also six functions which solve the inhomogeneous equation,
each one corresponding to one of the six coefficients remaining in the r.h.s. of
(\ref{psipsi}),
\be
F_{\psi\psi} = F_{\psi\psi}\Big|_{\rm homog} + F_{\psi\psi}\Big|_{A_{1}} +
F_{\psi\psi}\Big|_{B} + F_{\psi\psi}\Big|_{E_{1}} + F_{\psi\psi}\Big|_{F_{1}}
+ F_{\psi\psi}\Big|_{G} + F_{\psi\psi}\Big|_{H}.
\ee
These seven functions are given explicitly by
\bea
&& F_{\psi\psi}\Big|_{\rm homog} = I_{1} {x-2 \over \sqrt{1-x}} + 
I_{2}\Big\{ {x-2 \over \sqrt{1-x}} \log{1-\sqrt{1-x} \over 1+\sqrt{1-x}} - 
2\Big\},\\
&& F_{\psi\psi}\Big|_{A_{1}} = {4 - 4x - 3x^2 \over 4(1-x)},\\
&& F_{\psi\psi}\Big|_{E_{1}} = -{1 \over 3\sqrt{1-x}} 
\Big\{1 + (x-2) \log{1-x \over x}\Big\}, \quad
F_{\psi\psi}\Big|_{F_{1}} = {\sqrt{1-x} \over 3} - 
{1 \over 3}{x-2 \over \sqrt{1-x}} \log{x},\\
&& \lambda^{-1} F_{\psi\psi}\Big|_{G} = {1 \over 3\sqrt{1-x}} 
\Big\{2 - 3x + 2(x-2)\log{1-x \over x}\Big\}, \\
&& \lambda^{-1} F_{\psi\psi}\Big|_{H} = {1 \over 3\sqrt{1-x}} \Big\{
x + 2 - 2(x-2)\log{x}\Big\},\\
&& \lambda^{-2} F_{\psi\psi}\Big|_{B} = {2-16 \log{2} \over 9} - {2 - 2x + x^{2}
\over 3(1-x)} - {4 \over 9} {x-2 \over \sqrt{1-x}} \Big\{
{\rm Li}_{2}({1-\sqrt{1-x} \over 1+\sqrt{1-x}}) + 
{\rm Li}_{2}(-{1-\sqrt{1-x} \over 1+\sqrt{1-x}})\Big\} \nonumber\\
&& - {2 \over 9} \log{1-x \over x^{2}} + {10 \over 9} \Big\{
{\log{x} \over \sqrt{1-x}} + \sqrt{1-x} \log{x \over x-1}\Big\} + 
{8 \over 9} {x-2 \over \sqrt{1-x}} \log{x} \log{x \over x-1} \nonumber\\
&& - {1 \over 9} \Big\{4 + {x-2 \over \sqrt{1-x}} 
\log{\left({1-x \over x^{2}}{1-\sqrt{1-x} \over 1+\sqrt{1-x}}\right)}\Big\} 
\log{1-\sqrt{1-x} \over 1+\sqrt{1-x}},
\eea
where the polylogarithmic function ${\rm Li}_{2}(z) = \sum_{k=1}^{\infty} {z^{k}
\over k^{2}}$ is regular at the origin. 

Imposing now that the correlator $\la \mu \mu \psi \psi \ra$ itself is free of
logarithmic singularity at $z_{12} = 0$, and moreover that $\lim_{z_{12} 
\to 0} z_{12}^{-1/4} \, \la \mu(1) \mu(2) \psi(3) \psi(4) \ra$ depends on $z_{34}$
only, leads to five new conditions on the coefficients,
\be
B = 0, \qquad E_{1} = -F_{1} = -A_{1}, \qquad H = G, \qquad 
I_{2} = {2 \over 3}(A_{1} - \lambda G),
\label{param2}
\ee
leaving only three free parameters, $A_1, G$ and $I_1$. The explicit solution is
\bea
\la \mu(z_1) \mu(z_2) \psi(z_3) \psi(z_4) \ra &=& {z_{12}^{1/4} \over z_{34}^2} \Big\{
\Big(I_1-{A_1 \over 3}-{2\lambda G \over 3}\Big) {x-2 \over \sqrt{1-x}} + {A_1 \over
4} {4-4x-3x^2 \over 1-x} \nonumber\\
&& \hspace{-2.3cm} - {2 \over 3}(A_1-\lambda G) {x-2 \over \sqrt{1-x}}
\log{(1+\sqrt{1-x})^2 \over \sqrt{1-x}} - {4 \over 3}(A_1-\lambda G)
 - 2\lambda G {x-2 \over \sqrt{1-x}} \log{|z_{34}|} \nonumber\\
&& \hspace{-2.3cm} +\; A_1 \Big({1 \over z_{24}\sqrt{1-x}} - {\sqrt{1-x} \over
z_{23}}\Big)z_{34} + {A_1 \over 2} \Big({x \over z_{23}} - {x \over (x-1)z_{24}} +
{2z_{34} \over z_{13}z_{14}}\Big)z_{34}\Big\}.
\label{3param}
\eea

We have required that the correlator be regular when $z_1,z_2 \to 0$, namely when
the boundary is all open. The other extreme regime corresponds to an all closed
boundary, obtained by letting $-z_1,z_2 \to +\infty$. In this limit, the correlator
should reproduce the 1-site probabilities in presence of a closed boundary and thus
should also be regular. This is the case provided
\be
A_{1} = \lambda G.
\label{A1G}
\ee 

At this stage, the correlator $\la \mu \mu \psi(z,z^*) \ra = \la \mu \mu \psi(z)
\psi(z^*) \ra$ physically relevant to our situation depends on the two arbitrary
coefficients $A_1$ and $I_1$, and reads explicitly,
\bea
\la \mu(z_1) \mu(z_2) \psi(z_3,z_4) \ra &=& {z_{12}^{1/4} \over z_{34}^2} \Big\{
(I_1-A_1) {x-2 \over \sqrt{1-x}} + {A_1 \over 4} {4-4x-3x^2 \over 1-x} -
2A_1 {x-2 \over \sqrt{1-x}} \log{|z_{34}|} \nonumber\\
&& \hspace{-2cm} +\; A_1 \Big({1 \over z_{24}\sqrt{1-x}} - {\sqrt{1-x} \over
z_{23}}\Big)z_{34} + {A_1 \over 2} \Big({x \over z_{23}} - {x \over (x-1)z_{24}} +
{2z_{34} \over z_{13}z_{14}}\Big)z_{34}\Big\}.
\eea
Anticipating a little bit, we trade the coefficients $A_1$ and $I_1$ for two other
coefficients $a_2$ and $b_2$,
\be
A_1 = -b_2, \qquad I_1 = 2a_2 - b_2 ({1 \over 2} + 2 \log{2}).
\ee
In terms of those, the correlator becomes
\bea
\la \mu(z_1) \mu(z_2) \psi(z_3,z_4) \ra &=& {z_{12}^{1/4} \over z_{34}^2} \, {x-2
\over \sqrt{1-x}} \; \Big\{2b_2 \log\Big|{z_{34} \over 2}\Big| + 2a_2 + {b_2 \over 2}
- {b_2 \over 4}  {x-2 \over \sqrt{1-x}} \nonumber\\
\noalign{\medskip}
&+&  b_2 {z_{34}^2 \over z_{13} z_{24}} \Big[{1 \over x-2} + {1 \over 2\sqrt{1-x}}\Big]
\Big\}.\label{mumupsi}
\eea

The other correlator we need involves the non-chiral field $\phi(z,z^*) =
{1 \over \lambda}\,L_0 \psi(z,z^*) = {1 \over \lambda}\,\bar L_0 \psi(z,z^*)$, which
becomes $\phi(z,z^*) = \phi(z)\psi(z^*) = \psi(z)\phi(z^*)$ in terms of chiral
fields. The relevant correlator is thus given by (\ref{psiphi}) or (\ref{phipsi}) with
$B=C=0$ and $H=G={A_1 \over \lambda}=-{b_2 \over \lambda}$, namely
\be
\la \mu(z_1) \mu(z_2) \phi(z_3,z_4) \ra =  -{b_2 \over \lambda}\,{z_{12}^{1/4} \over 
z_{34}^2} {x-2 \over \sqrt{1-x}}.
\label{mumuphi}
\ee 

When we will come to the comparison of the conformal predictions with the numerical
simulations, we will also need two other correlators, namely $\la \mu(z_1) \mu(z_2)
\rho(z_3,z_4) \ra$ and $\la \mu(z_1) \mu(z_2) \bar\rho(z_3,z_4) \ra$. In terms of
a pair of chiral fields, $\rho(z,z^*) = L_1 \psi(z,z^*)$ is identified with
$\rho(z) \psi(z^*)$. The corresponding correlator is given in (\ref{rhopsi}) and
reads (we use the values of the constants found above, and $A_1 = -b_2$)
\be
\la \mu(z_1) \mu(z_2) \rho(z_3,z_4) \ra = -b_2 \, z_{12}^{1/4} \Big\{{z_{12} \over
z_{14}z_{24}} {\sqrt{1-x} \over x} + {1 \over 2z_{14}} + {1 \over 2z_{24}}\Big\}.
\label{mumurho}
\ee
Similarly, identifying the field $\bar\rho(z,z^*) = \bar L_1 \psi(z,z^*)$ with
$\psi(z) \rho(z^*)$, the correlator in (\ref{psirho}) yields
\be
\la \mu(z_1) \mu(z_2) \bar\rho(z_3,z_4) \ra = -b_2 \, z_{12}^{1/4} \Big\{-{z_{12}
\over z_{13}z_{23}} {\sqrt{1-x} \over x} + {1 \over 2z_{13}} + {1 \over
2z_{23}}\Big\}.
\label{mumurhobar}
\ee

The value of $\kappa$ in the transformations of $\rho$ and $\bar\rho$ can also be
fixed. The field $\kappa {\mathbb I} = \bar L_1\rho(z,z^*)$ should be associated with
$\rho(z)\rho(z^*)$, so that the correlator (\ref{rhorho}) yields
\be
\kappa \la \mu(z_1) \mu(z_2) \ra = -b_2 \, z_{12}^{1/4},
\ee
and $\kappa = -b_2$.


\subsection{Closed versus open UHP probabilities}

Let us now come back to the 1-site probabilities on the UHP to see how they are
related when the boundary is either fully open or fully closed. We set $\lambda =
-{1 \over 2}$ in the expressions of the previous subsection.

We first consider the height 1 probabilities, already discussed in \cite{piru1}.
According to (\ref{PopI}), the identification $h_1(z,z^*)=\phi(z,z^*)$ and
(\ref{mumuphi}), the probability $P_1^{{\rm op},I}(z)$ is equal to ($z_4 = z_3^* =
z^*$)
\be
P_1^{{\rm op},I}(z) - P_1 = z_{12}^{-1/4}\, \la \mu(z_1)\, \mu(z_2) \, \phi(z,z^*) \ra
= {2b_2 \over (z-z^*)^2} \, {x-2 \over \sqrt{1-x}}.
\ee

When $z_1$ and $z_2$ go to 0, the cross ratio $x = {(z_1-z_2)(z-z^*)
\over (z_1-z)(z_2-z^*)}$ goes to 0, while $\sqrt{1-x}$ goes to $+1$. In this limit,
$P_1^{{\rm op},I}(z) - P_1$ must go to the open boundary result. This fixes $b_2 =
a_1 = {P_1 \over 4}$, so that
\be
\lim_{z_1,z_2 \to 0} P_1^{{\rm op},I}(z) - P_1 = P_1^{\rm op}(m) - P_1 = 
{a_1 \over m^2}, \qquad \quad (z-z^*=2im).
\label{p1op}
\ee

If now $-z_1,z_2$ go to $+\infty$, the cross ratio still goes to 0 but $\sqrt{1-x}$
goes to $-1$ because $1-x$ circles around zero when $-z_1=z_2$ take the values from
0 to $+\infty$. Therefore we obtain the expected change of sign,
\be
\lim_{-z_1,z_2 \to +\infty} P_1^{{\rm op},I}(z) - P_1 = P_1^{\rm cl}(m) - P_1 = 
-{a_1 \over m^2}.
\label{p1cl}
\ee

The probability $P_2^{{\rm op},I}(z)$ that the site at $z$ has height 2 in
presence of an open boundary except on the interval $I = [z_1,z_2]$ which is
closed, is equal to the correlator (\ref{mumupsi}) divided by $z_{12}^{1/4}$ (there
was a misprint in the last term of the formula (8) of \cite{piru3})
\bea
P_2^{{\rm op},I}(z) - P_2 &=& {1 \over (z-z^*)^2} \, {x-2 \over \sqrt{1-x}} \;
\Big\{2b_2 \log\Big|{z-z^* \over 2}\Big| + 2a_2 + {b_2 \over 2} - {b_2 \over 4} 
{x-2 \over \sqrt{1-x}} \nonumber\\
\noalign{\medskip}
&+&  b_2 {(z-z^*)^2 \over (z_1-z)(z_2-z^*)} \Big[{1 \over x-2} + {1 \over
2\sqrt{1-x}}\Big] \Big\},
\label{p2opI}
\eea
with $x = {(z_1-z_2)(z-z^*) \over (z_1-z)(z_2-z^*)}$ as before (although $x$ is a
complex number, $x^*={x \over x-1}$, the probability is real).

Upon taking $z_1$ and $z_2$ to 0, the expression reduces to
\be
\lim_{z_1,z_2 \to 0} P_2^{{\rm op},I}(z) - P_2 = {1 \over m^2} [a_2 + {b_2 \over 2}
+ b_2 \log{m}] = P_2^{\rm op}(m) - P_2,
\ee
which confirms that the two coefficients $a_2,b_2$ are those derived in the lattice
calculations, and given in (\ref{a2b2}). It is now an easy check to see that the
other limit $-z_1,z_2 \to +\infty$, bearing in mind the monodromy factor of
$\sqrt{1-x}$, yields exactly the result for the closed boundary, namely
\be
\lim_{-z_1,z_2 \to +\infty} P_2^{{\rm op},I}(z) - P_2 = -{1 \over m^2} [a_2 +
b_2 \log{m}] = P_2^{\rm cl}(m) - P_2.
\ee

The same limits also hold for the height 3 and 4 probabilities. The 
relations (\ref{p3}) and (\ref{p4}) expressing $P_3(m)$ and $P_4(m)$ as linear
combinations of $P_1(m)$ and $P_2(m)$ suggest that the fields $h_3(z)$ and $h_4(z)$
are given by the same linear combinations of $h_1(z)$ and $h_2(z)$, as in
(\ref{h34}). Then the formula (\ref{p2opI}) holds for $P_i^{{\rm op},I}(z)$,
$i=1,2,3,4$, provided the coefficient $a_2,b_2$ are replaced by $a_i,b_i$, given in
Section VIII. Consequently the open and closed probabilities $P_i^{\rm op}(m)$ and
$P_i^{\rm cl}(m)$, for $i=3$ and $i=4$, are correctly related to each other by the
insertions of $\mu$ fields.

So the lattice results for the open and closed boundary conditions are
exactly compatible with the effect of a change of boundary condition computed in the
conformal setting. The formula (\ref{p2opI}) is however far more general, and gives
access to a probability that looks at present very hard to obtain from lattice
calculations. 

Beyond the two extreme cases of an all open or all closed boundary, this formula
may be confronted with numerical simulations in an intermediate situation. Before
doing this, we return to the pure closed boundary with a further comment. 

Relying on the operator product expansion (\ref{mumuop}), we have imposed that the
correlator $\la \mu(1) \mu(2) \psi(3) \psi(4) \ra$ should not produce
logarithmic singularities in the limit $z_1,z_2 \to 0$. On physical grounds, we have
then required the same regular behaviour in the limit $-z_1,z_2 \to +\infty$ since the
4-point correlator reduces to the closed boundary probabilities. In that
limit, the two fields $\mu$ come close to each other at infinity, so that the
operator product expansion may again be invoked. However the relevant expansion is
not the same as in the other limit, because it is now the open stretch which is
reduced to nothing, in such a way that the expansion must close on fields which
live on a closed boundary. The proper operator product expansion reads \cite{piru1}
\be
\mu^{\rm cl,op}(z_1) \mu^{\rm op,cl}(z_2) = z_{12}^{1/4} \,
[\omega_N(z_2) - {1 \over \pi} \, {\mathbb I} \, \log{z_{12}} ] + \ldots
\ee
where the boundary field $\omega_N$ has conformal weight 0, and forms with the
identity, a rank two Jordan block under dilations. The interpretation of $\omega(z)$
in the sandpile model is that it introduces dissipation at $z$ \cite{piru1}.

The insertion of the previous expansion in the 4-point function yields in the limit
\be
\la \psi(z,z^*) \ra_{\rm cl} = \lim_{-z_1,z_2 \to +\infty} \Big\{
\la \omega_N(z_2) \psi(z) \psi(z^*) \ra - {\log{z_{12}} \over \pi} \la \psi(z)
\psi(z^*) \ra \Big\}.
\ee
The absence of logarithmic singularity then implies that the chiral 2-point
function $\la \psi(z) \psi(z^*) \ra$, appropriate for the closed boundary, vanishes
identically, but also that
\be
\la \psi(z,z^*) \ra_{\rm cl} = \la \omega_N(\infty) \psi(z) \psi(z^*) \ra.
\ee
Though strange at first sight, this relation has a natural and physical origin. 
Probabilities on infinite lattices are limits of probabilities calculated on
finite lattices. The sandpile model is not well-defined if the lattice does
not have dissipative sites. In the model considered here, dissipation takes place at
the open boundary sites, namely at those sites which are connected to the sink, itself
located all around the lattice. In the case of the UHP, three boundaries have been
sent off to infinity. When the remaining boundary (the real axis) is open, the
sandpile model is well-defined without introducing further dissipation on the other
three boundaries. When it is closed, dissipation has to be introduced on the other
three boundaries by connecting them to the sink. The sink is, with the boundaries,
subsequently sent to infinity, which is where the dissipation is eventually located.
This is exactly what the field $\omega_N(\infty)$ does: it inserts at infinity the
dissipation needed to make the sandpile model well-defined.

A similar relation holds for the $\phi$ field, describing the height 1 variable,
\be
\la \phi(z,z^*) \ra_{\rm cl} = \la \omega_N(\infty) \phi(z) \phi(z^*) \ra.
\ee
The insertion of $\omega_N$ prevents $\la \phi(z,z^*) \ra_{\rm cl} \sim P_1^{\rm
cl}(m)$ from being identically zero, since the chiral 2-point function of $\phi$
vanishes.


\section{Finite size corrections versus simulations}

The simplest case after the pure open and pure closed boundary condition is when
the boundary is half closed and half open. It is also a case which is relevant for
a comparison with numerical simulations. 

We take the boundary of the UHP to be closed on the negative side of the real axis,
and open on the positive side. It clearly corresponds to taking $z_1=-\infty$ and
$z_2=0$. From (\ref{p2opI}), we obtain for $i=1,2,3,4$,
\be
P_i^{\rm cl|op}(z) - P_i = -{z + z^* \over |z|(z-z^*)^2} \, \Big\{
2b_i \log\Big|{z-z^* \over 2}\Big| + 2a_i + {b_i \over 2} + {b_i \over 4} \,
{z + z^* \over |z|} \Big\}.
\label{pmix}
\ee

The conformal map $w={L \over \pi} \log{z}$ transforms the upper half plane to an
infinite strip of width $L$. The negative and positive real axis are mapped
respectively to the line $\Im w = L$ and $\Im w = 0$, which are therefore
respectively closed and open.

In order to transform the probabilities from the UHP to the strip, we need the
conformal transformation law of the logarithmic field $\psi$ (the field $\phi$ is
primary). The integration of the infinitesimal conformal transformations
(\ref{Tpsi}) and (\ref{Tbarpsi}) yields
\bea
\psi_{\rm strip}(w,\bar w) &=& |z'(w)|^{2} \Big\{\psi_{\rm uhp}(z,\bar z) 
- {1 \over 2} \log |z'(w)|^2 \, \phi_{\rm uhp}(z,\bar z) \nonumber\\
&& \hspace{1cm} + \; {z''(w) \over 2z'^{2}(w)}\, \rho_{\rm uhp}(z,\bar z)
+ {\bar z''(\bar w) \over 2\bar z'^{2}(\bar w)} \, 
\bar \rho_{\rm uhp}(z,\bar z) + \kappa \, \Big|{z''(w) \over 2z'^{2}(w)}\Big|^2\Big\},
\eea
where the last piece comes from the transformations\footnote{The inhomogeneous
terms in the non-chiral transformation laws of $\rho$ and $\bar\rho$ have been
overlooked in \cite{piru3}, resulting in a number of missing terms in the equations
(10-12) and (14) of that reference. In particular the equation (14) of \cite{piru3}
misses a constant term proportional to $1/L^2$, see the equation (\ref{pistrip}) in
the present article. This however makes no numerical difference.} of $\rho$ and
$\bar\rho$.

Taking the expectation value, one obtains the corresponding 1-site probabilities on
the infinite strip in terms of those on the UHP and the expectation values of
$\rho,\bar\rho$. For the height 1 probability, the relation is
\be
P_1^{\rm strip}(w) - P_1 = |z'(w)|^{2} \, [P_1^{\rm cl|op}(z) - P_i],
\ee
whereas for the height 2 probability,
\bea
P_2^{\rm strip}(w) - P_2 &=& |z'(w)|^{2} \Big\{[P_2^{\rm cl|op}(z) - P_2]
- {1 \over 2} \log |z'(w)|^2 \, [P_1^{\rm cl|op}(z) - P_1] \nonumber\\
&& \hspace{-1.5cm} + \; {z''(w) \over 2z'^{2}(w)}\, \la \rho(z,
z^*)\ra_{\rm cl|op} + \Big({z''(w) \over 2z'^{2}(w)}\Big)^* \, 
\la\bar \rho(z,z^*)\ra_{\rm cl|op} + \kappa \, \Big|{z''(w) \over
2z'^{2}(w)}\Big|^2 \, \la {\mathbb I} \ra_{\rm cl|op}\Big\}.
\eea
The height 3 and 4 probabilities are simply obtained by linear combinations (see
(\ref{h34})). In these formulas, the expectation values refer to the UHP with the boundary
condition closed on the negative axis and open on the positive axis, {\it i.e.} $z_1
\to -\infty$ and $z_2 \to 0$. From (\ref{mumurho}) and (\ref{mumurhobar}), these are
given by 
\be
\la \rho(z,z^*)\ra_{\rm cl|op} = b_2 \, \Big\{-{\sqrt{z/z^*} \over z-z^*} + {1
\over 2z^*}\Big\}, \qquad
\la\bar \rho(z,z^*)\ra_{\rm cl|op} = b_2 \, \Big\{{\sqrt{z^*/z} \over z-z^*} + {1
\over 2z}\Big\},
\ee
and $\la {\mathbb I} \ra_{\rm cl|op} = 1$.

Putting everything together and setting $w=u + iv$, we find the following 1-site
probabilities on the strip, for $i=1,2,3,4$, as functions of $v$ only,
\be
P_i^{\rm strip}(w) - P_i = \Big({\pi\over L}\Big)^{\!2} {\cos(\pi v/L)\over\sin^2(\pi
v/L)} \Big\{ a_i + {b_i\over4} \Big[1+\cos\big({\pi v\over L}\big)\Big] 
+ b_i \log\!\Big({L\over\pi}\sin\big({\pi v\over L}\big)\Big) \Big\} + {b_i
\pi^2 \over 4L^2}.
\label{pistrip}
\ee

Should the field identifications proposed earlier be correct, these formulas 
represent the scaling limit of the probability that a site of an infinite strip,
located at a distance $v$ from the open boundary and $L-v$ from a closed boundary,
has a height equal to $i$. 

\begin{figure}[hbt]
\begin{center}
\includegraphics[scale=0.62]{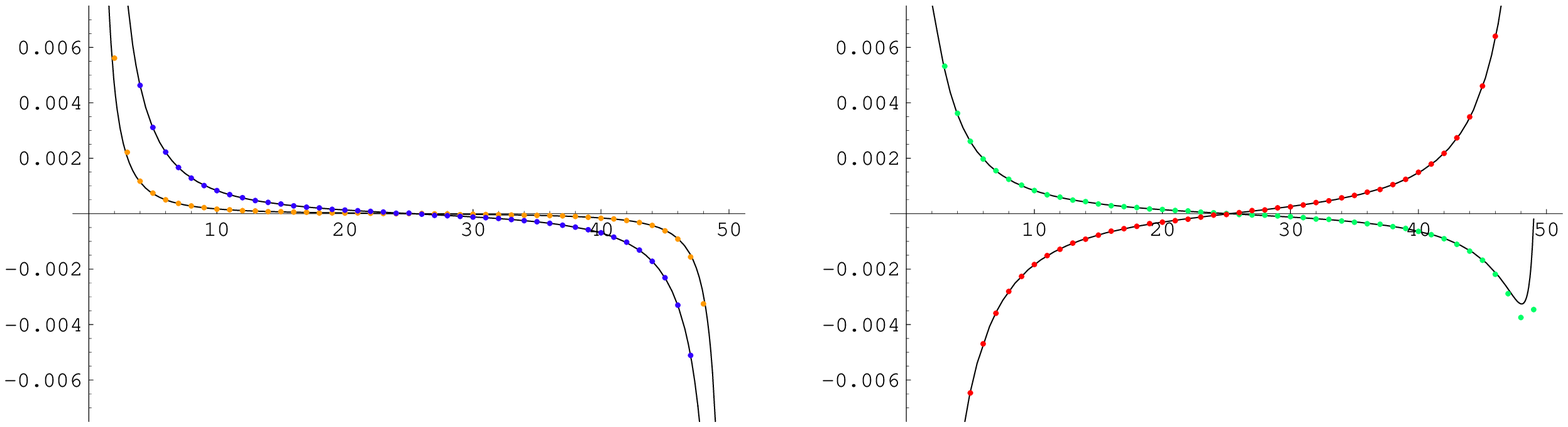}
\end{center}
\caption{Subtracted probabilities to find a height 1 (orange), 2 (in blue), 3 (green) or 4 
(red) at a site lying along the short medial line of a rectangle $M \times N$, and
going from an open boundary (left end of the graph) to a closed boundary (right end).
The dots correspond to data obtained from numerical simulations, with $M=200$ and
$N=50$, while the solid curves represent the conformal predictions (\ref{pistrip}),
with $L=N+\half$, for the limit $M=+\infty$.}
\label{Figure:simul}
\end{figure}

We have run numerical simulations on strips of width $N$ and height $M$ for various
values of $M$ and $N$ (details about these simulations have been given in
\cite{piru3}). For ratios $M \over N$ large enough, we expect that
the results on a finite height strip be very close to those on an infinite strip, so
that the numerical simulations should be reliable enough to be compared with the
exact scaling limit (\ref{pistrip}), provided by the conformal approach. For
simplicity, we chose the boundary conditions on the two sides separated by $M$ to
be open ---exactly what boundary conditions are imposed there should make no
difference---, and the other two were open and closed. The two sets of plots, the
numerical results and the conformal formulas, in each case for the four heights,
are shown in Fig. \ref{Figure:simul} (as explained in \cite{piru3}, the width $L$
in the continuum and the width $N$ of the lattice should be related by $L=N+\half$).
The simulation data are the same as in the figures of \cite{piru3}, but the conformal
curves are plotted here with the exact coefficients $a_i,b_i$, derived from the
lattice calculations, rather than with fitted coefficients. 

The agreement is very satisfactory for the four heights. In the previous section, we
had shown that our proposal for the field identifications were fully compatible with
the lattice results for homogeneous boundary conditions. Here we have an
independent and direct confirmation.


\section{Nature of the logarithmic field \pbm $\psi$}

The previous sections contain two major results: first, the computation of the
1-site probabilities for having a height 2, 3 or 4 on the UHP, and second, the
identification of the scaling limit of the corresponding height variables with a
logarithmic field $\psi$, belonging to a logarithmic conformal field theory with
$c=-2$. There exists a well-known free field realization of a local logarithmic theory
with $c=-2$, so that a natural question is whether our field $\psi$ can be
realized in this free field theory. Our answer to that question will be negative.

\subsection{Comparison of correlations}

The free field theory defined by the following action
\be
S = {1 \over \pi} \int \, \p\t \bp \tit,
\ee
where $\t, \tit$ are anticommuting scalar fields, is a logarithmic conformal field
theory, with central charge $c=-2$, and is a realization of the so-called triplet
theory \cite{gur,gk,k,gfn,gr}. Zero (constant) modes,
if any, play a special role, due to their grassmannian nature. Because the action does
not depend on them, their integration ``eats up'' one pair of $\t,\tit$ fields in
the expectation values if there is such a pair, and yields identically zero if not.
Apart from this peculiarity, correlators of local functionals of $\t$ and $\tit$ can
be computed in the usual way from Wick's theorem. 

Correlation functions are usually normalized by the partition function, with the
result that $\la \mathbb I \ra = 1$. When the fields $\t,\tit$ have zero modes, the
partition function vanishes identically, so that this normalization is not possible.
A convenient alternative choice is to set $\la \omega(z,\bar z) \ra = 1$, where $\omega =
\alpha_0 \NP{\t\tit} + \, \beta_0$ is a dimension zero scalar field, which forms with
the identity $\mathbb I$ a rank two Jordan cell, 
\be
L_0 \omega = - {\alpha_0 \over 2} \, \mathbb I.
\ee
On the other hand, when there are no zero modes, the usual normalization $\la
\mathbb I \ra = 1$ can be chosen. 

In \cite{piru1}, the bulk field $\omega$ and the boundary field $\omega_N$ describing the    
insertion of isolated dissipation at a bulk site and at a site of a closed
boundary respectively, have expectation value equal to 1, $\la \omega(z,\bar z)
\ra_{\rm plane} = \la \omega_N(x) \ra_{\rm cl} = 1$. They can be realized in terms of
the free fermions as
\be
\omega = {1 \over 2\pi} \NP{\t\tit} + {1 \over 2\pi}(\gamma + {3 \over 2} \log 2) + 1,
\qquad \omega_N = {1 \over 2\pi} \NP{\t\tit} + {2 \over \pi} \log{2}.
\ee
Their transformations under dilations are $L_0\omega = -{1 \over 4\pi}\mathbb I$ and
$L_0\omega_N = -{1 \over \pi}\mathbb I$.

On the infinite plane, the fields $\t,\tit$ have zero modes, $\xi$ and $\tilde \xi$.
The above normalization implies 
\be
\la \mathbb I \ra = 0, \qquad \la \t(z,\bar z) \tit(w,\bar w) \ra = \la \xi
\tilde\xi \ra = 2\pi.
\label{zm}
\ee
Higher correlators (and OPEs) are computed using the following elementary Wick
contraction,
\be
\Cont{\t(z,\bar z) \, \tit}(w,\bar w) = -\log{|z-w|}.
\label{wickpl}
\ee
For instance, the 4-point function on the plane takes the form \cite{k}
\be
\la \t(z_1) \tit(z_2) \t(z_3) \tit(z_4) \ra = 2\pi [\log{|z_1-z_2|} +
\log{|z_3-z_4|} - \log{|z_1-z_4|} - \log{|z_2-z_3|}].
\ee

On the UHP with closed (Neumann) boundary condition on the real axis, namely $\p\t -
\bp\t = \p\tit - \bp\tit = 0$ on $\mathbb R$, the fields also have zero modes. Thus
the relations (\ref{zm}) still hold, while the Wick contraction appropriate to the
closed boundary condition is
\be
\Cont{\t(z,\bar z) \, \tit}(w,\bar w) = -\log{|(z-w)(z-\bar w)|}. \hspace{1.5cm}
(\hbox{closed on } \mathbb R)
\label{wickcl}
\ee

If the open (Dirichlet) boundary condition is imposed, $\t = \tit = 0$ on $\mathbb
R$, the fields no longer have zero modes. In this case, one has $\la \mathbb I
\ra_{\rm op} = 1$ and the 2-point function $\la \t(z) \tit(w) \ra_{\rm op}$ is simply
given by a Wick contraction, equal to 
\be
\Cont{\t(z,\bar z) \, \tit}(w,\bar w) = -\log{\Big|{z-w \over z-\bar w}\Big|}.
\hspace{1.5cm} (\hbox{open on } \mathbb R)
\label{wickop}
\ee

The same is true for the mixed boundary condition we have considered earlier,
namely closed on $\mathbb R_-$ and open on $\mathbb R_+$. The fields do not have
zero modes, and the relevant Wick contraction reads (the phases of
$\sqrt{z},\sqrt{w}$ are between 0 and $\pi/2$)
\be
\Cont{\t(z,\bar z) \, \tit}(w,\bar w) = -\log{\Big|{(\sqrt{z}-\sqrt{w}) 
(\sqrt{z}+\sqrt{\bar w}) \over (\sqrt{z}-\sqrt{\bar w})(\sqrt{z}+\sqrt{w})}\Big|}.
\hspace{1cm} (\hbox{closed on } \mathbb R_-,\hbox{ open on } \mathbb R_+)
\label{wickmix}
\ee

Let us now come back to the 1-point functions on the UHP we have computed earlier
for the height variables, to see what they are in the free fermionic theory. 

The scaling field $\phi$ corresponding to the height 1 variable has been much
studied. It is a primary field with conformal weights (1,1), realized in the free
theory as \cite{mr}
\be
\phi(z,\bar z) = -P_1 \,\NP{\p\t \bp \tit + \bp \t \p\tit}\; = -P_1 \, \p\bp
\NP{\t\tit}.
\ee
In order to compute the expectation value of $\phi$ on the UHP, one
may use the OPE to make a point splitting, 
\be
\NP{\t\tit} \, = \lim_{w \to z} \: [\t(z) \tit(w) + \log{|z-w|}],
\ee
and then evaluate Wick contractions.

For an open or a mixed boundary condition, the 1-point function of $\phi$ is equal to 
\be
\la \phi(z,z^*) \ra_{\rm op} =
\la \phi(z,z^*) \ra_{\rm cl|op} = -P_1 \:\p\bp \; \lim_{w \to z} \: [\Cont{\t(z)
\tit}(w) + \log{|z-w|}].
\ee
For the closed boundary condition, the boundary dissipation field $\omega_N(\infty)$
is to be inserted, but is eaten up by the path integration over the zero modes
(the piece in $\omega_N$ proportional to the identity does not contribute since
$\phi$ contains derivative fields only). Once the zero modes are integrated out,
one simply compute Wick contractions for what remains, so that the previous formula
is also valid for the closed boundary. 

Using Eqs. (\ref{wickcl})-(\ref{wickmix}), one easily obtains
\bea
&& \la \omega_N(\infty) \phi(z,z^*) \ra_{\rm cl} = {P_1 \over (z-z^*)^2}, \qquad
\la \phi(z,z^*) \ra_{\rm op} = -{P_1 \over (z-z^*)^2},\\
&& \la \phi(z,z^*) \ra_{\rm cl|op} = -{P_1 \over 2}{z + z^* \over |z|(z-z^*)^2},
\eea
in agreement with our earlier results, Eqs. (\ref{p1op}), (\ref{p1cl}) and
(\ref{pmix}) with $a_1 = P_1/4$ and $b_1=0$.

The scaling fields corresponding to the heights 2, 3 and 4 involve the logarithmic
partner of the primary field $\phi$. In the fermionic theory, the logarithmic partner
of $\phi$ is proportional to 
\be
\psi_\t(z,\bar z) = \:\NP{\t\tit (\p\t\bp\tit + \bp\t\p\tit)} + I \, \p\bp
\NP{\t\tit} = \:\NP{\t\tit \; \p\bp \,\t\tit} + I \, \p\bp \NP{\t\tit},
\ee
and may contain an arbitrary multiple $I$ of its primary partner. The
expectation value of this term is given by the expressions we have just established.

The same point splitting procedure as for $\phi$ can be applied to $\psi_\t$, for
instance,
\be
\psi_\t(z) = - \lim_{w \to z} \: [\t\p\t(z) \, \tit\bp\tit(w) + \t\bp\t(z) \,
\tit\p\tit(w) - \hbox{singular terms}].
\ee
The subtraction terms are such that the 1-point function $\la \psi_\t(z,z^*)
\ra$ is equal to the limit $w \to z$ of the point splitted fields $\la \t\p\t(z) \,
\tit\bp\tit(w) + \t\bp\t(z) \, \tit\p\tit(w) \ra$, in which one removes the
singular part in each Wick contraction. 

Again for an open boundary or a mixed one, the 1-point function follows from Wick
contracting the point-splitted fields, resulting in 
\bea
\la \psi_\t(z,z^*) \ra_{\rm op} &=& \la \psi_\t(z,z^*) \ra_{\rm cl|op} = \lim_{w \to
z} \: \Big\{\Big[\Cont{\t(z) \tit}(w) + \log{|z-w|}\Big] \, \Cont{\p\t(z) \bp\tit}(w)
\nonumber\\
\noalign{\medskip}
&& - \Big[\Cont{\t(z) \bp\tit}(w) - {1 \over 2(z^*-w^*)}\Big] \;
\Big[\Cont{\p\t(z) \tit}(w) + {1 \over 2(z-w)}\Big] + \hbox{c.c.} \Big\}.
\label{prev}
\eea

For the closed boundary, the insertion of $\omega_N(\infty) = {1 \over 2\pi} \NP{\t\tit}
+ {2 \over \pi}\mathbb I \log{2}$ brings two contributions. One comes from the
identity piece, and reduces to the calculation we made above for the field $\phi$.
If we remember that the integration over the zero modes equals $2\pi$, one obtains
that this first contribution is
\be
{2 \over \pi} \log{2} \; \la \psi_\t(z,z^*) \ra_{\rm cl} = -{4 \log{2} \over (z-z^*)^2}.
\ee

The second contribution from the $\NP{\t\tit}$ piece produces logarithmically
divergent terms, because after the integration of the two zero modes,
one $\t$ (or $\tit$) in $\omega_N(\infty)$ may be contracted with a $\tit$ (or $\t$)
in $\psi_\t$. On the other hand a regularized, finite answer is obtained if we
decide to integrate the two zero modes contained in the $\NP{\t\tit}$ piece of
$\omega_N(\infty)$, since this leaves us with contractions of $\t,\tit$ within
$\psi_\t$. This regularized expectation value is then given by the same formula as
for the open or mixed boundary, Eq. (\ref{prev}), and yields, by using the
appropriate Wick contraction,
\be
\la {\NP{\t\tit} \over 2\pi} \psi_\t \ra_{\rm cl,reg} = {1 \over (z - z^*)^2} \, 
\Big\{\half + \log{|z-z^*|}\Big\}.
\ee

Computing the required Wick contractions for the open and mixed boundaries, and
adding the contribution from the term $I \, \p\bp \NP{\t\tit}$, we obtain the 1-point
function of $\psi_\t$ on the UHP for the three boundary conditions,
\bea
&& \hspace{-7mm} \la \psi_\t(z,z^*) \ra_{\rm op} = {1 \over (z - z^*)^2} \,
\Big\{I + \half + \log{|z-z^*|}\Big\}, \label{thetaop} \\
\noalign{\medskip}
&& \hspace{-7mm} \la \omega_N(\infty) \psi_\t(z,z^*) \ra_{\rm cl,reg} = {1 \over (z
- z^*)^2} \, \Big\{-I - 4\log{2} + \half + \log{|z-z^*|}\Big\}, \label{thetacl} \\
\noalign{\medskip}
&& \hspace{-7mm} \la \psi_\t(z,z^*) \ra_{\rm cl|op} = {z+z^* \over |z|(z-z^*)^2} \,
\Big\{{I \over 2} + {1 \over 2} \log{\Big|{4z(z-z^*) \over
(\sqrt{z}+\sqrt{z^*})^2}\Big|} - {(z-z^*)^2 - 4|z|^2 \over 8|z|(z+z^*)}\Big\}.
\label{thetamix}
\eea

We see that the first two expectation values of $\psi_\t$ are not related as they
are for the field $\psi$ associated with the height variables $h \geq 2$. In
particular, they do not show the characteristic change of sign of the logarithmic term.
Moreover, the third one $\la \psi_\t(z,z^*) \ra_{\rm cl|op}$ does not have the
form obtained in (\ref{pmix}) for $\la \psi(z,z^*) \ra_{\rm cl|op}$. This suggests
that the fields $\psi$ and $\psi_\t$ are distinct fields, and that $\psi$ has no
realization in the free fermionic theory. 

A further check of this statement is as follows. The two fields $\psi$ and $\psi_\t$
are different, but they have the same abstract conformal transformations. Namely, the
OPEs with the stress-energy tensors (\ref{Tpsi}) and (\ref{Tbarpsi}) are formally the
same for $\psi$ and $\psi_\t$. Consequently, the correlation functions $\la \mu(z_1)
\mu(z_2) \psi(z,z^*) \ra$ and $\la \mu(z_1) \mu(z_2) \psi_\t(z,z^*) \ra$ satisfy the
same differential equations, so that $\la \mu(z_1) \mu(z_2) \psi_\t(z,z^*) \ra$ must
be of the same general form (\ref{3param}), as derived in Section IX.A.

The general solution (\ref{3param}) involved three arbitrary parameters, $A_1,G$ and
$I_1$. For the $\psi$ corresponding to the heights 2, 3 and 4, we had required that
the limit $-z_1,z_2 \to \infty$, in which the boundary becomes fully closed, does not
produce the logarithmic singularity $\la \psi(z,z^*) \ra \log{z_{12}}$ coming from the
identity channel in the OPE $\mu(z_1)\mu(z_2)$,
\be
\lim_{-z_1,z_2 \to +\infty} z_{12}^{-1/4} \la \mu^{\rm op,cl}(z_1) \mu^{\rm
cl,op}(z_2) \psi(z,z^*) \ra = \lim_{-z_1,z_2 \to +\infty} \Big\{\la \omega_N(z_2)
\psi(z,z^*) \ra - {\log{z_{12}} \over \pi} \la \psi(z,z^*) \ra \Big\}.
\label{opecl}
\ee
This had implied $\la \psi(z,z^*) \ra = 0$, and the relation $A_1 = \lambda G$. This
condition cannot be imposed for $\psi_\t$ because the expectation value $\la
\psi_\t(z,z^*) \ra$ is not zero on the UHP. 

Thus for $\psi_\t$, the limit in (\ref{opecl}) does produce a logarithmic
singularity, but this singularity is in any case discarded since, according to our
prescription of inserting a field $\omega_N(\infty)$ for the closed boundary, we
are only interested in the $\omega_N$ channel of the OPE, namely the first term in
(\ref{opecl}). As it turns out, the limit (\ref{opecl}) also produces the divergent
terms $\log{|z_1-z|}$ and $\log{|z_2-z^*|}$, which correspond to the divergencies 
discussed above in terms of the free fermions. The rest, which forms the regularized
expectation value, should be a function of $z-z^*$, which, in order to conform with
the results for $\la \omega_N(\infty) \psi_\t(z,z^*) \ra_{\rm cl}$ and $\la
\psi_\t(z,z^*) \ra_{\rm op}$ established above, ought to contain a term $\log{|z-z^*|}$ with the same
sign as for the open boundary (corresponding to the limit $z_1,z_2 \to 0$). It is not
difficult to see that this condition imposes the relation $A_1 = -2\lambda G$,
reducing the general solution (\ref{3param}) to
\bea
\la \mu(z_1) \mu(z_2) \psi_\t(z_3,z_4) \ra &=& {z_{12}^{1/4} \over z_{34}^2} \Big\{
I_1 {x-2 \over \sqrt{1-x}} + {A_1 \over 4} {4-4x-3x^2 \over 1-x} 
- A_1 {x-2 \over \sqrt{1-x}} \log{\Big|{(1+\sqrt{1-x})^2 \over z_{34} \sqrt{1-x}}\Big|}
\nonumber\\
&& \hspace{-3cm} +\; A_1 \Big({1 \over z_{24}\sqrt{1-x}} - {\sqrt{1-x} \over
z_{23}}\Big)z_{34} + {A_1 \over 2} \Big({x \over z_{23}} - {x \over (x-1)z_{24}} +
{2z_{34} \over z_{13}z_{14}}\Big)z_{34} - 2 A_1 \Big\}.
\eea

The limit $z_1,z_2 \to 0$ yields, after the division by $z_{12}^{1/4}$, the
expectation value of $\psi_\t$ in front of an open boundary,
\be
\la \psi_\t(z,z^*) \ra_{\rm op} = {-1 \over (z-z^*)^2} \, \Big[2(I_1 - 2A_1\log{2})
+ A_1 + A_1 \log{|z-z^*|^2}\Big].
\ee
The limit $-z_1,z_2 \to +\infty$, after discarding the singular terms $\log{z_{12} 
\over z_{13}z_{24}}$ and dividing by $z_{12}^{1/4}$, yields the regularized
expectation value in front of a closed boundary,
\be
\la \psi_\t(z,z^*) \ra_{\rm cl,reg} = {-1 \over (z-z^*)^2} \, \Big[-2(I_1 -
2A_1\log{2}) - 8A_1 \log{2} + A_1 + A_1 \log{|z-z^*|^2}\Big].
\ee
Both expressions coincide with the results (\ref{thetaop}) and (\ref{thetacl})
obtained in the fermionic free theory if we choose $A_1 = -\half$ and $I = -2(I_1 -
2A_1 \log{2})$.

Moreover, the limit $z_1\to -\infty$ and $z_2 \to 0$ yields the expectation value
of $\psi_\t$ in front of a half-closed half-open boundary,
\be
\la \psi_\t(z,z^*) \ra_{\rm cl|op} = -{z+z^* \over |z|(z-z^*)^2} \,
\Big\{A_1 \log{\Big|{4z(z-z^*) \over (\sqrt{z}+\sqrt{z^*})^2}\Big|} - A_1
{(z-z^*)^2 - 4|z|^2 \over 4|z|(z+z^*)} + I_1 - 2A_1 \log{2}\Big\},
\ee
which, remarkably, exactly matches the form (\ref{thetamix}) obtained in the
$\t,\tit$ theory, for the same values of $A_1$ and $I$. Again we conclude that the
logarithmic field $\psi_\t$ in the fermionic theory and the field $\psi$ describing
the height variables $h>1$ correspond to two distinct fields.


\subsection{Representation theory}

In the previous subsection, we have shown that the field $\psi$ describing the
heights $h \geq 2$ in the sandpile model has no realization in terms of free fermions.
On the other hand, the fields $\phi$ and $\omega$ corresponding to the height 1
variable and to the insertion of dissipation in the bulk, do have such a
realization. Likewise the fields $\rho$ and $\bar\rho$ appearing in the conformal
transformations of $\psi$ can also be realized in the free fermionic theory. So all
the fields we need except the $\psi$ can be realized in terms of $\t,\tit$. This
peculiar situation can be understood from a representation theory point of view.
Indeed the conformal transformations typical of these fields allow for an infinite
number of inequivalent representations \cite{gk96}. We show here that the two
fields $\psi$ and $\psi_\t$ in fact belong to inequivalent representations. 

Let us first examine the situation in the free fermionic theory. For the sake of
clarity, we attach a suffix $\t$ to the fields constructed out from $\t$ and $\tit$.
So we take $\psi_\t = \t\tit (\p\t\bp\tit + \bp\t\p\tit)$ and its primary partner 
$\phi_\t = \p\t\bp\tit + \bp\t\p\tit$. The Wick contraction of $\t$ and $\tit$,
$\Cont{\t(z,\bar z) \, \tit}(w,\bar w) = -\log{|z-w|}$, implies the normalization of
the stress-energy tensor, $T_\t = 2\,\p\t\p\tit$.

{}From the OPE of $\psi_\t$ with $T$, we obtain 
\be
L_0 \psi_\t = \psi_\t - {1 \over 2} \phi_\t\,, \qquad L_1 \psi_\t = \rho_\t,
\ee
with $\rho_\t = {1 \over 2} (\t\bp\tit - \tit\bp\t)$ a field of weight $(0,1)$.
There is a second field $\bar\rho_\t = {1 \over 2} (\t\p\tit - \tit\p\t) = \bar L_1
\psi_\t$ arising from the right conformal transformations of $\psi_\t$. All these
fields belong to a reducible but indecomposable representation $\cal R$, which is
however a representation of an extended algebra \cite{gk}.

Clearly the two fields $\rho_\t$ and $\phi_\t$ are not independent but satisfy
\be
L_{-1} \rho_\t = {1 \over 2} \phi_\t.
\label{relrhophi}
\ee
The keypoint of this section is to show that, although a relation between $\rho_\t$ and 
$\phi_\t$ is to be expected in general, the value of the coefficient $\half$ is
particular to the triplet theory, of which the symplectic fermion theory is 
a realization.

In Section IX, we have computed the correlator $\la \mu(z_1) \mu(z_2) \psi(z_3,z_4)
\ra$ where $\psi$ is a field with the same conformal transformation as $\psi_\t$. The
general form of this correlator has been obtained on the sole basis of the conformal
transformations of the fields involved, namely (the first relation had been written
$L_0\psi = \psi + \lambda \phi$ but $\lambda$ was taken to be equal to $-\half$ later on)
\bea
L_0 \psi &=& \psi - {1 \over 2}  \phi\,, \qquad L_1 \psi = \rho, \qquad \bar L_1 \psi
= \bar \rho, \\
L_0 \rho &=& 0\,, \qquad L_1 \rho = 0\,, \qquad \bar L_0 \rho = \rho\,, \qquad \bar
L_1 \rho = \kappa {\mathbb I},
\eea
and similar relations for $\bar \rho$ (moreover $L_{-1}$ and $\bar L_{-1}$
act by differentiations). The important fact in these calculations is that we have not
used the identification of $L_{-1}\rho$ with a multiple of $\phi$.

The general solution for $\la \mu(z_1) \mu(z_2) \psi(z_3,z_4) \ra$ is given in 
(\ref{3param}). It depends on three arbitrary coefficients, $A_1,\, G,\, I_1$. The
derived three-point correlators $\la \mu(z_1) \mu(z_2) \rho(z_3,z_4) \ra$ and $\la
\mu(z_1) \mu(z_2) \phi(z_3,z_4) \ra$ are given in (\ref{rhopsi}) and (\ref{phipsi})
respectively, where we can use the values of the coefficients as given in
(\ref{param1}) and (\ref{param2}). One obtains
\bea
\la \mu(z_1) \mu(z_2) \rho(z_3,z_4) \ra &=& {z_{12}^{5/4} \over z_{14}z_{24}} 
\Big\{A_1 \, {\sqrt{1-x} \over x} + {A_1 \over 2} + A_1 \, {z_{24} \over
z_{12}}\Big\},\\
\noalign{\medskip}
\la \mu(z_1) \mu(z_2) \phi(z_3,z_4) \ra &=& G \, {z_{12}^{1/4} \over z_{34}^2}
{x-2 \over \sqrt{1-x}}.
\label{mumuphiG}
\eea

To see whether $L_{-1}\rho$ can be consistently identified with $\phi$, we simply
compute the partial derivative $\p_3$ of the first correlator to obtain
\be
\p_3 \, \la \mu(z_1) \mu(z_2) \rho(z_3,z_4) \ra = {A_1 \over 2} \, {z_{12}^{1/4}
\over z_{34}^2} {x-2 \over \sqrt{1-x}}.
\ee
A simple comparison with (\ref{mumuphiG}) implies that the general solution allows the 
following identification
\be
L_{-1} \rho = \beta \phi \qquad \hbox{with} \quad \beta = {A_1 \over 2G}.
\ee

In the interpretation of the height 2,3 and 4 variables of the sandpile model in
terms of the $\psi$ field, we have set $A_1 = \lambda G = -{G \over 2}$ in order
to avoid a logarithmic singularity when the two $\mu$ fields are sent off to infinity (see
(\ref{A1G})). This gives
\be
\beta_{ASM} = -{1 \over 4}.
\ee

In the previous subsection, where we made contact with the fermionic theory, we argued
that the correct choice to reproduce the fermion correlators is $A_1 = -2\lambda G =
G$. This then gives
\be
\beta_\t = {1 \over 2},
\ee
in agreement with the above relation (\ref{relrhophi}) obtained from straight calculations.

Since the normalizations of $\rho$ and $\phi$ are completely fixed by that of $\psi$, the
parameter $\beta$ cannot be scaled away, and this implies that different values of $\beta$
label inequivalent conformal representations \cite{gk96}. So there is a one-parameter
family of inequivalent Virasoro representations from which a field $\psi$ transforming
as above can be chosen. The specific $\psi$ that is relevant to the sandpile model
belongs to the representation with $\beta=-{1 \over 4}$, while the one realized in the
free fermion theory corresponds to $\beta=\half$. Because the coefficient $G$ can be
traded for $\beta$, the general solution of $\la \mu(z_1) \mu(z_2) \psi_\beta(z_3,z_4)
\ra$ for a fixed given value of $\beta$, depends on two arbitrary coefficients $A_1,I_1$, 
as expected (one for the norm of $\psi$, the other for the multiple of $\phi$ that
can be freely added to it). 

This explains why our general correlators involving $\rho$, $\bar\rho$ and $\phi$ (and
$\omega$) are separately consistent with what these fields would be in the
fermionic theory. But when we bring in the $\psi$, it automatically adjusts the
relative normalization of $\rho$ and $\phi$ in a way that is not consistent with the
fermionic picture, except if the parameter $\beta$ is fine-tuned, and taken equal
to $\half$.


\section{Bulk correlations of height variables}

In this last Section, we build on the knowledge we have of the scaling description
to assess further properties of the lattice height variables in the sandpile model. 
We will focus on the 2-site correlations for height variables on the plane, and
deduce their large distance limits. On the lattice, these correlations have never been
computed, so that the formulas which follow can be considered as predictions, or
conjectures.

If $P_{ij}(z_{12})$ denotes the joint probability that the height at $z_1$ be $i$ and
the height at $z_2$ be $j$, the 2-site correlations are given by
$P_{ij}(z_{12}) - P_iP_j$ for $i,j=1,2,3,4$. In the scaling regime, when the
distance $z_{12}$ is large, these correlations should be equal to expectation
values of pairs of fields $h_i(z_1,\bar z_1) h_j(z_2,\bar z_2)$. However the plane
has no boundary where sand can leave the system, so that the prescription we used
earlier requires to insert a bulk dissipation field $\omega(\infty)$ at infinity.
Thus one should have, in the scaling limit, that
\be
P_{ij}(z_{12}) - P_iP_j = \la h_i(z_1,\bar z_1) h_j(z_2,\bar z_2) \omega(\infty)
\ra, \qquad |z_{12}| \gg 1.
\ee
As the four height variables $h_i(z,\bar z)$ are linear combinations of the fields
$\phi$ and $\psi$, we clearly need to compute the (non-chiral) 3-point functions $\la
\psi \psi \omega\ra$, $\la \phi \psi \omega\ra$ and $\la \phi \phi \omega\ra$. We
concentrate on the first one, and let first the $\omega$ field be inserted at $z_3$.

The general form of $\la \psi(z_1,\bar z_1) \psi(z_2,\bar z_2) \omega(z_3,\bar z_3)
\ra$ can be computed from the Ward identities, but the nested procedure as we used
it in Section IX, is rather cumbersome. Instead it is simpler to use the finite
M\"obius transformations directly, by seeing the insertion points $z_1,z_2,z_3$ as
the images of a fixed triplet, say $w_1=-1, w_2=+1, w_3=0$. 

The conformal tranformation $w \to z$ of the plane onto itself which brings
$(w_1,w_2,w_3)=(-1,+1,0)$ onto three arbitrary points $(z_1,z_2,z_3)$ is
\be
w(z) = -{z_{12}(z - z_3) \over z_{23} (z-z_1) + z_{13}(z-z_2)},
\ee
under which the conformal transformations of $\psi$ and $\omega$ read
\bea
\psi(z,\bar z) &=& |w'(z)|^{2} \Big\{\psi(w,\bar w) 
- {1 \over 2} \log |w'(z)|^2 \, \phi(w,\bar w) \nonumber\\
&& \hspace{1cm} + \; {w''(z) \over 2w'^{2}(z)}\, \rho(w,\bar w)
+ {\bar w''(\bar z) \over 2\bar w'^{2}(\bar z)} \, 
\bar \rho(w,\bar w) + \kappa \, \Big|{w''(z) \over 2w'^{2}(z)}\Big|^2\Big\},\\
\omega(z,\bar z) &=& \omega(w,\bar w) - {1 \over 4\pi} \log |w'(z)|^2.
\eea

It is straightforward to compute the various derivatives, and express
$\psi(z_1,\bar z_1)$ as a linear combination of $\psi(-1,-1), \phi(-1,-1), \rho(-1,-1),
\bar\rho(-1,-1)$ and the identity, and similarly for $\psi(z_2,\bar z_2)$ and
$\omega(z_3,\bar z_3)$. Inserting these expressions into the 3-point function
yields $\la \psi(z_1,\bar z_1) \psi(z_2,\bar z_2) \omega(z_3,\bar z_3) \ra$ in
terms of fifty constants, given by various correlators of fields inserted at
$-1,+1,0$. In the limit $z_3 \to \infty$ however, most of these fifty constants
combine or drop out. Specifically, when $z_3 \to \infty$, the transformations of the
fields read
\bea
\psi(z_1, \bar z_1) &=& {4 \over |z_{12}|^2}\Big\{\psi(-1) + \log{\Big|{z_{12} \over
2}\Big|}\, \phi(-1) - \rho(-1) - \bar\rho(-1) - \kappa \Big\},\\
\psi(z_2, \bar z_2) &=& {4 \over |z_{12}|^2}\Big\{\psi(1) + \log{\Big|{z_{12} \over
2}\Big|}\, \phi(1) + \rho(1) + \bar\rho(1) + \kappa \Big\},\\
\omega(z_3,\bar z_3) &=& \omega(0) - {1 \over 2\pi} \log{\Big|{z_{12} \over
2z_{13}z_{23}}\Big|}.
\eea

Using these, one sees that the 3-point function retains a logarithmic singularity
in the limit $z_3 \to \infty$, unless the following identities hold (among others)
\be
\la \psi(z_1, \bar z_1) \psi(z_2, \bar z_2) \ra = 
\la \phi(z_1, \bar z_1) \psi(z_2, \bar z_2) \ra = 
\la \phi(z_1, \bar z_1) \phi(z_2, \bar z_2) \ra = 0.
\ee
The 3-point function then takes the form
\bea
\la \psi(z_1,\bar z_1) \psi(z_2,\bar z_2) \omega(\infty) \ra &=& 
{16 \over |z_{12}|^4} \left\la \Big[\psi(-1) + \log{\Big|{z_{12} \over
2}\Big|}\, \phi(-1) - \rho(-1) - \bar\rho(-1) - \kappa \Big]\right.\nonumber\\
&& \hspace{-1cm} \left. \Big[\psi(1) + \log{\Big|{z_{12} \over
2}\Big|}\, \phi(1) + \rho(1) + \bar\rho(1) + \kappa \Big] \, \omega(0)
\right\ra \\
&& \hspace{-3cm} \equiv {1 \over |z_{12}|^4} \Big\{C + 2B \log{|z_{12}|} + A
\log^2{|z_{12}|} \Big\},
\eea
where the coefficients $A$ and $B$ are given by
\bea
A &=& 16\, \la \phi(-1) \phi(1) \omega(0) \ra,\\
B &=& 16\, \la \phi(-1) [\psi(1) - \phi(1) \log{2} + \rho(1) + \bar\rho(1)]
\omega(0) \ra.
\eea

The same procedure with one $\psi$ replaced by a $\phi$, or the two $\psi$'s
replaced by $\phi$'s, yields
\bea
\la \phi(z_1,\bar z_1) \psi(z_2,\bar z_2) \omega(\infty) \ra &=& 
{16 \over |z_{12}|^4} \left\la \phi(-1) \Big[\psi(1) + \log{\Big|{z_{12} \over
2}\Big|}\, \phi(1) + \rho(1) + \bar\rho(1) \Big] \, \omega(0)
\right\ra \nonumber\\
&=& {1 \over |z_{12}|^4} \Big\{B + A \log{|z_{12}|} \Big\},
\eea
and
\be
\la \phi(z_1,\bar z_1) \phi(z_2,\bar z_2) \omega(\infty) \ra = 
{16 \over |z_{12}|^4} \la \phi(-1) \phi(1) \omega(0) \ra = {A \over |z_{12}|^4}.
\label{ppo}
\ee

{}From these results and the way the height variables are expressed in terms of $\phi$
and
$\psi$, namely $h_i(z,\bar z) = \alpha_i \psi(z,\bar z) + \beta_i \phi(z,\bar z)$,
one easily obtains the scaling form of the 2-site correlations,
\bea
P_{ij}(z_{12}) - P_iP_j &=& {1 \over |z_{12}|^4} \Big\{\a_i\a_jC +
(\a_i\b_j+\b_i\a_j)B + \b_i\b_j A \nonumber\\
&& \hspace{.5cm} +\; [2\a_i\a_j B + (\a_i\b_j+\b_i\a_j)A] \log{|z_{12}|} +
\a_i\a_j A \log^2{|z_{12}|}\Big\}.
\label{pijgen}
\eea

The value of $A$ is related, through (\ref{ppo}), to the correlation of the two height 1 
variables on the plane, whose dominant term was computed in \cite{majdhar},
\be
P_{11}(r) = P_1^2 - {P_1^2 \over 2r^4} + ...
\ee
This fixes the constant $A=-P_1^2/2$. The other two constants $B$ and $C$ cannot be
determined purely from conformal theoretic arguments, and require further input
from lattice calculations.

The value of $A$ alone however determines the dominant terms of all correlations,
and yields
\be
P_{1i}(r) - P_1P_i \simeq -{\a_i P_1^2 \over 2r^4} \, \log{r}, \qquad
P_{ij}(r) - P_iP_j \simeq -{\a_i\a_j P_1^2 \over 2r^4} \, \log^2{r}, \qquad
i,j=2,3,4,
\ee
where the values of $\a_i$ have been given in Section VIII, namely $\a_2=1$,
$\a_3=2.12791$ and $\a_4=-3.12791$. The heights are found to be anticorrelated except
the height 4 variable, which has a positive correlation with the other three (but is
anticorrelated with itself).

In order to check these behaviours and to assess the values of $B$ and $C$, we have
run further numerical simulations of the model. The main problem here is that the
correlation term $\sim r^{-4}$ in the joint probabilities $P_{ij}(r)$ is largely
subdominant with respect to the trivial constant piece $P_iP_j$. It means that we
cannot test these scaling behaviours in the regime where they are supposed to hold
($r \gg 1$), since that would require a huge numerical precision on the data. We have
nevertheless tested the scaling formulas for small and intermediate values of $r$ ($r
\leq 20$), and found very good agreement, in some cases for as small distances
as $r=2$. Also the fitted values of $B,C$ are consistent for all fits.

We have run the simulations on a cylindrical lattice, of height $L=50$ and perimeter
$P=200$. We have sampled a total of $3.5 \times 10^9$ recurrent configurations, and
for each configuration, we have measured the heights on the six circles located at
mid-height. On each circle, and for each value of $r$ between 2 and 20, we have
recorded all pairs of heights which are $r$ sites apart, and subsequently computed
the occurrence probabilities $P_{ij}(r)$. Thus $r$ is a distance which was always
directed along the perimeter of the cylinder. 

\begin{figure}[htb]
\psfrag{P13}[Bl][Bl][.8][0]{$P_{13}-P_1P_3$}
\psfrag{P34}[Bl][Bl][.8][0]{$P_{34}-P_3P_4$}
\psfrag{P33}[Bl][Bl][.8][0]{$\!P_{33}-P_3^{\;2}$}
\psfrag{P44}[Bl][Bl][.8][0]{$P_{44}-P_4^{\;2}$}
\mbox{\hspace{-15pt} \includegraphics[scale=0.83]{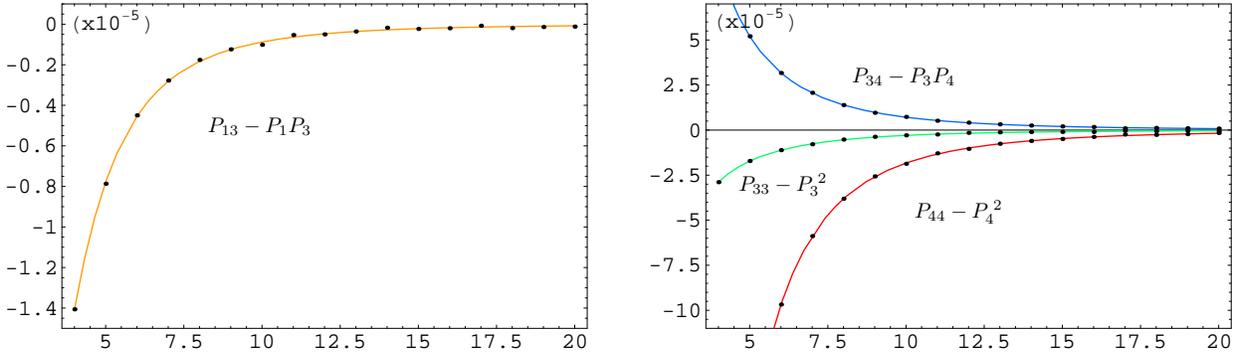}}
\caption{Two-site correlations of height variables on a cylindrical lattice (height
$L=50$, perimeter $P=200$), measured along the perimetrical circles located at mid-height.
The abscissa represents the separation distance, in lattice sites. The dots correspond
to data obtained from numerical simulations, and the solid curves represent the fitted
functions, given in (\ref{pijgen}).}
\label{Figure:pijsimul}
\end{figure}

Once the joint probabilities $P_{ij}(r)$ are obtained, we fit them (leaving out the
data for $r=2,3$) to the form given by (\ref{pijgen}), that is, a constant
contribution $P_iP_j$ and terms $\log^k{r}/r^4$, for $k=0,1,2$ depending on the values
of $i,j$. Then, using the exact values of the coefficients $\a_i$ and $\b_i$, the
values of $P_iP_j,A,B$ and $C$ could be extracted (note that, as a further check, we
did not use the known value of $P_iP_j$ and $A$). As expected, better results are
obtained for larger values of $i,j$. The results for $P_{13}, P_{33}, P_{34}$ and
$P_{44}$, as well as the fitted curves, are shown in Fig. \ref{Figure:pijsimul}.

\bigskip

As a first qualitative check, we see that the signs of the correlations are as
predicted, namely negative, except for the height 4 with a strictly smaller height
($P_{34}$ here). 

As mentioned above, the agreement is excellent, and is so, somewhat unexpectedly,
down to very small values of $r$ ($r=4$ and even smaller). For $P_{13}(r)$, the fitted
values of the subtraction term $P_1P_3$ and of $A$ and $B$ were found to be (no $\log^2
r$ term was included in this case)
\be
P_1P_3 = 0.02261, \qquad A=-0.002648, \qquad B=-0.004861.
\ee
The first two are consistent with the exact values $P_1P_3\Big|_{\rm exact}=0.02255$
and $A_{\rm exact}=-0.002711$.

The fits of $P_{33}(r), P_{34}(r), P_{44}(r)$ yield the following values,
\bea
P_3P_3 &=& 0.09410, \qquad A=-0.002821, \qquad B=-0.004718, \qquad C=-0.008883,\\
P_3P_4 &=& 0.1366, \qquad A=-0.002517, \qquad B=-0.004329, \qquad C=-0.008773,\\
P_4P_4 &=& 0.1982, \qquad A=-0.002699, \qquad B=-0.004065, \qquad C=-0.009608.
\eea
Again the values of the subtraction terms compare well with the exact values,
$P_3^2\Big|_{\rm exact}=0.09381,\, P_3P_4\Big|_{\rm exact}=0.1367,\, 
P_4^2\Big|_{\rm exact}=0.1991$, and the fitted values for $A$ are also comparable
to the exact value quoted above. The values of $B$ and $C$ obtained from these four
(or three) independent fits are all close to each other, from which we infer that the
exact values should be around
\be
B \sim -0.0045, \qquad C \sim -0.009.
\ee
As far as we can see, these values can only be confirmed by exact calculations of
correlations on the lattice. 


\section{Conclusions}

Let us first summarize our main results. As a preliminary but crucial step, we have
carried out the calculation of the 1-point probabilities of the four height variables
on the discrete upper-half plane. This was long and technical, but gave us an exact
formula for their scaling term $m^{-2}$, where $m$ is the distance to the boundary.
As 1-point functions on the upper-half plane are 2-point correlators, this 
enabled us to obtain our second main result, namely, the field identification of the
four bulk height variables in the scaling limit, in terms of conformal fields of a
$c=-2$ logarithmic conformal theory. The picture that emerges is that the height 1
variable turns out to be a primary field, whereas the heights 2, 3 and 4
are all proportional to the logarithmic partner of this primary field (in a rank 2
Jordan cell). 

This should constitute a major step towards the exact solution of the Abelian
sandpile model, since, if the conformal picture we propose is confirmed, it
basically solves the aspects of the model concerned with the height variables, which
are the most natural microscopic variables. (Indeed the simpler problem of
identifying the scaling fields corresponding to the boundary height variables on open
and closed boundaries was solved in \cite{jeng2,piru2}.) It however does not solve the
part of the model which is related to other non-local variables, like avalanche
observables, whose exact solution remains a challenge. 

Before getting there, we believe that a number of issues deserve further
investigations, which lie in the context of the present article, and from which the
sandpile model but also logarithmic conformal field theory in general could benefit.

One of them concerns the unexpected nature of the logarithmic field describing the
heights 2, 3 and 4. We have argued that it has no realization in the free symplectic
fermion field theory, widely considered as the canonical local realization of a
$c=-2$ conformal theory. One certainly would like to know to what extent the sandpile
realization and the symplectic fermion realization differ, and what the specific
properties of the theory realized by the sandpile model are. 

\acknowledgments
We would like to thank Matthias Gaberdiel for very useful and encouraging discussions. 
The hospitality of the ETH Z\"urich is also gratefully acknowledged. M.J. was 
supported by a Southern Illinois University Edwardsville Summer Research
Fellowship. P.R. is financially supported by the Belgian Fonds National de la
Recherche Scientifique.


\end{document}